\documentclass[sn-mathphys,Numbered]{sn-jnl}
\usepackage{hyperref}
\usepackage{listings}%
\usepackage{graphicx}%
\usepackage{xcolor}%
\usepackage{multirow}%
\usepackage{amsmath,amssymb,amsfonts}%
\usepackage{amsthm}%
\usepackage{mathrsfs}%
\usepackage[title]{appendix}%
\usepackage{textcomp}%
\usepackage{manyfoot}%
\usepackage{booktabs}%
\usepackage{algorithm}%
\usepackage{algorithmicx}%
\usepackage{algpseudocode}%
\usepackage{comment}%
\usepackage{ulem}

\usepackage[scaled=0.85]{beramono}
\usepackage[caption=false]{subfig}
  \graphicspath{{./figures/}}
 \title[Article Title]{Image-To-Mesh Conversion for Biomedical Simulations}
 \author*{\fnm{Fotis} \sur{Drakopoulos*}}\email{fdrakopo@gmail.com}
  \author{\fnm{Kevin} \sur{Garner}}\email{kgarn006@odu.edu}
  \author{\fnm{Christopher} \sur{Rector}}\email{chrisrector14@gmail.com}
  \author*{\fnm{Nikos} \sur{Chrisochoides*}}\email{npchris@gmail.com}
  
 \affil{\orgdiv{Center for Real-Time Computing, Department of Computer Science}, \orgname{Old Dominion University}, \orgaddress{
 \city{Norfolk}, 
 \state{VA}, \country{United States of America}}}

\begin{document}

\abstract{Converting a three-dimensional medical image into a 3D mesh that satisfies both the quality and fidelity constraints of predictive simulations and image-guided surgical procedures remains a critical problem. Presented is an image-to-mesh conversion method 
called CBC3D. It first discretizes a segmented image by generating an adaptive Body-Centered Cubic (BCC)
mesh of high-quality elements. Next, the tetrahedral mesh is converted into a mixed-element mesh of tetrahedra, pentahedra, and hexahedra to decrease element count while maintaining quality. Finally, the mesh surfaces are deformed to their corresponding physical image boundaries, improving the mesh's fidelity. The deformation scheme builds upon the ITK open-source library and is based on the concept of energy minimization, relying on a multi-material point-based registration. It uses non-connectivity patterns to implicitly control the number of extracted feature points needed for the registration and, thus, adjusts the trade-off between the achieved mesh fidelity and the deformation speed. We compare CBC3D with four widely used and state-of-the-art homegrown image-to-mesh conversion methods from industry and academia. Results indicate that the CBC3D meshes (i) achieve high fidelity, (ii) keep the element count reasonably low, and (iii) exhibit good element quality.
}

\keywords{Medical imaging, Image-to-mesh conversion, Segmentation, Mesh generation}

\maketitle

\section{Introduction}\label{introduction}
Image-to-mesh conversion algorithms are widely used for the quantitative analysis of patient-specific images 
using the Finite Element (FE) method. This has significant implications in many areas, 
such as image-guided therapy \cite{2006IntegrationPatientModeling, TalosAndArchip2007, ArcBRSIMRI07, frontiersAPBNRR}, interactive surgery simulation for training young clinicians \cite{CBC3D2015},
and endovascular flow diversion for aneurysm geometries \cite{Ronak1, Ronak2, Kazakidi1}. The FE method is essential in modeling tissue deformation for these applications. 
An intrinsic difficulty of generating meshes from isosurface-based data (i.e., parametric surfaces/volumes, level-sets, segmented multi-labeled images) 
is the processing and recovery of the object geometry. General-purpose mesh generators
(for solid and geometric modeling applications) expect that the object boundary is either parameterized (i.e., it is defined using constructive solid geometry primitives) or 
explicitly defined (e.g., through the boundary discretization) as a collection of patches. 
To convert the isosurface data into a tetrahedral mesh, one can 
either (i) recover the parametrized object surface and follow up with a conventional mesh generation 
technique, or (ii) use a mesh generation method, which operates directly on this data type.

We propose an image-to-mesh conversion method that has been tested in generating anatomically correct models of cerebral aneurysms (to be used in predicting the efficacy of endovascular flow diversion designs) and in creating Arteriovenous Malformation (AVM) models for surgical simulations. Medical surgical simulations are classified into two main categories - predictive and interactive. A predictive simulation predicts and optimizes the outcome of an intervention by using patient-specific, pre-operative image data. Predictive simulations require high geometric, topologic, and material fidelity, where mesh generation likely takes place offline for the intervention. On the other hand, interactive simulations offer training within virtual environments to allow surgical residents to achieve certain skill levels without imposing risks on patients. Interactive simulations involved with deforming solids are widely used in physically-based simulations. 
In particular, the design and development of surgical simulations require accurate and efficient simulation tools due to the real-time computation requirements and physics fidelity. However, because of the complexity of the problem, these simulations are computationally intensive, making interactive and real-time constraints 
a challenging problem. Broadly, the main components for most interactive medical simulators require knowledge in the following areas: biomechanical and 
anatomical modeling, collision detection, haptics, and visualization. A virtual anatomical model constitutes a precise computerized description of a human organ, commonly generated from medical images, like Magnetic Resonance Imaging (MRI) or Computed Tomography (CT). 
For example, an AVM is a pathology containing draining vessels, feeding arteries, the nidus of vessel bundles, and surrounding structures.
The anatomical model should contain the minimal elements to describe the small vessel features. We focus on mesh generation requirements of predictive and interactive simulations - fidelity and quality. Both smoothness and the real-time requirement will be addressed elsewhere in the future. Fidelity concerns the degree to which the mesh surface aligns with an image boundary. Quality is determined by the shape and size of elements, as this also affects the accuracy of solutions for Computational Fluid Dynamics (CFD) simulations \cite{babuvska1976angle, meshEvaluationNRR}.

Image-to-mesh (I2M) conversion is an inherently challenging problem as it must balance trade-offs between several criteria: 
(i) minimizing the mesh size and maximizing the element quality by utilizing proper mesh gradation and tissue-specific multi-resolution, 
(ii) without sacrificing geometric and topologic fidelity, (iii) while maintaining a smooth surface to reflect a certain degree of visual reality, 
and (iv) simultaneously allowing the upstream numerical computations for collision detection and biomechanical FE analysis to be completed within the time constraints imposed by a surgical simulation.

We present a method called CBC3D that directly converts segmented multi-labeled image data 
into adaptive multi-tissue isotropic tetrahedral meshes. This method builds upon previous work \cite{CBC3DFedorov, CBC3DYixun, CBC3D2015}, maintaining the ability to generate meshes of good quality that were shown to enhance non-rigid registration solver performance and reduce error when compared to other image-to-mesh conversion methods \cite{meshEvaluationNRR}. CBC3D initially relabels image voxels to manage image noise and eliminate non-manifold voxel connectivity. It then discretizes the segmented, labeled image with a uniform BCC lattice of high-quality tetrahedra. Red-green templates subdivide the lattice near the segmented boundaries while ensuring mesh conformity (i.e., manifold connectivity between mesh regions representing different tissues).
Finally, the generated surfaces are deformed to their corresponding tissue boundaries to improve fidelity while maintaining quality.

The presented procedure focuses on several of the above simulation requirements, offering the following:
(i) It provides an accurate geometric and topological representation of the medical cases in our evaluation,
(ii) it provides material-dependent mesh resolution to reduce element count (i.e., fidelity can be specified per material, which will determine the level of adaptivity near mesh boundaries, allowing for a localized refinement in materials of interest)
(iii) it maintains good element quality during mesh deformation,
(iv) it further reduces memory and CPU requirements for the solver by introducing mixed elements,
and (v) it improves the overall reliability and portability of the code as it builds 
upon the ITK open-source, cross-platform system\footnote{https://itk.org}.
A single-material version of CBC3D is available within the 3D Slicer\footnote{https://www.slicer.org} package for visualization and image analysis.
A previous multi-material version of CBC3D \cite{CBC3D2015} is integrated within an interactive simulator for neurosurgical procedures involving 
brain AVM developed in SOFA\footnote{https://www.sofa-framework.org}, 
a framework for real-time medical simulations.

\section{Related Work} \label{related_work}
Several methods use medical data as input to create either isotropic or anisotropic 2-dimensional (2D) or 3D meshes. 2D mesh generation from medical imaging data \cite{CurvilinearDiscretization2015, AutomaticCurvilinear2014} is outside the scope of this paper. We focus on 3D medical images and 3D meshes. For a comprehensive overview of these 3D methods, see \cite{ZhangBook2016}.
Anisotropic mesh generation methods \cite{ClericiAnisotropic2020, AnisotropicFluidSolidI2M2010} are well suited for capturing directional information and high curvature geometries (e.g., hemodynamics). In certain cases, anisotropy offers an improvement over isotropy due to its lower number of (high aspect ratio) elements that provide high accuracy for a segmented image, compared to isotropic meshes, which may require more elements. Anisotropic meshing techniques are also suitable for flow problems like modeling cerebral aneurysms for fluid-structure interaction (FSI) simulations \cite{AnisotropicCerebralAneurysms2013}. However, this particular method processes geometries generated from the medical images elsewhere and do not specifically process image data. Qualitative and performance evaluations of sequential methods that process surface meshes (also generated elsewhere from medical image data) as input can be found in \cite{AssessmentMeshGen2008}, with regards to suitability for surgical simulations and non-rigid registration applications \cite{MeshGenNRR2008}. Although CBC3D generates isotropic meshes, it reduces mesh size (element count) by converting its tetrahedral meshes into mixed-element meshes (made of tetrahedra, pentahedra, and hexahedra). CBC3D is tested on both CT and MRI data, where different types of isotropic methods like Advancing Front \cite{ParallelAdvancingFront2007} or Terminal-Edge Bissection Methods \cite{ Rivara04P, ParallelTerminalEdge2006} were considered. 

Some 3D isotropic methods utilize a Delaunay refinement scheme \cite{Boltcheva, Pons2007,Fuchs20011400, Shewchuk:1998,AnisotropicFluidSolidI2M2010, rineau2007generic, cheng2000silver,foteinos2014high, Feng2017Scalable3H, 4DPODM2014}. In our evaluation, presented in section \ref{evaluation}, we compare two such methods to CBC3D - CGAL's 3D Mesh Engine\footnote{https://doc.cgal.org/4.5.2/Manual} and Parallel Optimistic Delaunay Method~\cite{Foteinos10G, Foteinos11H, Foteinos12H, Foteinos12D,foteinos13high, foteinos2014high} (referred to throughout this paper in short as PODM). The code prototype for CGAL, including its initial design and specifications, is presented in \cite{rineau2007generic}. CGAL first constructs a boundary mesh from the image and then generates a volume mesh given the boundary mesh as input. Before refinement, a mechanism of protecting balls is set up on 1-dimensional features, if any, to ensure a fair representation of those features in the mesh. This mechanism also guarantees the termination of the refinement process, independently of the input geometry. 
The criteria concerning the size or shape of mesh cells and surface facets drive the refinement process. It terminates when no more mesh cells or surface facets violate the criteria.  A mesh optimization phase follows the refinement \cite{cheng2000silver} that attempts to remove slivers.

On the other hand, PODM constructs the isosurface of the biological object with geometric and topological guarantees while simultaneously providing good-quality surface and volume elements. PODM implements a tightly coupled, shared-memory, parallel speculative 
execution approach using carefully designed contention managers, load balancing, synchronization, and optimization schemes. In \cite{4DPODM2014}, the robustness of PODM was extended to 4D segmented images, providing faithful geometric and topological approximations (when the hyper-surface is a closed smooth manifold). However, it suffered in runtime since its space-time meshing method is sequential and requires increased memory space as opposed to the 3D meshing method. A problem with Delaunay refinement is that almost flat tetrahedra, called slivers, can survive known heuristics to remove them \cite{Boltcheva, cheng2000silver, Chew:1997}. Sliver removal for Delaunay-based methods is still an open problem. On the other hand, CBC3D utilizes a lattice-based approach to generate high-quality elements rather than a Delaunay scheme.


A large number of meshing methods are based on lattice space-tree (octree) decomposition \cite{LatticeCleaving, ZhangAutomaticMultimaterial2010, Zhang3DFinite2005, Labelle:2007:ISF, Zhang20055083, LD_Chernikov, ZhangChallenges2013, Zhang2012166, ZhangDualContour2014, DualContour3DSegmentationAnisotropic2018, ParallelDC2013, Radovitzky2000543, Molino03acrystalline}, which provides adaptively-sized, high-quality elements for several medical applications. Two of these methods, CLEAVER \cite{LatticeCleaving} and Lattice De-refinement (LD) \cite{LD_Chernikov}, are compared to CBC3D in our evaluation. CLEAVER first constructs a high-quality Body-Centered Cubic (BCC) lattice that covers the input image. Then, it recovers the surface by finding the points where the lattice intersects the object. Subsequently, it warps the mesh vertices to the model surface or inserts new vertices on the surface. The resultant mesh quality depends on the thresholds used to determine whether to warp or 
insert a vertex. Lattice De-refinement (LD) applies a mesh decimation step to an octree-based mesh to improve the element shape and/or modify the mesh size. This method allows for guaranteed bounds on the smallest dihedral angle and the distance between the segmented surface and the internal tissues' (materials') boundaries. The number of tetrahedra in the mesh is as few as possible, provided the quality and distance requirements are satisfied. Instead of performing a decimation step, other approaches improve element quality by optimizing an objective function based on tetrahedral shape measures, such as the Dual Contouring technique \cite{Freitag02acomparison, Zhang2012166, ZhangDualContour2014, DualContour3DSegmentationAnisotropic2018, ParallelDC2013}. Additionally, some methods warp the surfaces of the lattice to 
the boundaries of the object using either a mass-spring system, an FE constitutive model, or an optimization scheme \cite{Molino03acrystalline, CBC3DYixun, Fuchs20011400}.

CBC3D builds upon previous work \cite{CBC3DFedorov, CBC3DYixun, Molino03acrystalline}. The presented method extends earlier work that was used and tested only on fairly regular multi-material geometries. Both the current extension and earlier work employ a physics-based mesh deformation scheme to warp the surfaces of a Body-Centered Cubic (BCC) mesh to the physical image boundaries. The deformation scheme optimizes for fidelity and quality. Additional improvements for the reliable and accurate representation of multiple materials and their corresponding resolutions were presented in \cite{CBC3D2015}, an earlier version of CBC3D. The current version of  CBC3D improves the geometric and topologic accuracy of the methods presented in \cite{CBC3DFedorov, CBC3DYixun, CBC3D2015}.  Lattice subdivision uses a Euclidian Distance Transform (EDT) \cite{maurer2003linear} instead of a segmented image, allowing smoother refinement within a threshold from the segmented image boundaries.
The method eliminates non-manifold voxel connectivities in the segmented image to provide a more accurate construction of the object. 
When the lattice subdivision is complete, the disconnected mesh regions or non-manifold mesh connectivities are detected, and the method subdivides the lattice further (locally or globally) to resolve the image features.

The presented method also provides a material-dependent lattice subdivision to focus only on materials of interest, thus 
reducing the number of generated elements. Alternatively, global subdivision criteria can be specified. 
A physics-based deformation scheme eliminates poorly shaped elements with heuristics and quality criteria, such as a 
minimum dihedral angle or a scaled Jacobian metric. The trade-off between mesh fidelity and deformation time is balanced 
with the extraction of different amounts of image features needed for deformation. Mixed elements, including tetrahedra, pentahedra, and hexahedra, are introduced to reduce the number of vertices in the adaptive isotropic tetrahedral mesh. The evaluation shows that a mixed mesh can reduce the number of vertices by up to 30\% without compromising the tetrahedral mesh's fidelity. Finally, this work utilizes parallel computing (for some compute-intensive kernels) to improve overall performance and allow for faster processing of large data (e.g., micro-CT imaging of stents). Although multi-threading is used to improve the performance of the method, CBC3D does not quality (yet) as a parallel mesh generation method. 

Section \ref{method} describes the CBC3D method in detail. 
We evaluate CBC3D in section \ref{evaluation} and compare it to four common image-to-mesh conversion methods found in industry and academia: CGAL's 3D  Mesh Engine\footnote{https://doc.cgal.org/4.5.2/Manual} (v4.5.2), CLEAVER \cite{LatticeCleaving} (v1.5.4),
Lattice De-refinement (LD) \cite{LD_Chernikov}, and PODM \cite{foteinos2014high, Feng2017Scalable3H}. We provide a discussion and address future work to further improve CBC3D in section \ref{discussion}. We finally conclude in section \ref{conclusion}.

\section{Method} \label{method}
CBC3D uses a segmented, multi-labeled image as input and creates an adaptive tetrahedral or mixed element mesh as an output. 
Subsection \ref{segmentation} describes the segmentation algorithm used to obtain the labeled image.
Subsection \ref{preprocessing} describes pre-processing algorithms used to improve the quality of the input image.
Subsection \ref{BCCMeshGenerationandRefinement} describes the generation of the adaptive BCC lattice.
Subsection \ref{MixedMesh} describes the conversion of the adaptive tetrahedral mesh into a mixed element mesh.
Subsection \ref{MeshDeformation} describes mesh deformation.

\subsection{Segmentation}\label{segmentation}
\begin{figure*}[htb]
\centering	
\subfloat[]{
\label{StentOne}
\includegraphics [width=0.203\textwidth]{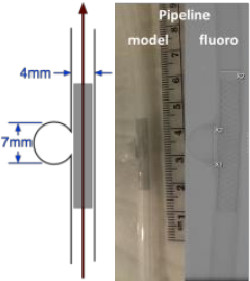}}
\subfloat[]{
\label{StentTwo}
\includegraphics [width=0.22\textwidth]{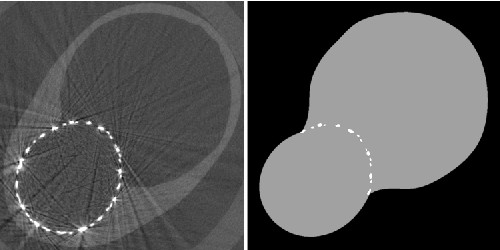}}
\subfloat[]{
\label{StentThree}
\includegraphics [width=0.22\textwidth]{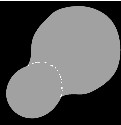}}
\subfloat[]{
\label{StentFour}
\includegraphics [width=0.303\textwidth]{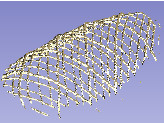}}
\caption{Raw micro-CT slice for pipeline stent model \cite{Ronak2} (a) before (b) and after segmentation with the device wires cropped (c). (d) depicts a volume rendering of the segmented stent. The segmented image has a spacing of $0.012\times0.012\times0.024\; \text{mm}^3$ and a size of $1001\times1001\times4421\; \text{voxels}^3$.}
\label{Stent}
\end{figure*}

The level set segmentation algorithm for the vessel structures uses the Vascular Modeling Toolkit (VMTK) \cite{Antiga}. 
Level sets are a kind of deformable model in which the deformable surface is represented by a 3D function whose contour at level zero is 
the surface in question. To extract the surface from the image, the output image is run through Marching Cubes \cite{Lorensen:1987} with level zero.
A fast marching initialization is used, consisting of placing a set of seeds and a set of targets in the image.
A front is then propagated from the seeds until the first target is met, at which point the region covered by the front is the initial deformable model. 
This type of initialization is effective when it is necessary to segment round objects such as 
aneurysms (e.g. by simply placing one seed at the center and one target on the wall, the volume will be initialized).
The output segmented image contains labels, where each one represents a single tissue (Figure \ref{Stent}).

The Materialise Mimics\footnote{http://www.materialise.com/en/medical/software/mimics} software is used 
to perform image segmentation on the micro-CT data of commercially available neurovascular stents \cite{Ronak1}. 
Two different labels are tagged onto the lumen of the sidewall aneurysm model and the flow diverter.
During segmentation, the device is cropped to only retain the portion providing neck coverage of the aneurysm (Figure \ref{Stent}).

\begin{figure}[htb]
\centering	
\subfloat[]{
\label{DoeOne}
\includegraphics [width=0.259\textwidth]{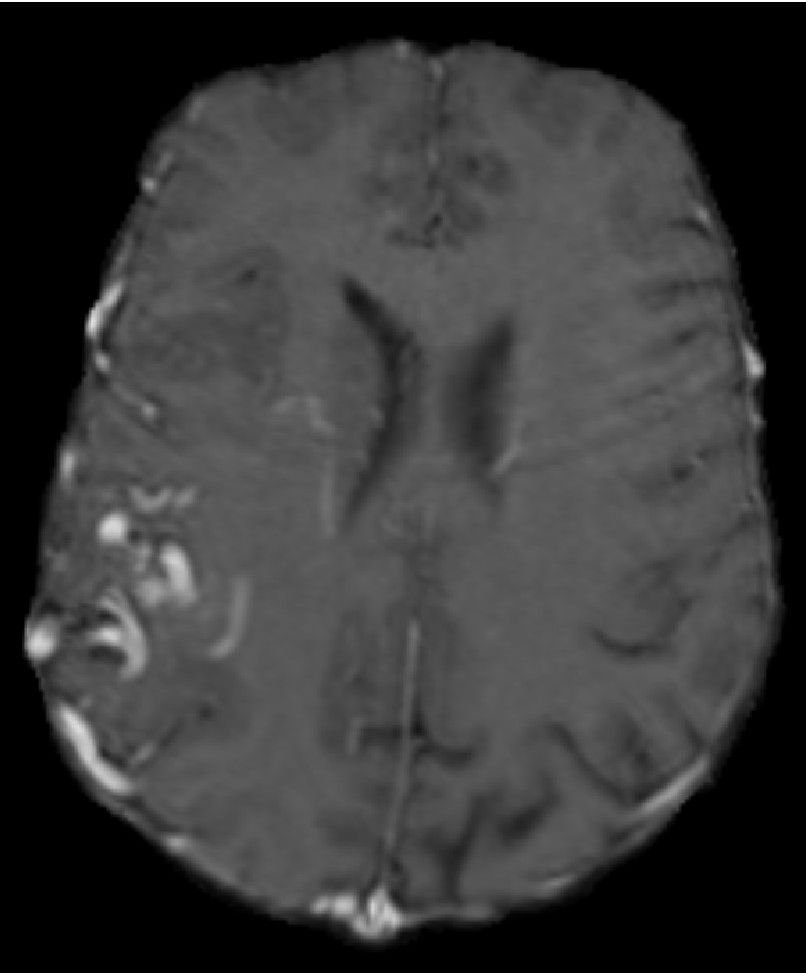}}
\subfloat[]{
\label{DoeTwo}
\includegraphics [width=0.259\textwidth]{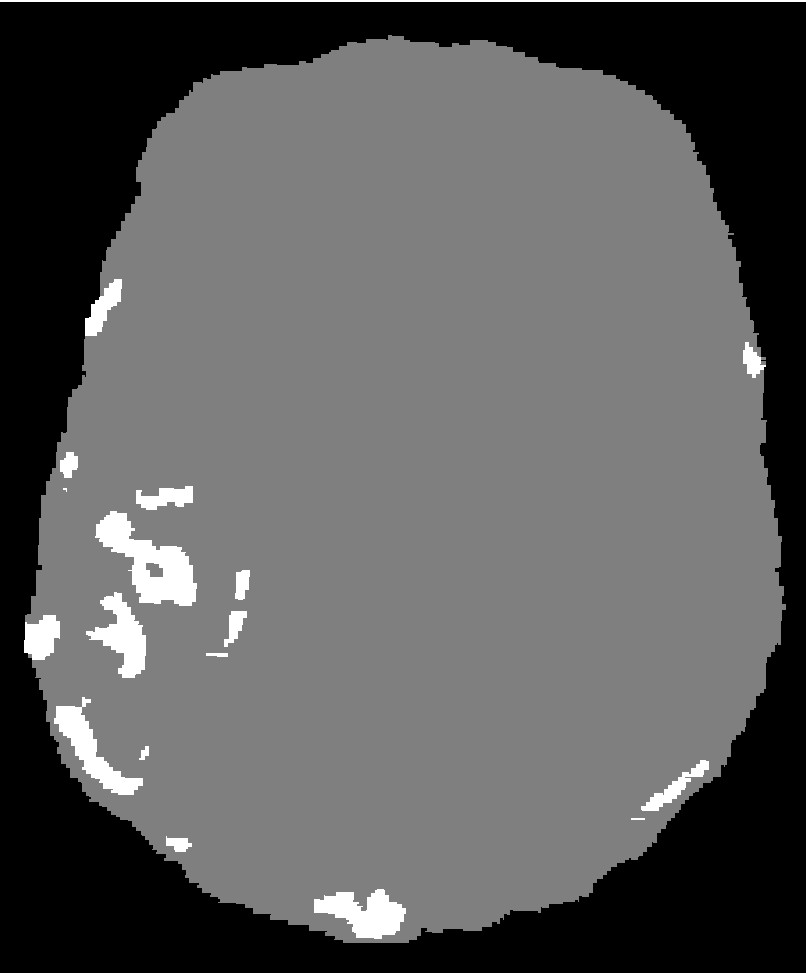}}
\subfloat[]{
\label{DoeThree}
\includegraphics [width=0.42\textwidth]{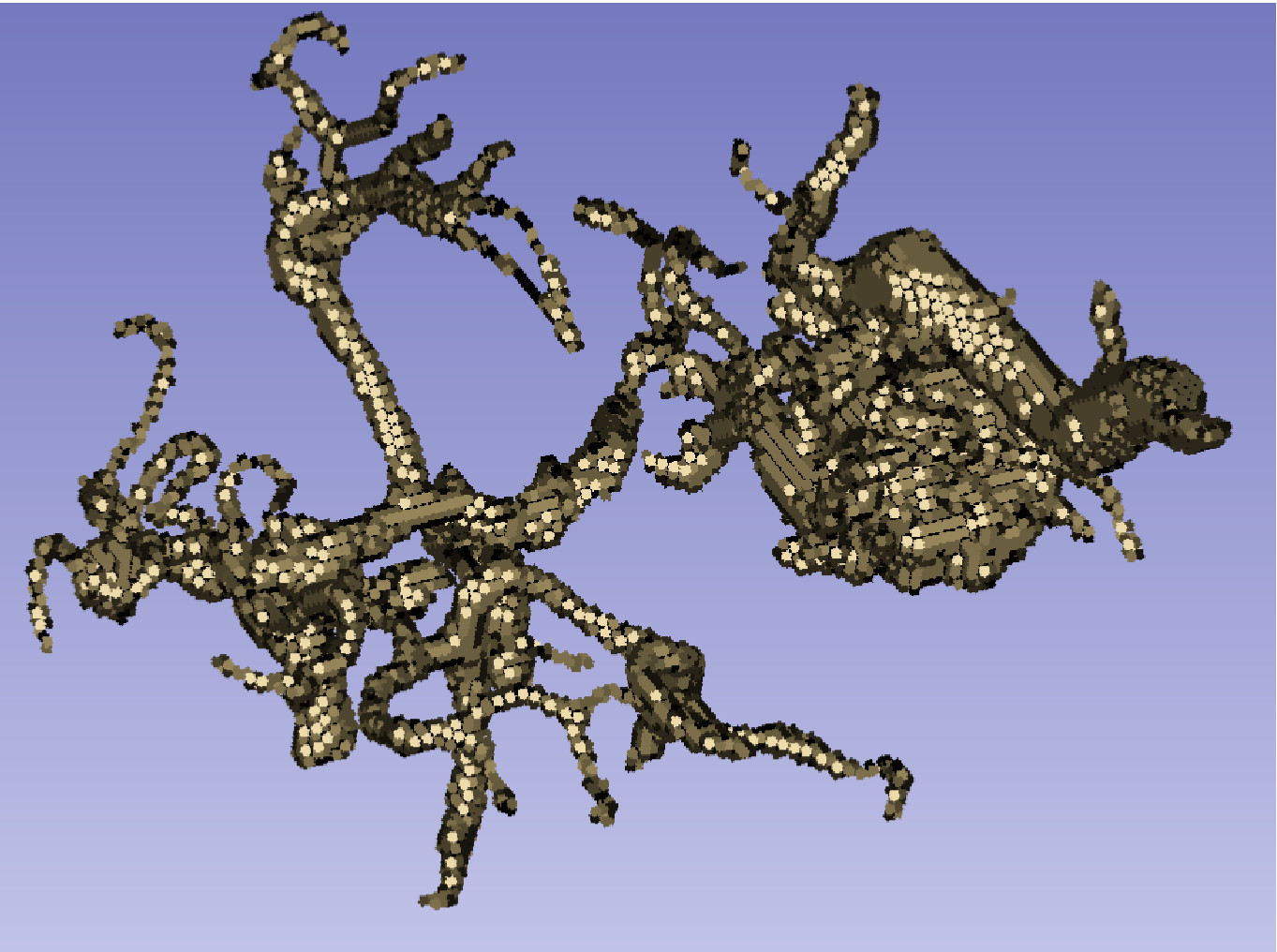}}
\caption{Brain MRI slice with AVM after skull stripping \cite{CBC3D2015} \protect\subref{DoeOne} and after segmentation \protect\subref{DoeTwo}.
\protect\subref{DoeThree} depicts a volume rendering of the segmented AVM. The segmented image has a spacing of $0.7\times0.7\times1.6\; \text{mm}^3$ and a size of $320\times320\times100\; \text{voxels}^3$.}
\label{Doe}
\end{figure}

\subsection{Image pre-processing}\label{preprocessing}
\begin{figure}[htb]
	\centering	
	\subfloat[Before down-sampling (spacing: $0.879\times0.879\times0.879\; \text{mm}^3$)]{
 	\label{VesselsDownSamplingOne}
	\includegraphics [width = 0.48\textwidth]{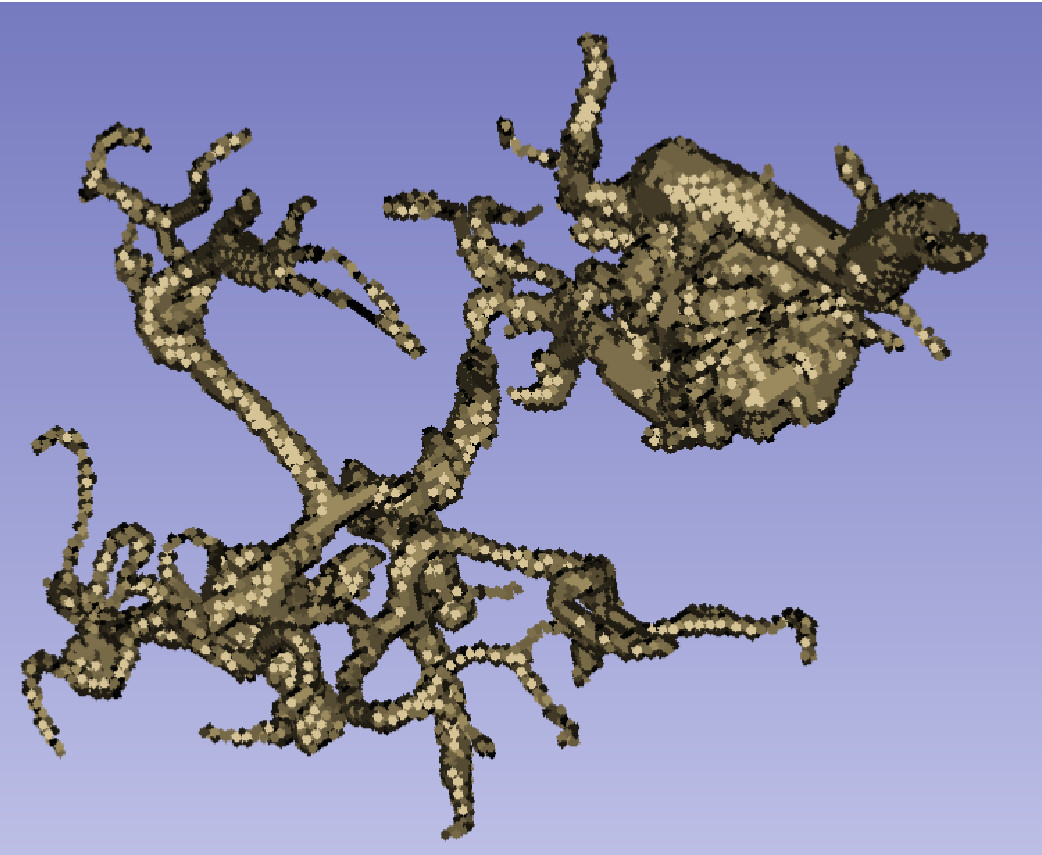}}
	\subfloat[After down-sampling (spacing: $1\times1\times1\; \text{mm}^3$)]{
 	\label{VesselsDownSamplingTwo}
	\includegraphics [width = 0.48\textwidth]{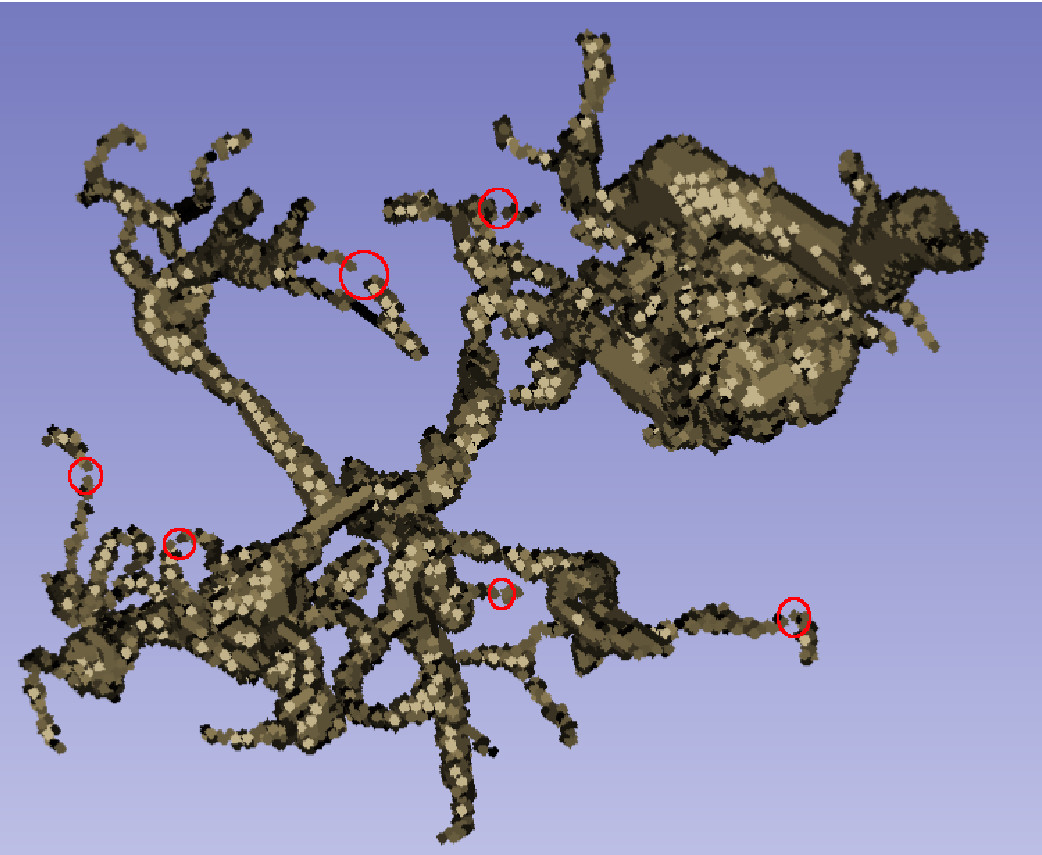}}
	\caption{Brain Arteriovenous Malformation (AVM) segmentation, before and after down-sampling. 
	The red circles indicate the problematic regions after down-sampling (i.e., disconnected voxels or non-manifold voxel connectivity).}
	\label{VesselsDownSampling}
	\end{figure}

The previous work \cite{CBC3DFedorov, CBC3DYixun, CBC3D2015} re-samples the segmented image to an isotropic unit-spaced image to avoid the transformation 
of the index to physical coordinates and vice versa. This approach is sufficient for fairly regular geometries and improves the speed 
of the mesh generation. However, in the case of complex data (i.e., AVM), image down-sampling can deteriorate the quality of the segmentation and may 
result in disconnected image regions or non-manifold connectivity (voxels that are connected to each other via an edge or a vertex). 
Figure \ref{VesselsDownSampling} illustrates such an example. The presented method does not perform image down-sampling.

\subsubsection{Relabeling noisy voxels}
\begin{figure}[htb]
\centering	
\subfloat[]{
\label{noisyVoxelOne}
\includegraphics [width=0.32\textwidth]{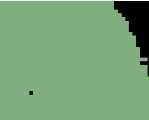}}
\subfloat[]{
\label{noisyVoxelTwo}
\includegraphics [width=0.32\textwidth]{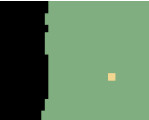}}
\subfloat[]{
\label{noisyVoxelThree}
\includegraphics [width=0.25\textwidth]{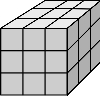}}
\caption{Candidate isolated voxels for relabeling (a)-(b) and 26-neighborhood region (c).}
\label{noisyVoxel}
\end{figure}

A semi-automatic or manual segmentation of poor quality may contain isolated voxels that do not correspond to an anatomical image feature. 
CBC3D identifies two types of such voxels and relabels them based on their neighbors.
The first type has a zero label (background) and all its neighbors have a non-zero label (Figure \ref{noisyVoxelOne}). 
The second type has a non-zero label and all its neighbors have a different label (Figure \ref{noisyVoxelTwo}). 
Not all neighbors necessarily have the same label. The neighbors are checked using a 26-neighborhood region (Figure \ref{noisyVoxelThree}).
After the noisy voxels are identified, they are relabeled to one of its neighbors' labels.
Either the first, second, or both types are relabeled.

\subsubsection{Relabeling disconnected regions} \label{relabeling_disconnected_regions}
CBC3D controls the level of refinement in each material by using a global or a material-specific fidelity (subsection \ref{BCCMeshGenerationandRefinement}). 
Complex segmentations (i.e, tangled bundles of abnormal and arbitrary thin blood vessels connecting arteries and veins in the brain) 
may contain materials with multiple disconnected regions. CBC3D implements a relabeling algorithm to handle each of those regions as a different material. Therefore, 
the present method allows for a more localized refinement which leads to a lower element count.

\begin{figure}[htb]
\centering	
\subfloat[Before relabeling]{
\label{DisconnectedMaterialsOne}
\includegraphics [width=0.45\textwidth]{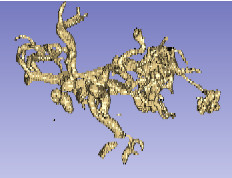}}
\subfloat[After relabeling]{
\label{DisconnectedMaterialsTwo}
\includegraphics [width=0.45\textwidth]{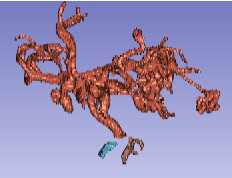}}
\caption{Relabeling an AVM segmented image with disconnected vessels. Before processing, the image contains one material (yellow) with five disconnected regions (a). 
After processing, the image contains three materials (red, cyan, and brown), each of which is a single region (b). 
Two small disconnected regions are relabeled to a background value.}
\label{DisconnectedMaterials}
\end{figure}

An ITK connected threshold image filter is employed to detect the disconnected regions in each material.
A seed is first computed based on the material's label, and then small regions on the image with the same label are merged iteratively to compute a connected region. 
A face connectivity is used for merging. Assume that $S_i$ is the set of voxels in material $i$, and the filter computes a set of voxels $S_{ij}$ in a region $j$ of material $i$.
If $|S_i| = |S_{ij}|$ (where $|*|$ is the number of voxels in $*$), then material $i$ is a single region. 
Otherwise, material $i$ consists of disconnected regions.
If $\frac{\|S_i\|-\|S_{ij}\|}{\|S_i\|} < s_{tol}$ then region $j$ is too small compared to region $i$ to be considered as tissue, hence it is relabeled to zero (background).
Otherwise, region $j$ is relabeled to a new label. A tolerance $s_{tol}=10^{-4}$ detects the small disconnected regions.
Figure \ref{DisconnectedMaterials} depicts an example before and after the relabeling process. 

\subsubsection{Eliminating non-manifold voxel connectivity}\label{EliminatingNonManifoldImageTopology}
\begin{figure}[htb]
\centering	
\subfloat[Before relabeling]{
\label{DoeExampleOne}
\includegraphics [width=0.48\textwidth]{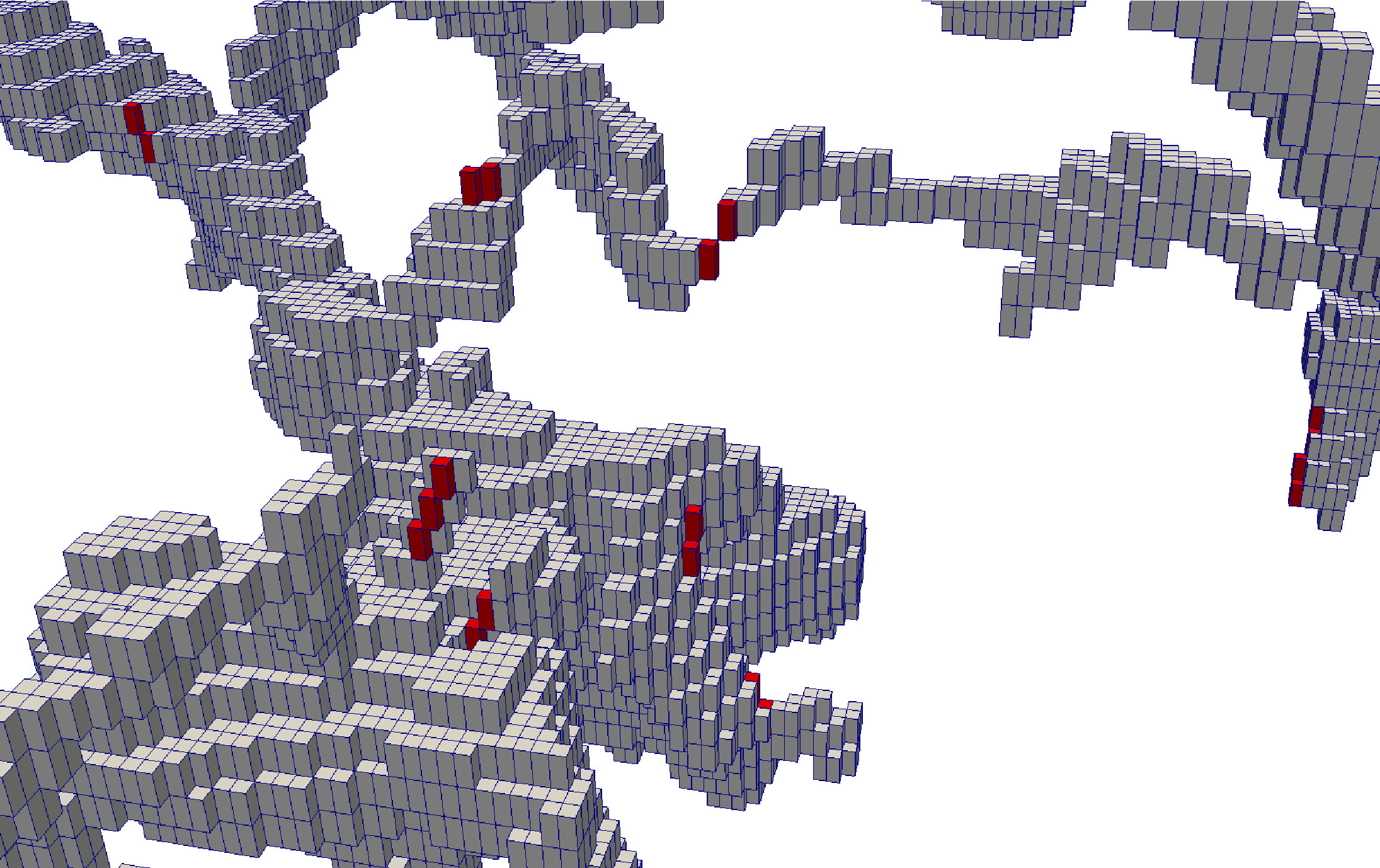}}
\subfloat[After relabeling]{
\label{DoeExampleTwo}
\includegraphics [width=0.48\textwidth]{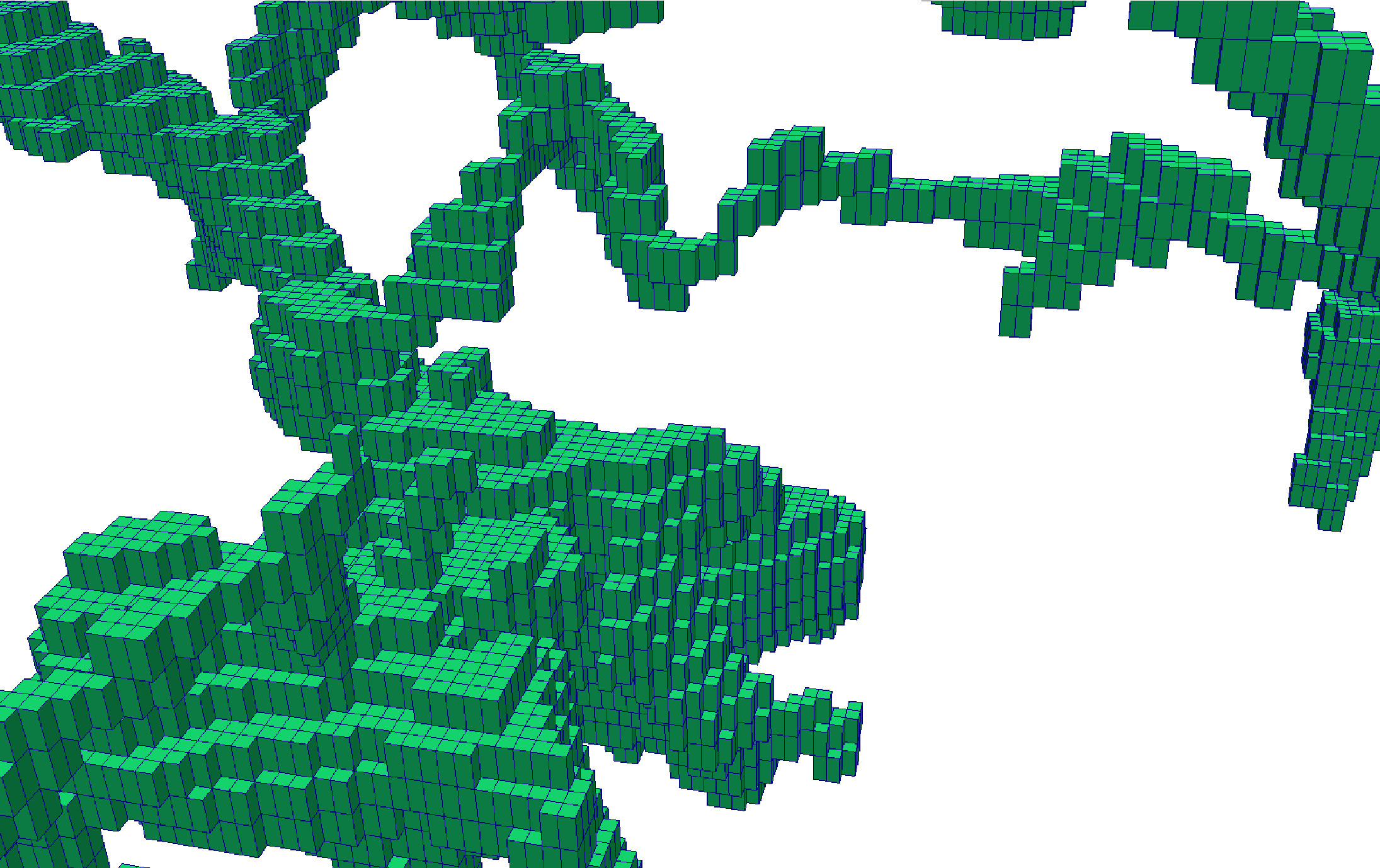}}
\caption{AVM anisotropic segmented image $(0.7\times0.7\times1.6\;mm^3)$ before (a) and after (b) relabeling to eliminate the 
non-manifold voxel connectivity. The voxels which are connected via an edge or a vertex are depicted with red color.}
\label{DoeExample}
\end{figure}

A segmentation with poor resolution may contain non-background voxels that are connected by a single edge or a vertex (Figure \ref{DoeExampleOne}).
This type of connectivity can lead to a non-manifold mesh (i.e., tetrahedra that are connected only by a mesh edge or a mesh vertex), especially 
when refinement is inadequate to resolve the small image features. A non-manifold mesh deteriorates the solution accuracy in blood flow simulations within stented 
arterial segments or AVM surgical simulations (Figures \ref{Stent}, \ref{DoeExample}).

\begin{figure}[htb]
    \centering	
    \subfloat[]{
        \label{vertexToVertexZero}
        \includegraphics [width=0.13\textwidth]{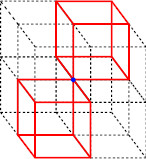}}
    \subfloat[]{
    \label{vertexToVertexOne}
    \includegraphics [width=0.13\textwidth]{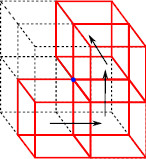}}
    \subfloat[]{
    \label{vertexToVertexTwo}
    \includegraphics [width=0.13\textwidth]{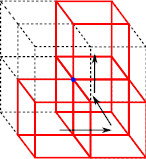}}
    \subfloat[]{
    \label{vertexToVertexThree}
    \includegraphics [width=0.13\textwidth]{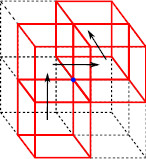}}
    \subfloat[]{
    \label{vertexToVertexFour}
    \includegraphics [width=0.13\textwidth]{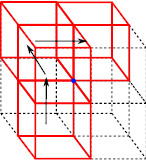}}
    \subfloat[]{
    \label{vertexToVertexFive}
    \includegraphics [width=0.13\textwidth]{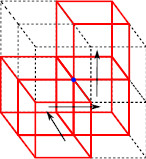}}
    \subfloat[]{
    \label{vertexToVertexSix}
    \includegraphics [width=0.13\textwidth]{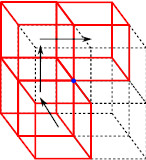}}
    
    \subfloat[]{
    \label{edgeToedgeZero}
    \includegraphics [width=0.13\textwidth]{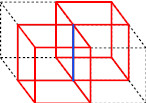}}
    \subfloat[]{
    \label{edgeToedgeOne}
    \includegraphics [width=0.13\textwidth]{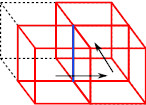}}
    \subfloat[]{
    \label{edgeToedgeTwo}
    \includegraphics [width=0.13\textwidth]{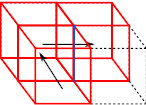}}
    \caption{Templates to eliminate a vertex-to-vertex connectivity (a) or an edge-to edge connectivity (h) in a segmented labeled image.
    In the case of a vertex-to-vertex connectivity, one of the six templates is randomly selected to relabel two voxels within a cluster of eight voxels (b)-(g).
    In the case of an edge-to-edge connectivity, one of the two templates is randomly selected to relabel a single voxel within a cluster of four voxels (i)-(j).
    The arrows illustrate the path of the transformation via face connected voxels after relabeling.}
    \label{imageTemplates}
\end{figure}

CBC3D incorporates relabeling operations with specific templates to eliminate the non-manifold voxel connectivities.
First, all vertex-to-vertex connectivities are detected (Figure \ref{vertexToVertexZero}) and eliminated by randomly selecting one of the six 
templates (Figures \ref{vertexToVertexOne}-\ref{vertexToVertexSix}).
All templates result in face-to-face connectivity; however, each template relabels different voxels.
When all vertex-to-vertex connectivities are eliminated, the edge-to-edge 
connectivities are detected (Figure \ref{edgeToedgeZero}) and eliminated by randomly selecting one of the two templates in Figures \ref{edgeToedgeOne}-\ref{edgeToedgeTwo}.
 Similarly, the two templates relabel different voxels. 
The two steps repeat until all non-manifold connectivities are eliminated. Randomness is introduced in the selection of these 
templates to achieve convergence in the case of relabeling neighbor clusters.
 Figure \ref{DoeExample} depicts an example of an AVM segmented image before and after elimination. 

\subsection{Adaptive Lattice Refinement}\label{BCCMeshGenerationandRefinement}
\subsubsection{BCC lattice construction}
\begin{figure}[htb]
\centering	
\subfloat[Cut section]{
\label{latticeOne}
\includegraphics [width=0.38\textwidth]{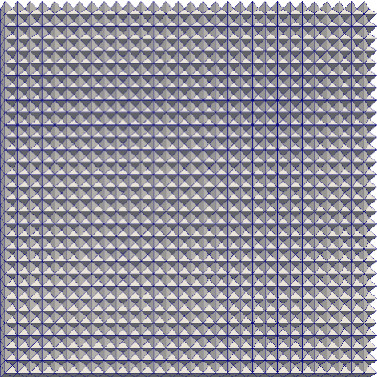}}
\subfloat[Lattice detail]{
\label{latticeTwo}
\includegraphics [width=0.35\textwidth]{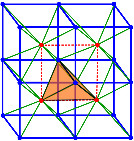}}
\caption{Uniform Body-Centered Cubic (BCC) lattice. The green edges lace the two lattices together. Each vertex is surrounded by 14 edges and 24 tetrahedra.}
\label{lattice}
\end{figure}

The pre-processed segmented image is discretized with a regular Body Centred Cubic (BCC) lattice. This structured lattice is generated as a result of two interlaced lattices (Figure \ref{lattice}). The first lattice is created by setting its
vertices with an input-defined distance between them ($BCC\_size$ parameter in Table \ref{inputParametersCBC}), 
then the edges connecting those vertices are added, consequently creating the tetrahedral elements. 
The elements of the second lattice are created with consideration to the first lattice so that their vertices correspond to the centroids of the initial lattice cells. 
As a final step, the tetrahedra located completely outside of the object are discarded. 
The resulting elements are of the best quality possible (the minimum dihedral angle is $60^\circ$) with the regular 
space tiling \cite{Fuchs20011400}.

\subsubsection{Adaptive refinement} \label{adaptive_refinement}
\begin{figure}[htb]
\centering	
\subfloat[Segmented image]{
\label{EDTOne}
\includegraphics [width=0.4\textwidth]{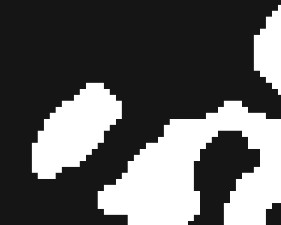}}
\subfloat[EDT of (a)]{
\label{EDTTwo}
\includegraphics [width=0.4\textwidth]{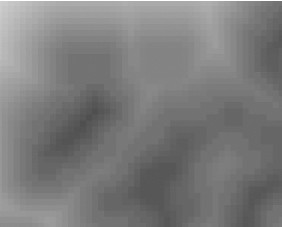}}
\caption{Euclidean Distance Transform (EDT) computed from a labeled image. The EDT calculates the minimum distance of a voxel in the image from its closest material boundary. 
Voxels inside a material have a positive distance value, voxels outside a material have a negative distance value and voxels on the boundary have a zero distance value.}
\label{EDT}
\end{figure}

 To subdivide the elements, some criterion must be considered. The current implementation uses
a Euclidean Distance Transform (EDT) to decide whether an element needs to be subdivided (Figure \ref{EDT}).
An EDT provides more information compared to a segmented image and it thus conducts a smoother refinement near the material boundaries.
In the presence of $n$ materials, $n$ EDTs will be calculated in order to exclusively subdivide the elements within each material.
ITK's Signed Maurer Distance Map Image Filter \cite{maurer2003linear} is employed to compute the EDT in parallel and reduce the running time.

\begin{figure}[htb]
	\centering	
	\subfloat[$1:8$]{
 	\label{BCCRefinementOne}
	\includegraphics [width = 0.23\textwidth]{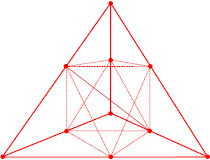}}
	\subfloat[$1:2$]{
 	\label{BCCRefinementTwo}
	\includegraphics [width = 0.23\textwidth]{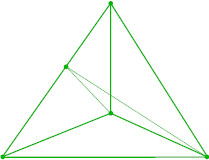}}
	\subfloat[$1:4$]{
 	\label{BCCRefinementThree}
	\includegraphics [width = 0.23\textwidth]{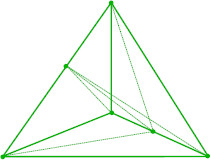}}
	\subfloat[$1:4$]{
 	\label{BCCRefinementFour}
	\includegraphics [width = 0.23\textwidth]{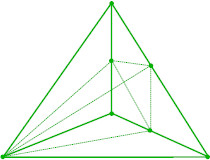}}
	\caption{Red-Green templates for lattice subdivision. 8, 2, and 4 is the number of tetrahedra after subdivision.}
	\label{BCCRefinement}
\end{figure}

The lattice is refined using red (regular) - green (irregular) templates (Figure \ref{BCCRefinement}).
An element is subdivided if the EDT to which the centroid of the element belongs changes sign in at least one of the four element vertices.
Initially, all tetrahedra are marked as red. A red tetrahedron is always subdivided into eight children (1:8 regular refinement), 
and each child is marked red, as shown in Figure \ref{BCCRefinementOne}. 
There are three choices for the internal edge of the red tetrahedron. If the shortest one is 
selected, the resulting eight-child tetrahedra are exactly the same as the parent 
except the size is one-half of the parent's size. The red subdivision will lead to T-junctions at the newly created edge midpoints where 
neighboring tetrahedra are not refined to the same level. To remove
the T-junctions, a green (irregular) subdivision is performed, shown in the three cases depicted in 
Figures \ref{BCCRefinementTwo}-\ref{BCCRefinementFour}.

To quantitatively evaluate the similarity between a mesh that corresponds to a single material (sub-mesh) and the
image region of this material (sub-region), $S_1$ is defined as the set of all voxels in the sub-mesh, 
$S_2$ is the set of all voxels in the sub-region, and $S_1\cap S_2$ is the point-set shared by the sub-mesh and the sub-region (common region).
Ratio $F_1 = \frac{|S_1\cap S_2|}{|S_1|}$ corresponds to the similarity between the common region and the sub-mesh, 
and ratio $F_2 = \frac{|S_1\cap S_2|}{|S_2|}$ corresponds to the similarity between the 
common region and the sub-region. The refinement criterion is defined as: $ \text{Refine the sub-mesh if} \;\; F > F_1 \;\; \text{or} \;\; F > F_2$, 
where $F \in (0,1]$ is a global input fidelity or a material-specific fidelity (Table \ref{inputParametersCBC}). 
The higher the input fidelity, the finer the mesh will be near the boundaries.
 The advantage of a material-specific fidelity compared to a global fidelity is that the former allows for a more localized refinement in the materials of interest, 
thereby reducing total element count.

\subsubsection{Candidate mesh selection} 
\begin{figure}[htb]
\centering	
\subfloat[Uniform lattice]{
\label{pipelineOne}
\includegraphics [width=0.32\textwidth]{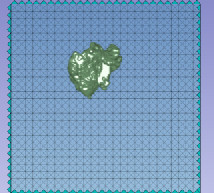}}
\subfloat[Uniform lattice after removals]{
\label{pipelineTwo}
\includegraphics [width=0.32\textwidth]{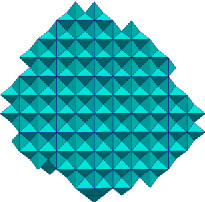}}
\subfloat[Adaptive lattice]{
\label{pipelineThree}
\includegraphics [width=0.32\textwidth]{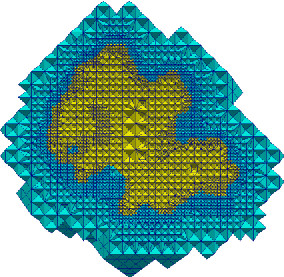}}

\subfloat[Adaptive lattice after redistribution]{
\label{pipelineFour}
\includegraphics [width=0.3\textwidth]{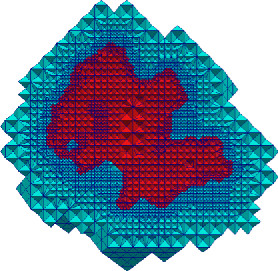}}
\subfloat[Final lattice]{
\label{pipelineFive}
\includegraphics [width=0.3\textwidth]{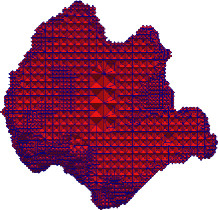}}
\caption{Pipeline of the BCC lattice construction and adaptive refinement.}
\label{pipeline}
\end{figure}

For a surface to be suitable for deformation, each non-background element can only be connected via at most one face with a background element. 
Otherwise, the elements on the surface can easily become distorted or inverted during deformation.
To avoid this type of connectivity, the labels of those problematic surface elements and those of 
the surrounding elements are redistributed. Every tetrahedron is examined by observing the labels of its four adjacent tetrahedra. 
If at least three of the adjacent tetrahedra have the same label as the examined tetrahedron, then there is no need for relabeling. However, if there
is more than one adjacent tetrahedron with a different label, then the examined tetrahedron needs
to be relabeled to one of its neighboring labels. Between these labels, the one chosen for relabeling
the tetrahedron depends on whether or not a considered label has already been examined and finalized in
the previous examinations. If a neighboring label has not been fully inspected yet, the tetrahedron is relabeled to 
this one. At the end of the relabeling process, the background elements are discarded and the final mesh is obtained.
Figure \ref{pipeline} depicts the basic steps of the adaptive mesh generation.

\subsubsection{Mesh topological checks}\label{TopologicalChecks}
The goal is to obtain a manifold topology, meaning that every tetrahedron must be connected to one another through a face. 
After the lattice subdivision is completed (i.e., all input fidelities are satisfied) and the candidate mesh is selected, 
the mesh topology is examined to ensure the absence of non-manifold (i.e. connected via an edge or a vertex) tetrahedral connections 
or disconnected tetrahedra in each sub-mesh. 

A "mesh connected"-threshold filter was developed to detect the non-manifold topology or the disconnected tetrahedral regions in each sub-mesh.
An arbitrary seed tetrahedron is first computed based on the label of each sub-mesh, and then neighbor tetrahedra are merged iteratively to compute a connected mesh region. 
A vertex connectivity is used for merging. The non-manifold edges and the non-manifold vertices are identified by checking the 
labels of the neighbor elements before merging.

When the connected threshold filter is completed, the number of tetrahedra in the connected mesh region is compared with the number of tetrahedra in the sub-mesh.
If they are equal, the sub-mesh is a single connected region. However, if non-manifold topologies exist, then the tetrahedra surrounding all non-manifold vertices and all non-manifold edges are marked 
and they are locally subdivided in the next refinement iteration. Otherwise, the refinement of the sub-mesh is completed.

If the number of tetrahedra in the connected mesh region is not equal to the number of tetrahedra in the sub-mesh, then the sub-mesh contains disconnected regions.
In this case, all disconnected regions are first checked for relabeling. A disconnected region is relabeled if its volume is too small $(< 1\%)$ compared to the volume of the sub-mesh.
Relabeling is performed using a label of a neighbor region (background or not). Relabeling is canceled if it produces a non-valid manifold connectivity.
If all disconnected regions are eliminated and all connectivities are manifold, then the refinement of this sub-mesh is completed.
If disconnected regions exist, then the whole sub-mesh is globally refined in the next iteration.
Finally, if non-manifold topologies exist, then the tetrahedra surrounding the non-manifold vertices and the non-manifold edges are marked 
and locally subdivided in the next refinement iteration. Although sufficient for the cases presented in our evaluation, it should be noted that this subsequent subdivision of tetrahedra surrounding non-manifold vertices and edges does not always guarantee that the sub-mesh will eventually become manifold.

The mesh topological checks should be performed only if the non-manifold voxel connectivities have been previously eliminated (subsubsection \ref{EliminatingNonManifoldImageTopology}).
Mesh topological checks can be time-consuming; therefore, they should only be employed when the application requires a manifold mesh or when the 
image contains small features which are harder to resolve (e.g., vessels or stents).

\subsection{Mixed Element Mesh}\label{MixedMesh}
Experimental evaluation in this study showed that a mixed mesh (with tetrahedra, pentahedra, and hexahedra) may contain up to $30\%$ fewer vertices compared to a tetrahedral mesh of the same input, 
without compromising the fidelity. Therefore, it can reduce the subsequent memory and CPU requirements for the solver without any loss of accuracy. 

CBC3D generates a mixed mesh from an adaptive tetrahedral lattice by merging clusters of tetrahedra into hexahedra.
A valid transition between the tetrahedra and the hexahedra is guaranteed with the use of prismatic elements (i.e., pyramids).
Merging is performed only within the homogeneous regions where the lattice is uniform.
Therefore, the topology of the generated surfaces or interfaces between materials is preserved.

\begin{figure}[htb]
\centering	
\subfloat[Tetrahedral mesh]{
\label{mixedMeshOne}
\includegraphics [width=0.48\textwidth]{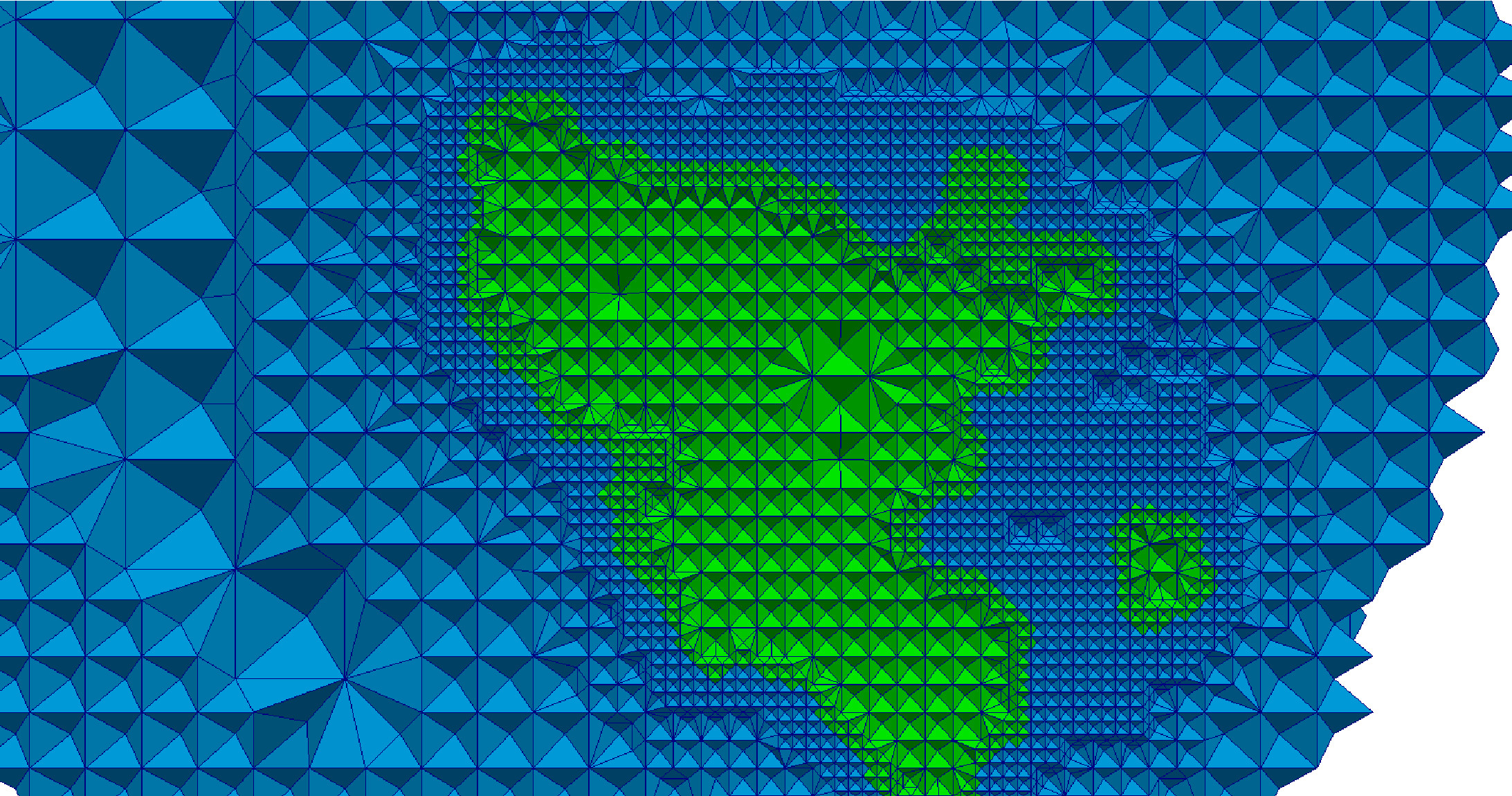}}
\subfloat[Mixed element mesh]{
\label{mixedMeshTwo}
\includegraphics [width=0.48\textwidth]{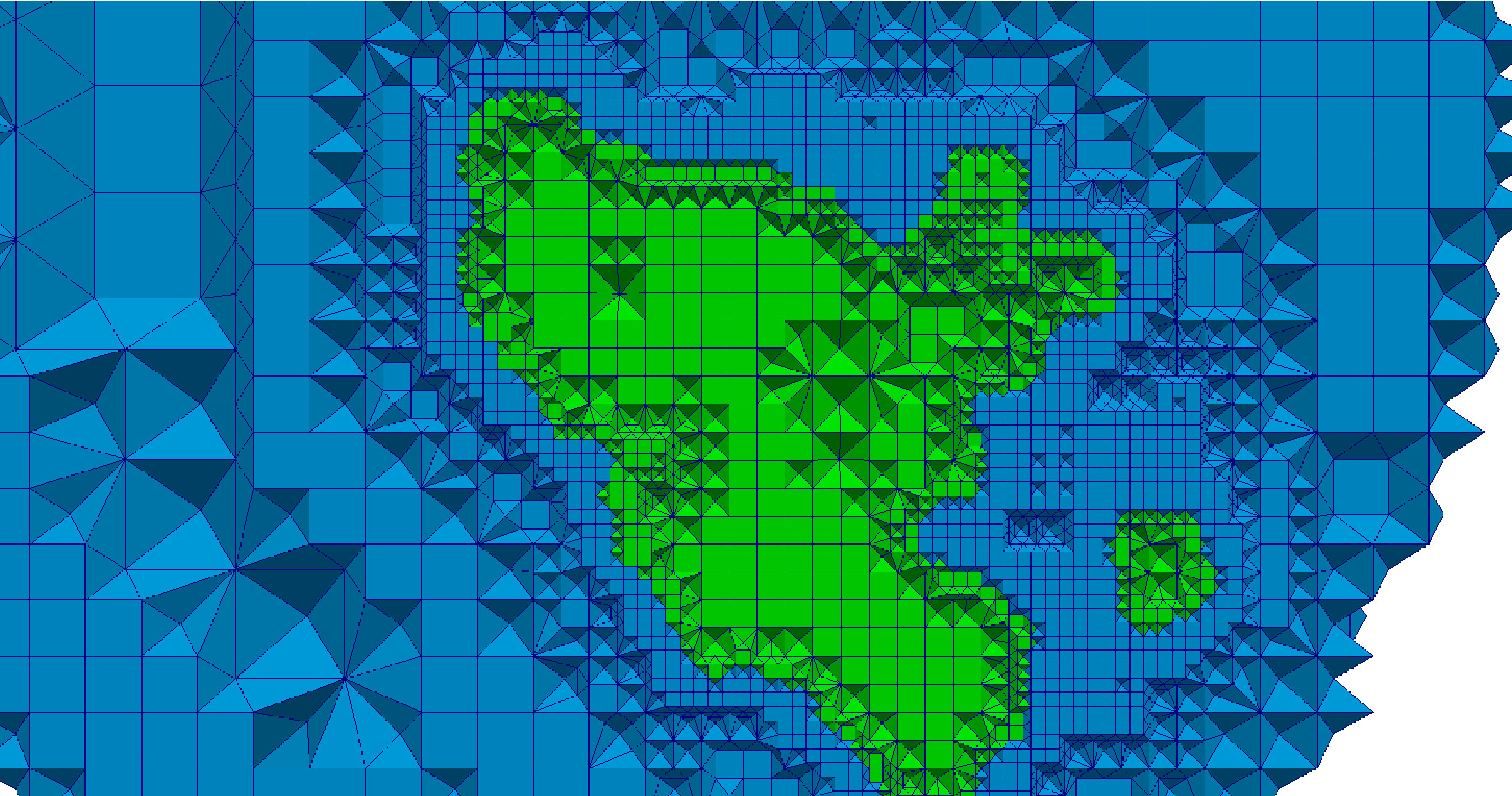}}
\caption{Adaptive BCC lattice before (a) and after (b) it is converted into a mixed element mesh.}
\label{mixedMesh}
\end{figure}

In the BCC lattice, the cardinality of a vertex in a uniform region is always 14, meaning that each vertex
has 14 edges to check in the process of merging (Figure \ref{latticeTwo}). 
Beginning from the generated adaptive BCC mesh, 
the inner tetrahedral vertices are located to start the transformations. For every vertex, its
adjacent tetrahedra constitute a polyhedron. To transform each one of these polyhedra to hexahedra, 
their inner tetrahedra must be red (according to the red-green refinement) and their labels need to be the same. 
If indeed all the tetrahedra of the polyhedron are of the same label and none of them are green, then for each
adjacent orthogonal edge of the vertex (Figure \ref{latticeTwo}), its four attached tetrahedra are computed. 
Their four orthogonal edges form the quadrilateral face of the hexahedron.
However, if an attached tetrahedron between these four is green or has a different label than the
others, then a transition pyramid is created. This element has its base perpendicular to the orthogonal edge and by being adjacent to the hexahedral face, its vertices and edges are calculated.
After creating each hexahedron, the tetrahedra inside the hexahedron are removed and the mesh topology is updated.
Figure \ref{mixedMesh} compares a tetrahedral mesh and a mixed mesh of the same input.

\subsection{Mesh Deformation}\label{MeshDeformation}
Red-green templates guarantee the high quality of the adaptive mesh. Indeed, the worst minimum dihedral angle achieved in our evaluation is $30^\circ$ and the worst maximum dihedral angle is $116.5^\circ$.
However, the surface of this mesh can be relatively bumpy, so it may not be suitable for applications such as AVM surgical simulations or CFD modeling of brain aneurysms/stents for endovascular flow diversion.
For example, a surgical simulation requires a smooth surface so that the mesh will reflect a certain degree of visual reality. A bumpy surface can deteriorate the accuracy of a 
CFD solution.

An approach that deforms the mesh surfaces to their corresponding image boundaries by minimizing an energy function was presented in \cite{CBC3DYixun}.
The present study utilizes and improves upon this method in several ways. First, the new implementation is built upon the ITK toolkit, therefore improving the overall reliability 
and portability of the software. Second, it employs heuristics and quality control to eliminate highly distorted elements during deformation.
Third, it allows meshes with mixed elements to undergo deformation. Fourth, it controls the trade-off between deformation time and mesh fidelity. 
Finally, it reduces deformation time using parallel computing.

This study formulates the deformation problem as an energy minimization problem represented by the objective function
\begin{equation}\label{deformationEq}
W = U^T KU + (HU - D)^T (HU -D)
\end{equation}
where $U$ is the unknown mesh displacement vector, $K$ is the mesh stiffness matrix, $H$ is a linear interpolation matrix, 
and $D$ is the vector of correspondences computed between two point sets: a source and a target point set.
The source point set contains the coordinates of the surface/interface vertices of the mesh to be deformed. 
The target point set contains the coordinates of the voxels on the surface/interface of the materials in the image.
The $K$ matrix depends on the element type and the mechanical properties of the brain model. 
First, the stiffness matrix of each element (e.g., tetrahedron, pyramid, hexahedron) is calculated in each integration point using Gaussian quadrature.
The global matrix $K$ is then assembled from the individual stiffness matrices \cite{Bathe}. 

\begin{figure}[htb]
	\centering	
	\subfloat{
 	\label{nidusDeformationZero}
	\includegraphics [width=0.35\textwidth]{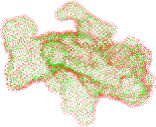}}
	
	\subfloat{
 	\label{nidusDeformationOne}
	\includegraphics [width=0.24\textwidth]{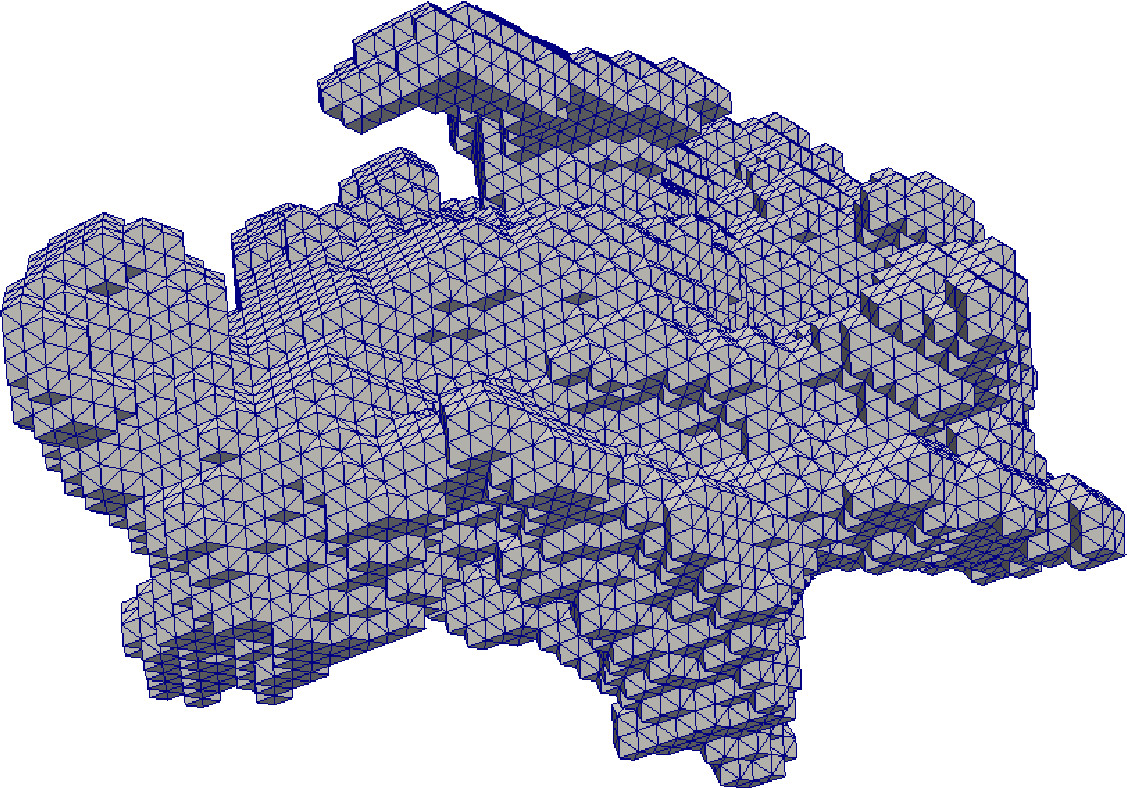}}
	\subfloat{
 	\label{nidusDeformationTwo}
	\includegraphics [width=0.24\textwidth]{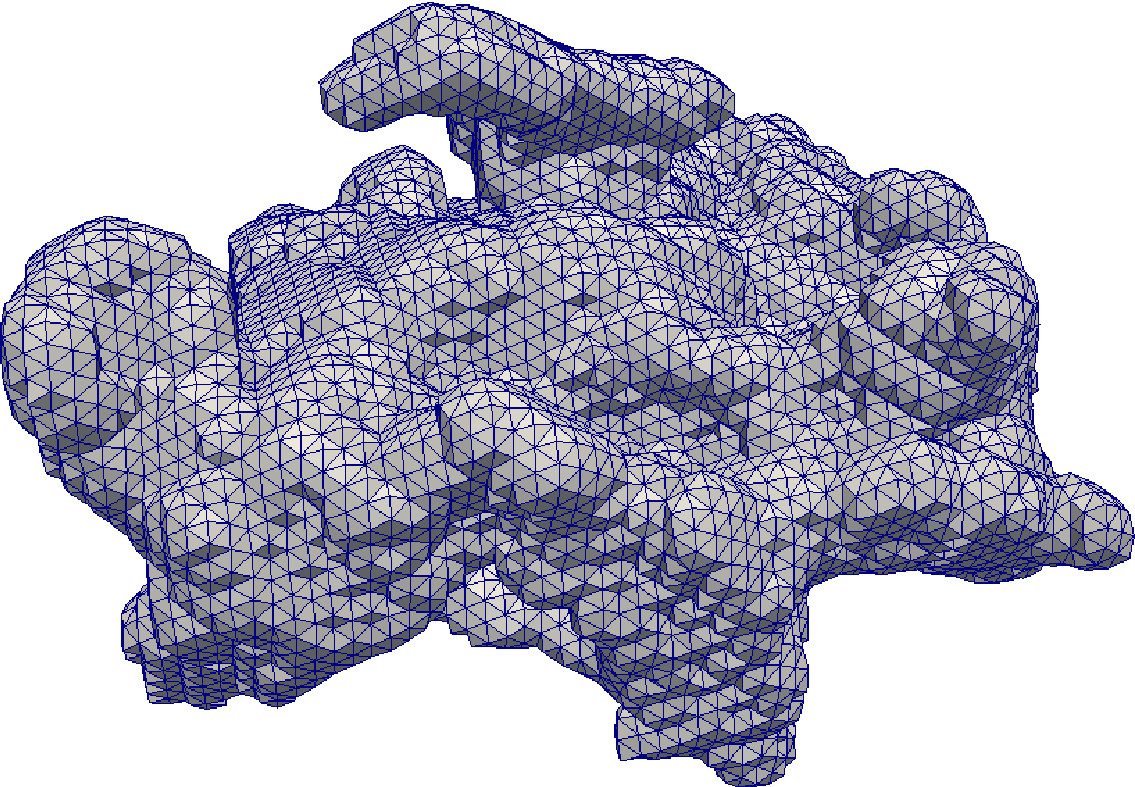}}
	\subfloat{
 	\label{nidusDeformationThree}
	\includegraphics [width=0.24\textwidth]{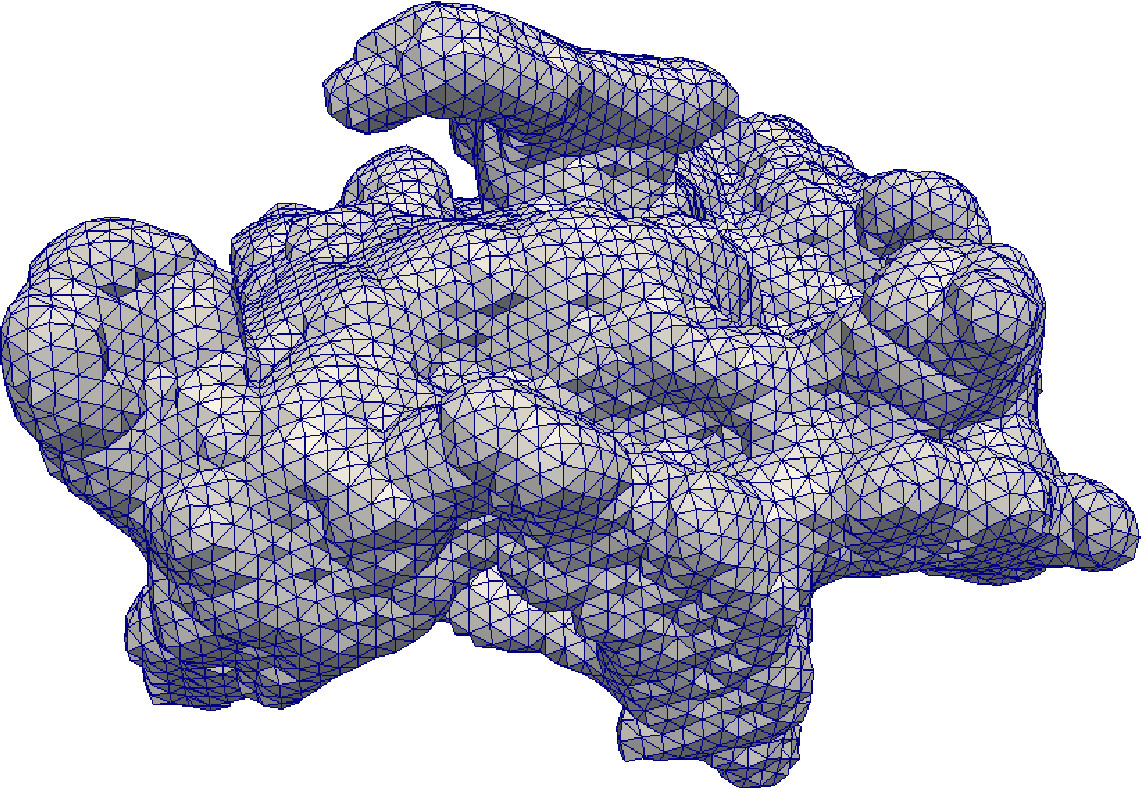}}
	\subfloat{
 	\label{nidusDeformationFour}
	\includegraphics [width=0.24\textwidth]{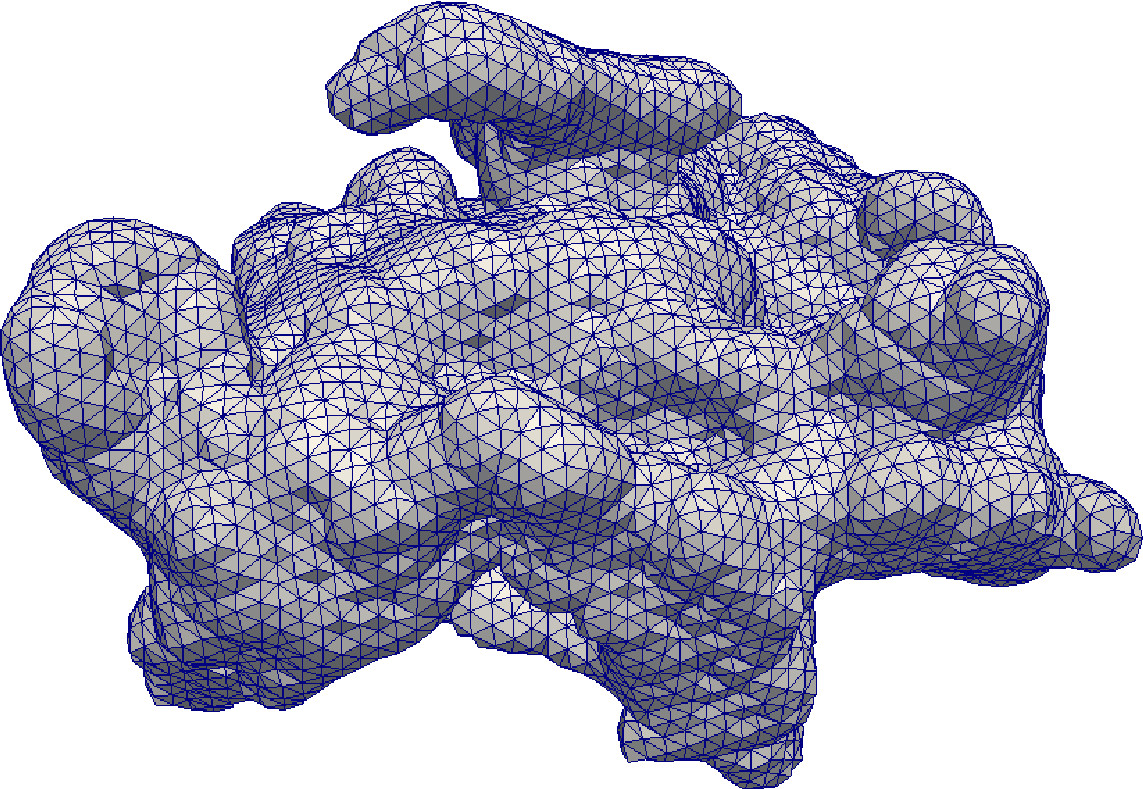}}
	
	\subfloat[$\text{HD}_0 = 1.79$ mm]{
 	\label{nidusDeformationFive}
	\includegraphics [width=0.24\textwidth]{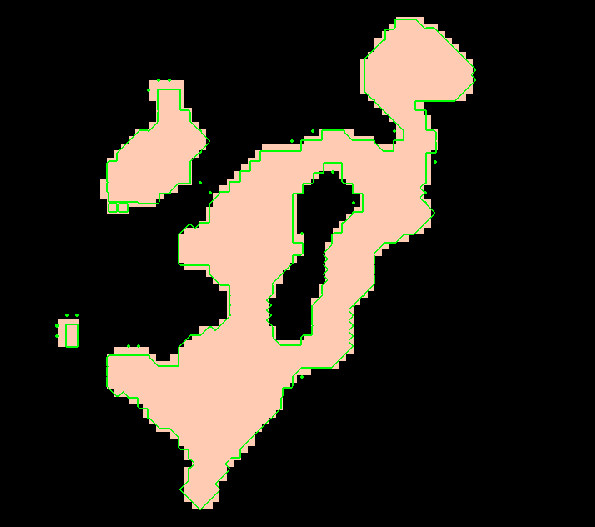}}
	\subfloat[$\text{HD}_3 = 1.31$ mm]{
 	\label{nidusDeformationSix}
	\includegraphics [width=0.24\textwidth]{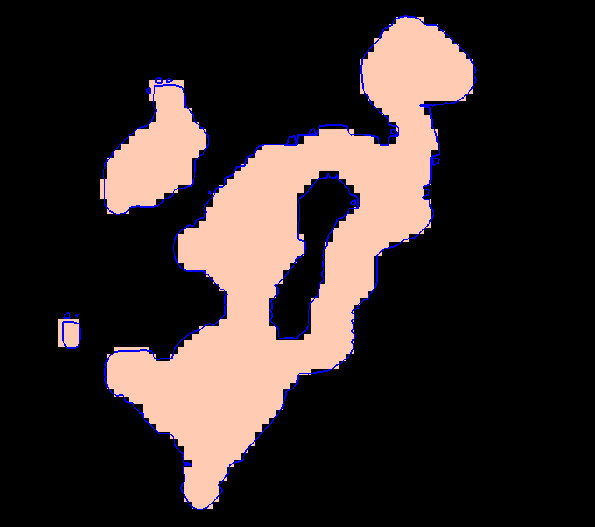}}
	\subfloat[$\text{HD}_7 = 0.96$ mm]{
 	\label{nidusDeformationSeven}
	\includegraphics [width=0.24\textwidth]{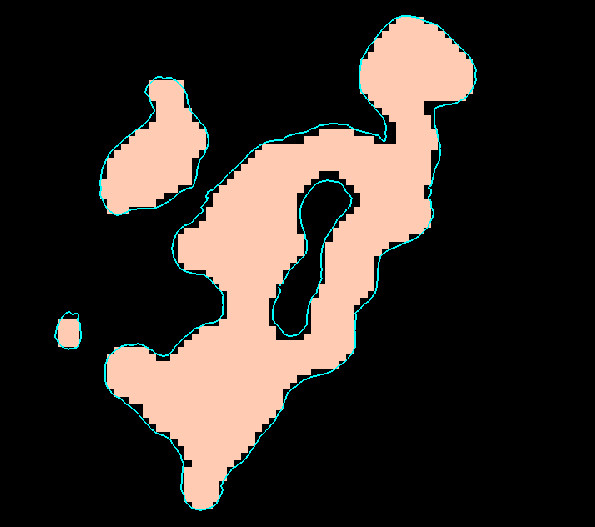}}
	\subfloat[$\text{HD}_{10} = 0.91$ mm]{
 	\label{nidusDeformationEight}
	\includegraphics [width=0.24\textwidth]{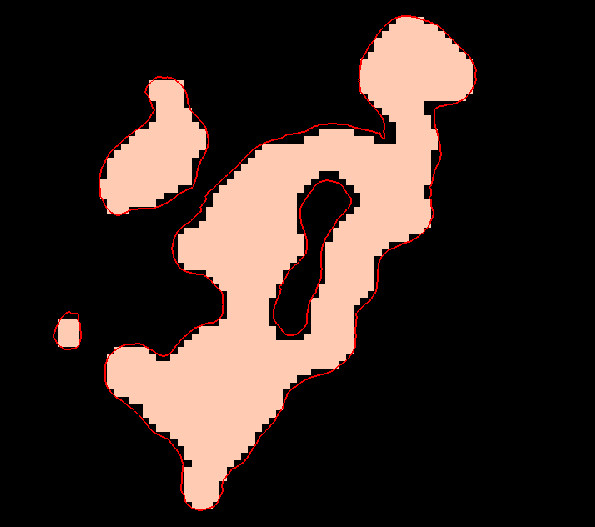}}
	\caption{Nidus mesh during deformation with 10 iterations. The figure on the top depicts the extracted source (green) and target (red) points used for deformation.
	Each column depicts the deformed mesh and an intersection between the mesh surface and the image plane at iterations $i=0,3,7,10$ (from left to right). 
	$\text{HD}_i$ denotes the mesh fidelity in terms of a Hausdorff Distance metric, at iteration $i$. 
	The smaller the HD value, the higher the fidelity. As the number of iterations advances, the mesh exhibits a smoother surface.}
	\label{nidusDeformation}
\end{figure}

To minimize \eqref{deformationEq}, the gradient with respect to $U$ must be zero, leading to a linear system of equations:
\begin{equation}\label{deformationSystemEq}
(K + H^TH) U = H^T D
\end{equation}
An iterative solver is employed \cite{Kincaid} to solve the system.
A linear assumption is made for the material stiffness and the displacements of the biomechanical model.
Once $U$ has been found, the mesh is updated by adding the computed displacements to the current
coordinates of the vertices, and the procedure is repeated until the iterations reach a maximum
number. Figure \ref{nidusDeformation} illustrates an example of the procedure.

\subsubsection{Extraction of source and target points using a multicore processor}
The source and the target points are extracted from the mesh to be deformed and the segmented image, respectively. 
The source points are the vertices on the mesh surface or on the interface between sub-meshes of different materials.
The target points are the physical centers of the voxels on the segmented image boundaries.

The extraction is performed before deformation begins. 
For the extraction of the source points, the mesh is first decomposed into $k$ sub-meshes so that the number of elements in each 
sub-mesh is approximately the same ($k$ is the number of threads). Each thread extracts points from a single sub-mesh.
The labels of the elements surrounding each source point are stored in a label set. 
This is because the deformation essentially registers pairs of source and target points with same label sets.
For the extraction of the target points, the segmented image is first partitioned into a number of $k$ sub-regions using an 
ITK threaded image region partitioner filter. Points are extracted from the segmented boundaries of each sub-region with the help of EDTs (EDTs are previously calculated at the refinement step).
Additionally, the labels within a 26-neighborhood region (Figure \ref{noisyVoxelThree}) around the target point are stored.
Figure \ref{nidusDeformation} depicts the extracted source and target points from a Nidus tetrahedral mesh and image, respectively.

\subsubsection{Quality control} \label{quality_control}
For each source point, an average displacement vector is calculated from the corresponding target points within 
a 3D region around the source point \cite{CBC3DYixun}. To improve element quality after 
each deformation iteration, the size of this 3D region is limited by a local element size, i.e., the average length of 
the element edges surrounding the source point. Then vector $D$ is assembled from the calculated displacement vector of each source point.
After each deformation iteration, a quality metric (i.e., minimum dihedral angle for tetrahedra or scaled Jacobian for hexahedra and pyramids \cite{knupp2000achieving})
is computed for each element. The elements that do not satisfy a minimum quality (e.g., $5^\circ$ for dihedral angles or $0.2$ for scaled Jacobian) are marked,
and the displacement vectors of the source points surrounded by at least one marked element are scaled with a factor of $0.2$.
$D$ is assembled again and the system in \eqref{deformationSystemEq} is solved. 
If the quality metric is not satisfied after three consecutive attempts, 
the deformation stops and the procedure recovers the mesh coordinates of the previous deformation iteration. The default minimum quality parameters can be changed by the user if desired. The default values strike a balance between generating a good quality mesh and reducing mesh size (i.e., the number of elements) for the cases in our evaluation. We also determine these values based on past experience \cite{LD_Chernikov} and on a study involving a deep learning model that found the optimal input parameter values for an adaptive non-rigid registration application \cite{FotisAPBNRRDeepLearning2021,AngelosDeepLearning2019}.

\subsubsection{Adjustable number of target points} 
\begin{figure}[htb]
	\centering	
	\subfloat[Segmentation]{
 	\label{nonConnectivityOne}
	\includegraphics [width=0.19\textwidth]{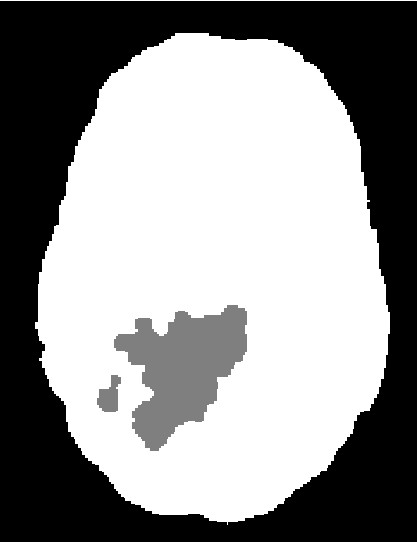}}
	\subfloat[``Vertex'']{
 	\label{nonConnectivityTwo}
	\includegraphics [width=0.19\textwidth]{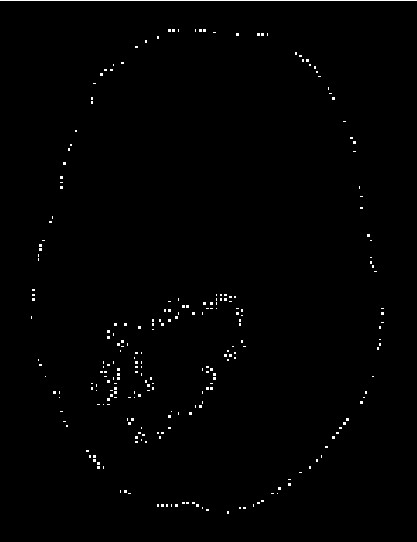}}
	\subfloat[``Edge'']{
 	\label{nonConnectivityThree}
	\includegraphics [width=0.19\textwidth]{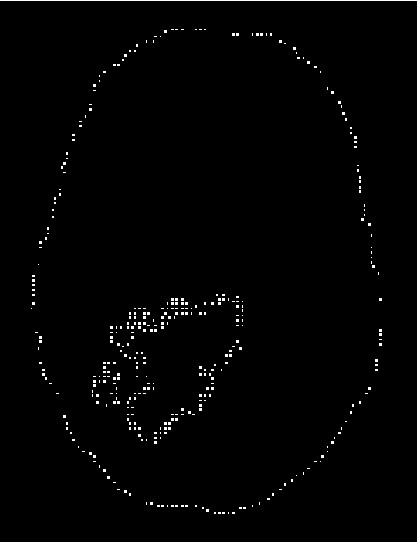}}
	\subfloat[``Face'']{
 	\label{nonConnectivityFour}
	\includegraphics [width=0.19\textwidth]{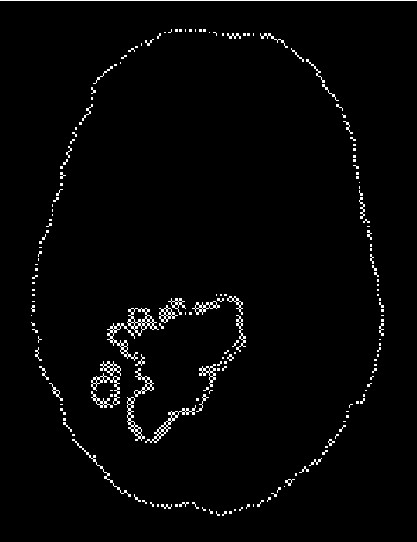}}
	\subfloat[``No'']{
 	\label{nonConnectivityFive}
	\includegraphics [width=0.19\textwidth]{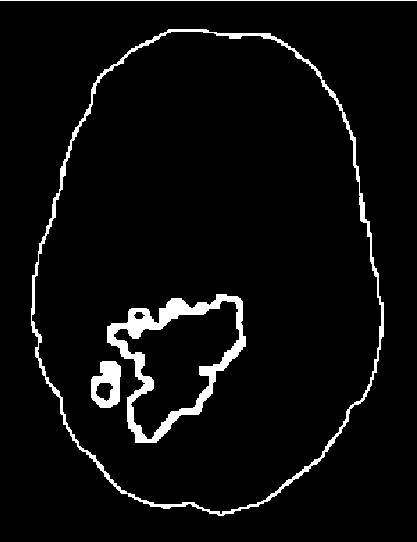}}%
	\caption{Extracted target points from a brain-nidus segmented image using the available connectivity patterns.
	Each pattern results in a different number of points. The number of points for ``Vertex,'' ``Edge,'' ``Face,'' and ``No'' patterns is 21510, 26387, 47306, and 91906, respectively. 
	For simplicity, the points in one volumetric slice are depicted.}
	\label{nonConnectivity}
\end{figure}

The number of source points is fixed during deformation because the mesh remains topologically the same; 
however, the number of target points can be adjusted to improve the accuracy of the displacement vector of a source point. 
Connectivity patterns are used to control the number of extracted target points.
The connectivity patterns prohibit the selection of points that are too close to each other.
A ``vertex'' (26-connectivity), an ``edge'' (18-connectivity), or a ``face'' (6-connectivity) pattern avoids the selection 
of neighboring voxels connected via a ``vertex'', an ``edge'' or a ``face,'' respectively.
The ``no'' option disables the connectivity patterns, hence maximizing the number of selected points. 
Figure \ref{nonConnectivity} depicts the extracted target points using the available patterns. 
Table \ref{meshDeformationPerformance} presents qualitative results on mesh deformation using those patterns. 
The higher the number of target points, the more accurate but slower the computation. 
For the experiments in this study, a ``face'' pattern is employed to balance the trade-off between accuracy and speed.

\begin{table}[htb]
\caption{Performance of deformation for a nidus geometry (Figure \ref{nidusDeformation}) using the available non-connectivity patterns. 
HD is a Hausdorff Distance metric. $\text{HD}_{10}$ corresponds to the mesh fidelity after a deformation step of 10 iterations. 
$\text{HD}_0 = 1.79$ mm is the mesh fidelity before the deformation. The smaller the HD value, the higher the fidelity. 
The experiment was conducted on a machine with an Intel i7-2600@3.40 GHz CPU, and 16 GB of RAM.}
\centering
\begin{tabular*}{\hsize}{@{\extracolsep{\fill}}cccc@{}}
\hline
Pattern 	& \#Target Points & $\text{HD}_{10}$ (mm) & Time (sec)\\\hline
``Vertex'' 	&$2770$	& $0.97$&	$21.01$\\
``Edge''	&$3498$	& $0.91$&	$22.57$\\
``Face''	&$6213$	& $0.88$&	$26.19$\\
``No''		&$11755$	& $0.84$&	$35.27$\\
\hline
\end{tabular*}
\label{meshDeformationPerformance}
\end{table}

\section{Evaluation Results} \label{evaluation}
The CBC3D software is evaluated on four segmented images. Table \ref{imagePropertiesCBC} lists the image data. Cases 1 and 4 are obtained from the Neurosurgery Department at Stony Brook University. Cases 2 and 3 are obtained from Kitware\footnote{https://www.kitware.com}.
In case 4, the volume image is obtained after combining 4182 bitmap slices using an ITK Tile Image Filter. The experiments for cases 1-2 are performed on a DELL workstation with 8 hardware cores, Intel(R) Core(TM) i7-2600 CPU @ 3.40 GHz, and 16 GB RAM.
The experiments for cases 3-4 are performed on a DELL workstation with 24 hardware cores, Intel(R) Xeon(R) CPU E5-2697v2@2.70 GHz, and 757 GB RAM.

\begin{table}[htb]
\caption{Segmented imaging data for experimental evaluation.}
\centering
\begin{tabular}{ccccc}
\hline
Case & Type & Materials& Image Spacing $({\text{mm}}^3)$ & Image Size $({\text{voxels}}^3)$\\\hline
$1$ & Isotropic &Aneurysm & $1.00 \times 1.00 \times 1.00$ &  $512 \times 512 \times 508$\\
$2$ & Anisotropic&Brain-Tumor & $0.48 \times 0.48 \times 1.00$ &  $384 \times 512 \times 176$\\
$3$ & Anisotropic&Brain-AVM& $0.70 \times 0.70 \times 1.60$ &  $320 \times 320 \times 100$\\
$4$ & Anisotropic&Lumen-LVIS Stent & $0.012 \times 0.012 \times 0.024$ &  $1001 \times 1001\times 4182$\\
\hline
\end{tabular}
\label{imagePropertiesCBC}
\end{table}

CBC3D is compared with four image-to-mesh conversion codes: CGAL's 3D Mesh Engine\footnote{https://doc.cgal.org/4.5.2/Manual} (v4.5.2), CLEAVER \cite{LatticeCleaving} (v1.5.4),
Lattice-Derefinement (LD) \cite{LD_Chernikov}, and PODM \cite{foteinos2014high, Feng2017Scalable3H}.
CGAL and CLEAVER are open-source codes. LD and PODM are codes previously developed by CRTC\footnote{https://crtc.cs.odu.edu}.
CBC3D, CLEAVER, and LD are lattice-based methods. CGAL and PODM are Delaunay-based methods. 

Table \ref{inputParametersCBC} lists the parameters used for each software in our evaluation. For PODM, parameter $\delta > 0$ specifies the size of the mesh. 
The smaller the $\delta$, the higher the element count. As mentioned previously, CLEAVER initially constructs a high-quality BCC lattice that covers the input image. 
Next, it locally warps or cuts the lattice so that it will conform to multi-material arbitrary surfaces. 
The violation parameters $\alpha_{short}, \alpha_{long}$ decide the trade-off between snapping/warping and stencil cleaving; therefore,
they implicitly control the size of the mesh. In this evaluation, the default violation parameters are used, thus a worse case 
minimum dihedral angle of $2.76^\circ$ and worst case maximum dihedral angle of $175.42^\circ$ is achieved. LD allows for guaranteed bounds on the smallest dihedral angle and on the distance between the boundaries
of the mesh and the boundaries of the materials \cite{LD_Chernikov}. The highest possible fidelity and a minimum dihedral angle of $15^\circ$ is 
specified for all the experiments.

\begin{table}[htb]
\caption{Input parameters for experimental evaluation.}
\centering
\begin{tabular}{llll} 
\hline
Method & Parameter& Value & Description\\\hline
\multirow{6}{*}{CBC3D}& \multirow{2}{*}{$lattice_{sp}$} & $10$ mm (Cases 1-3) & \multirow{2}{*}{\footnotesize lattice spacing}\\
& & $0.15$ mm (Case 4) & \\
&$F$&	$0.95$&	\footnotesize fidelity \\
&$connectivity$& ``no''& \footnotesize pattern for target point extraction \\
&\multirow{2}{*}{$N_{iter}$} &	$5$ (Cases 1-3) &\multirow{2}{*}{\footnotesize number of smoothing iterations}\\
& &	$7$ (Case 4) &\\\hline
\multirow{5}{*}{CGAL} & $facet\_angle$&	$30^\circ$ & \footnotesize lower angle bound for surface facets\\
&$facet\_size$&	$2$&	\footnotesize  radii upper-bound for Delaunay balls\\
&$facet\_distance$&	$2$&	\footnotesize face distance upper bound \\
&$cell\_radius\_edge\_ratio$&	$1.5$&	\footnotesize radius-edge ratio upper bound\\
&\multirow{3}{*}{$cell\_size$}&	$2.5$ (Cases 1-2)&	\multirow{3}{*}{\footnotesize circumradii upper-bound} \\
& &	$1.5$ (Case 3)& \\
& &	$0.2$ (Case 4)& \\\hline
\multirow{2}{*}{CLEAVER} & $\alpha_{short}$ &	$0.357$ &	\footnotesize diagonal edge threshold for edge-cuts\\
& $\alpha_{long}$ &	$0.203$ &	\footnotesize axis-aligned threshold for edge-cuts\\\hline
\multirow{3}{*}{LD} & $i2m$ &	$0$ &	\footnotesize image-to-mesh distance\\
& $m2i$ &	$0$ &	\footnotesize mesh-to-image distance\\
& $angle$&	$15^\circ$ &	\footnotesize minimum dihedral angle\\\hline
\multirow{3}{*}{PODM} & \multirow{3}{*}{$\delta$} & $1.2$ (Cases 1-2) & \multirow{3}{*}{\footnotesize element size}\\
& & $0.6$ (Case 3) & \\
& & $0.02$ (Case 4) & \\
\hline
\end{tabular}
\label{inputParametersCBC}
\end{table}

\begin{figure}[htb]
\centering	
\subfloat{
\label{aneurysmCBC3D}
\includegraphics[width=0.185\textwidth]{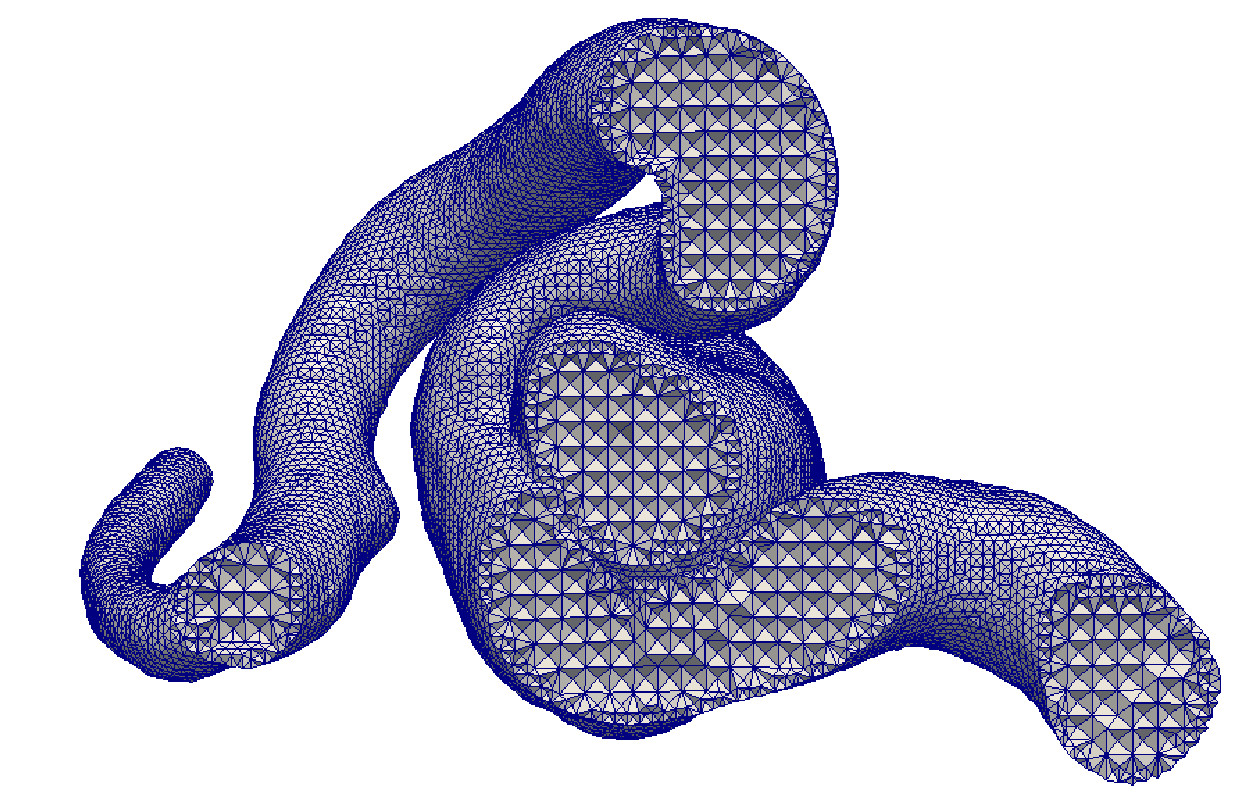}}
\subfloat{
\label{aneurysmCGAL}
\includegraphics[width=0.185\textwidth]{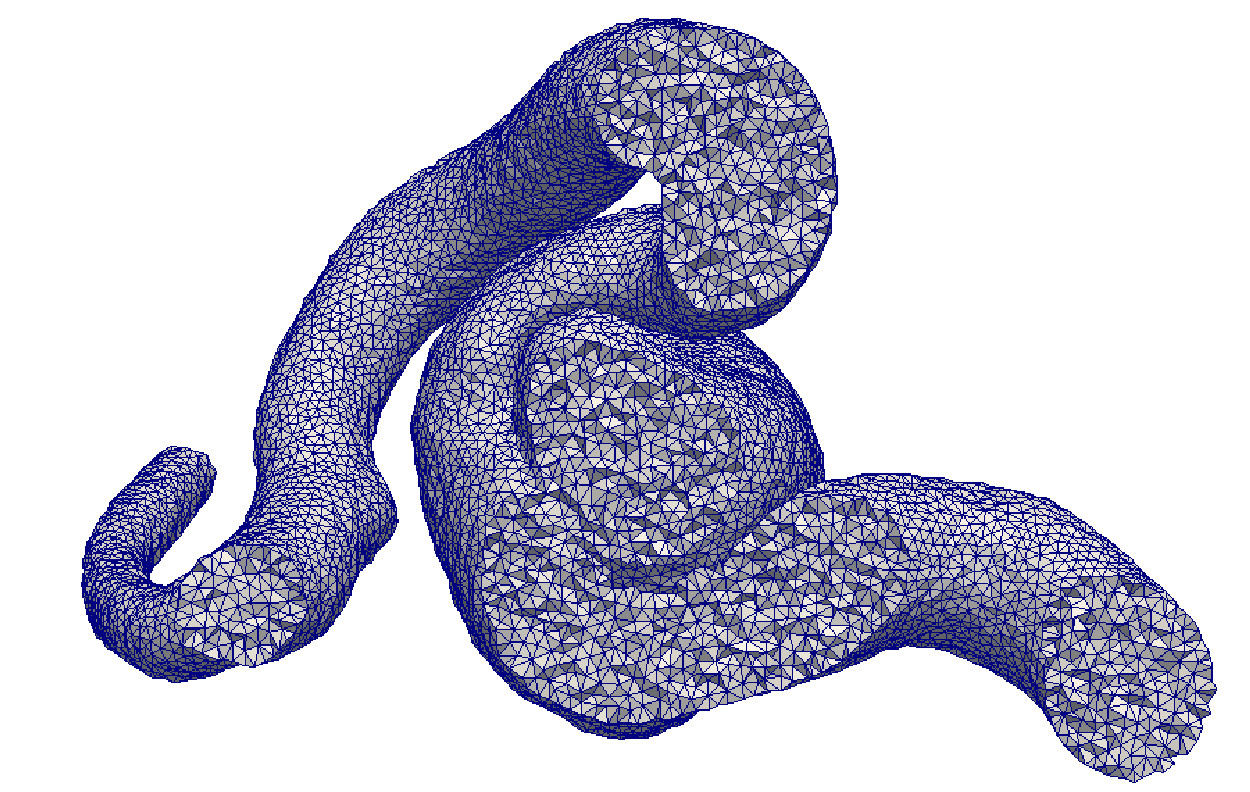}}
\subfloat{
\label{aneurysmCLEAVER}
\includegraphics[width=0.185\textwidth]{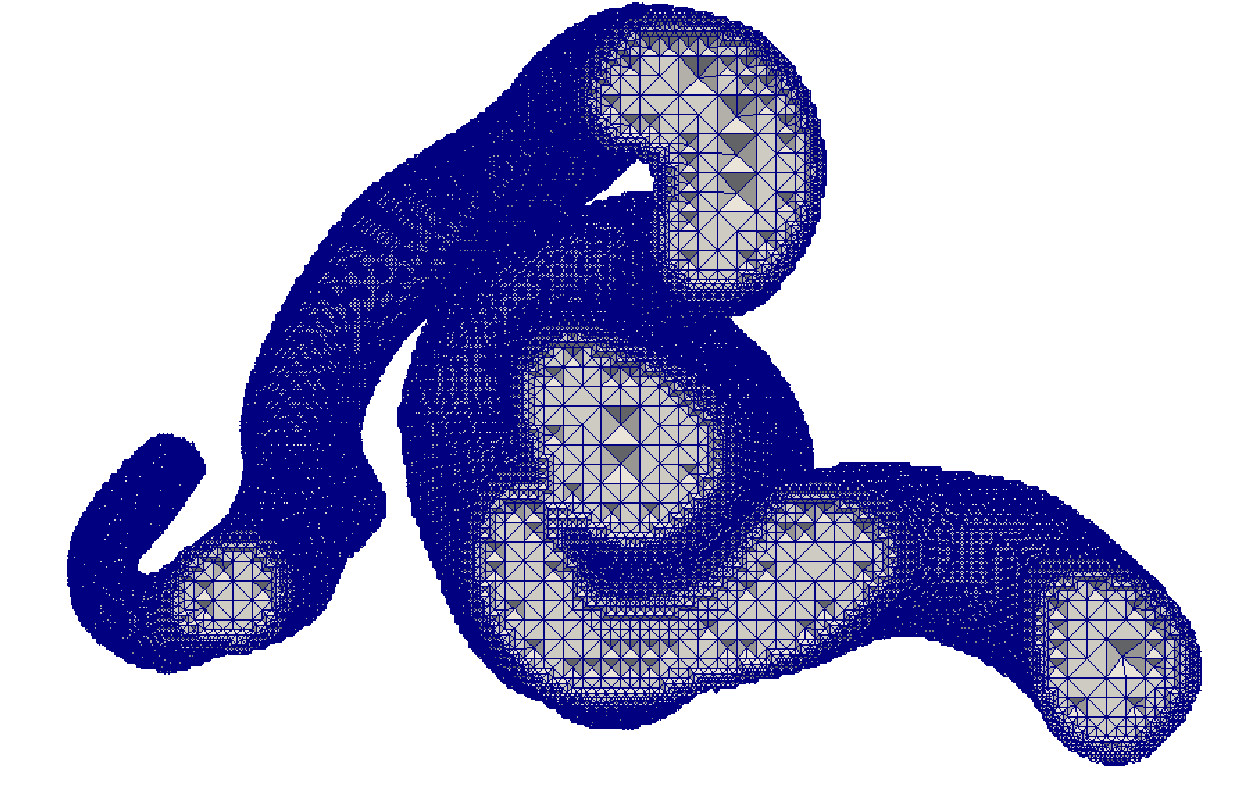}}
\subfloat{
\label{aneurysmLD}
\includegraphics[width=0.185\textwidth]{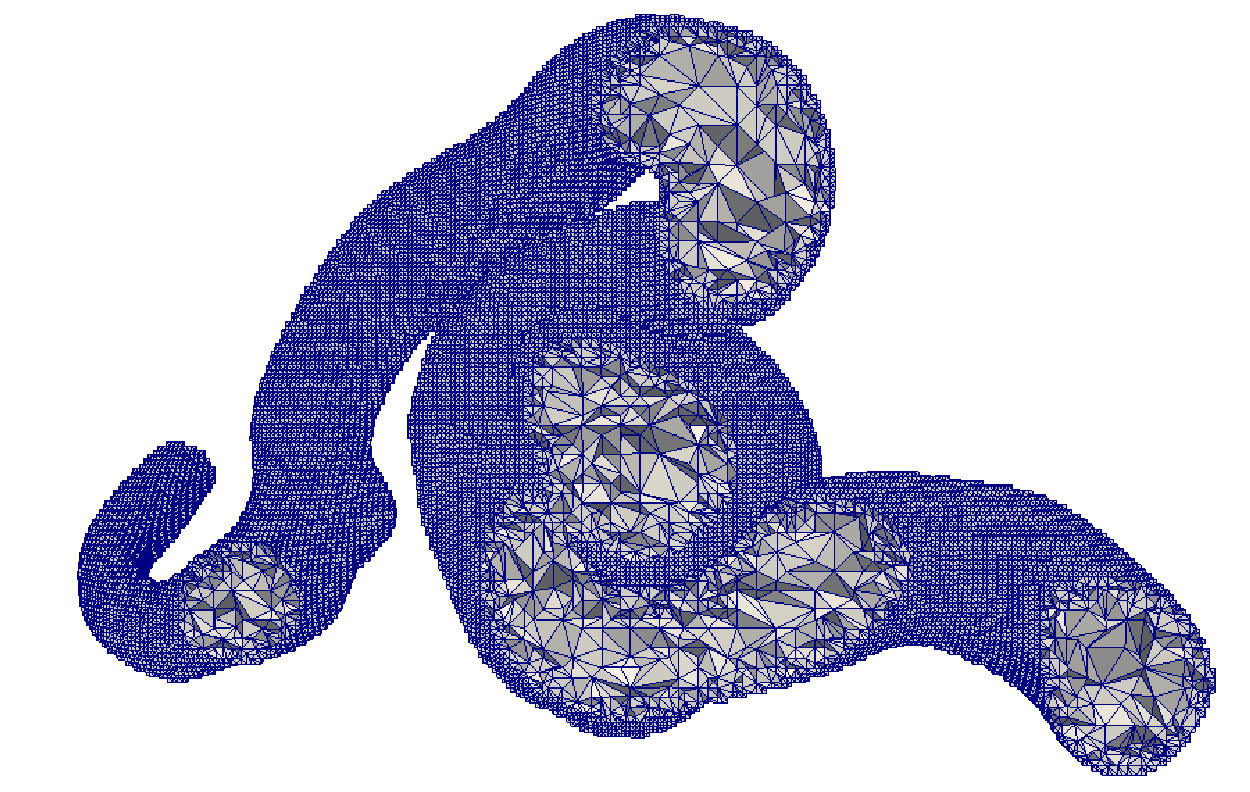}}
\subfloat{
\label{aneurysmPODM}
\includegraphics[width=0.185\textwidth]{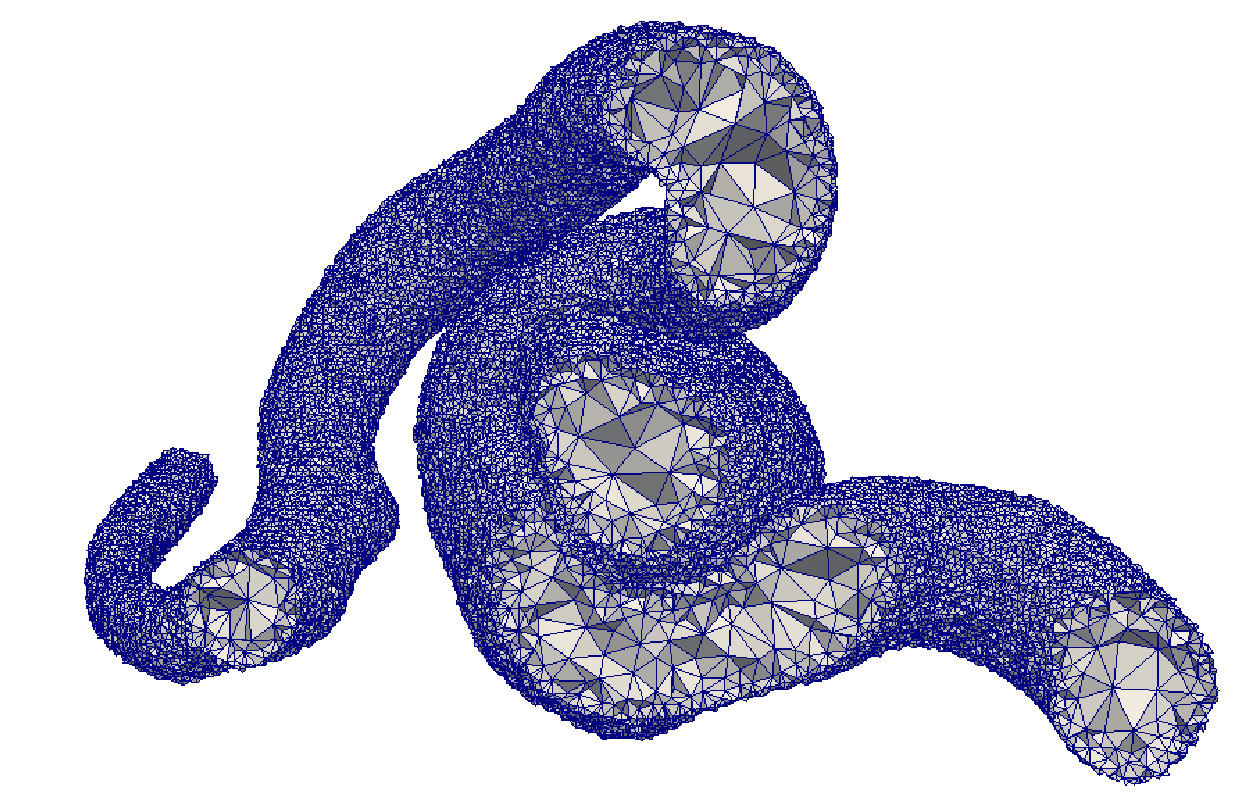}}

\subfloat{
\label{brainTumorCBC3D}
\includegraphics[width=0.185\textwidth]{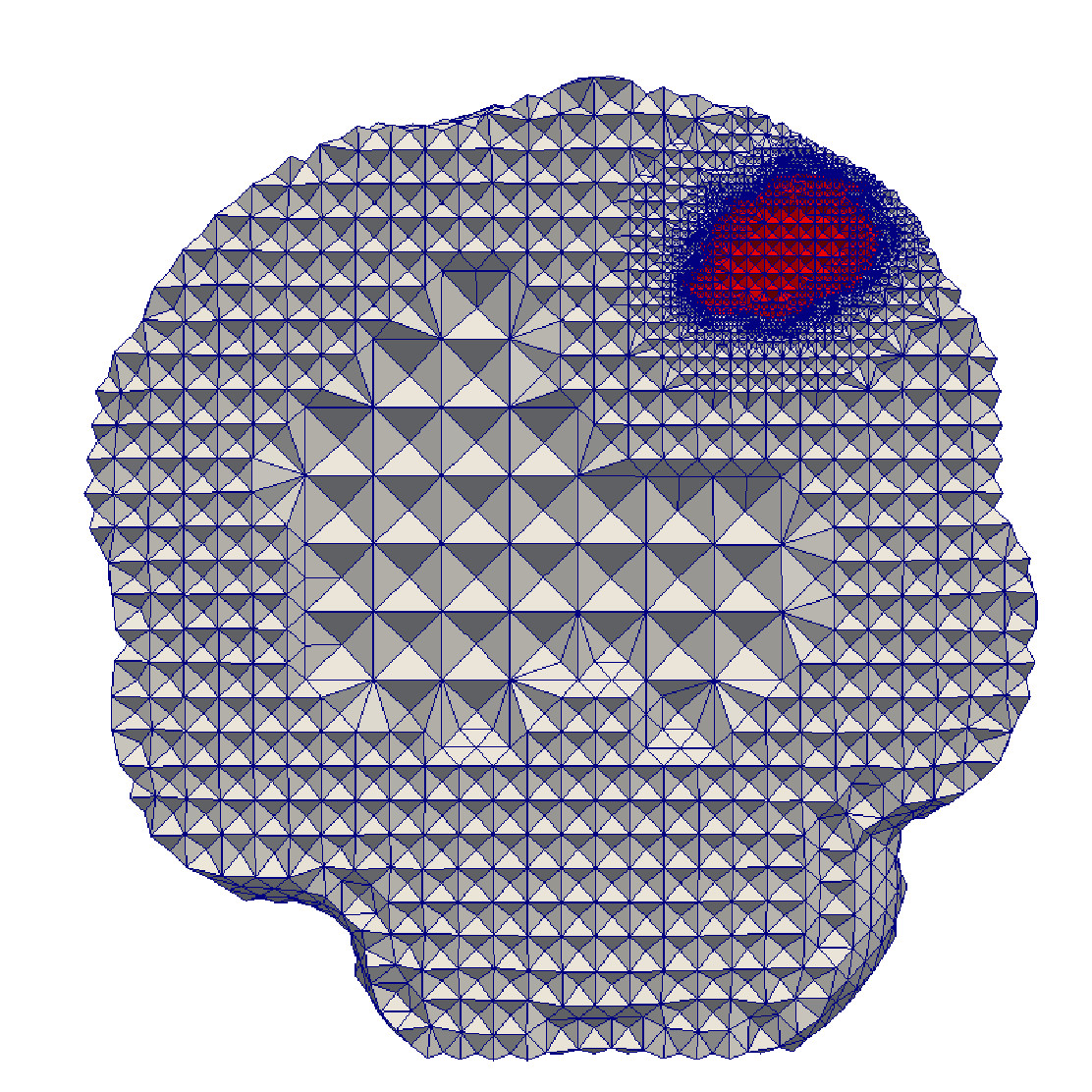}}
\subfloat{
\label{brainTumorCGAL}
\includegraphics[width=0.185\textwidth]{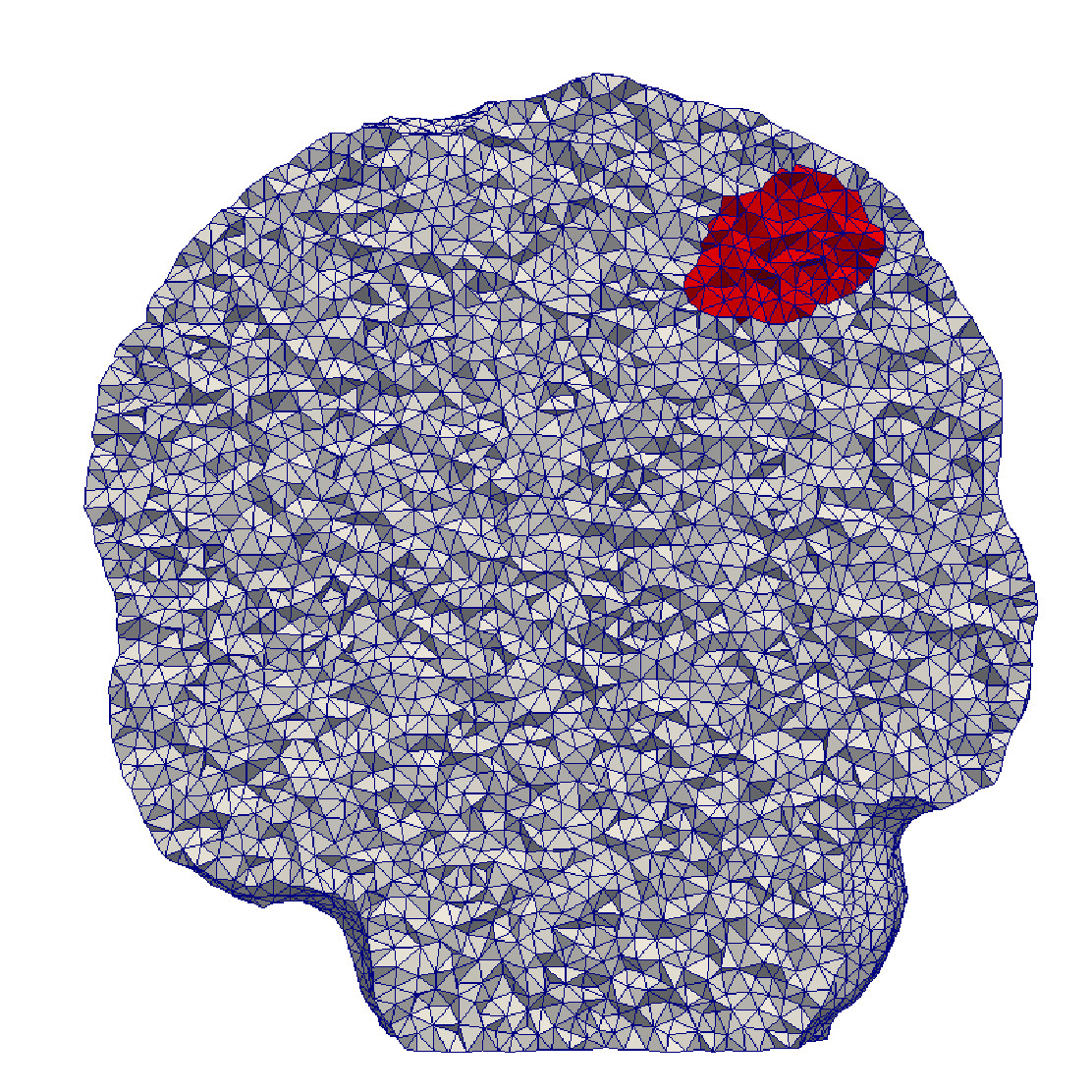}}
\subfloat{
\label{brainTumorCLEAVER}
\includegraphics[width=0.185\textwidth]{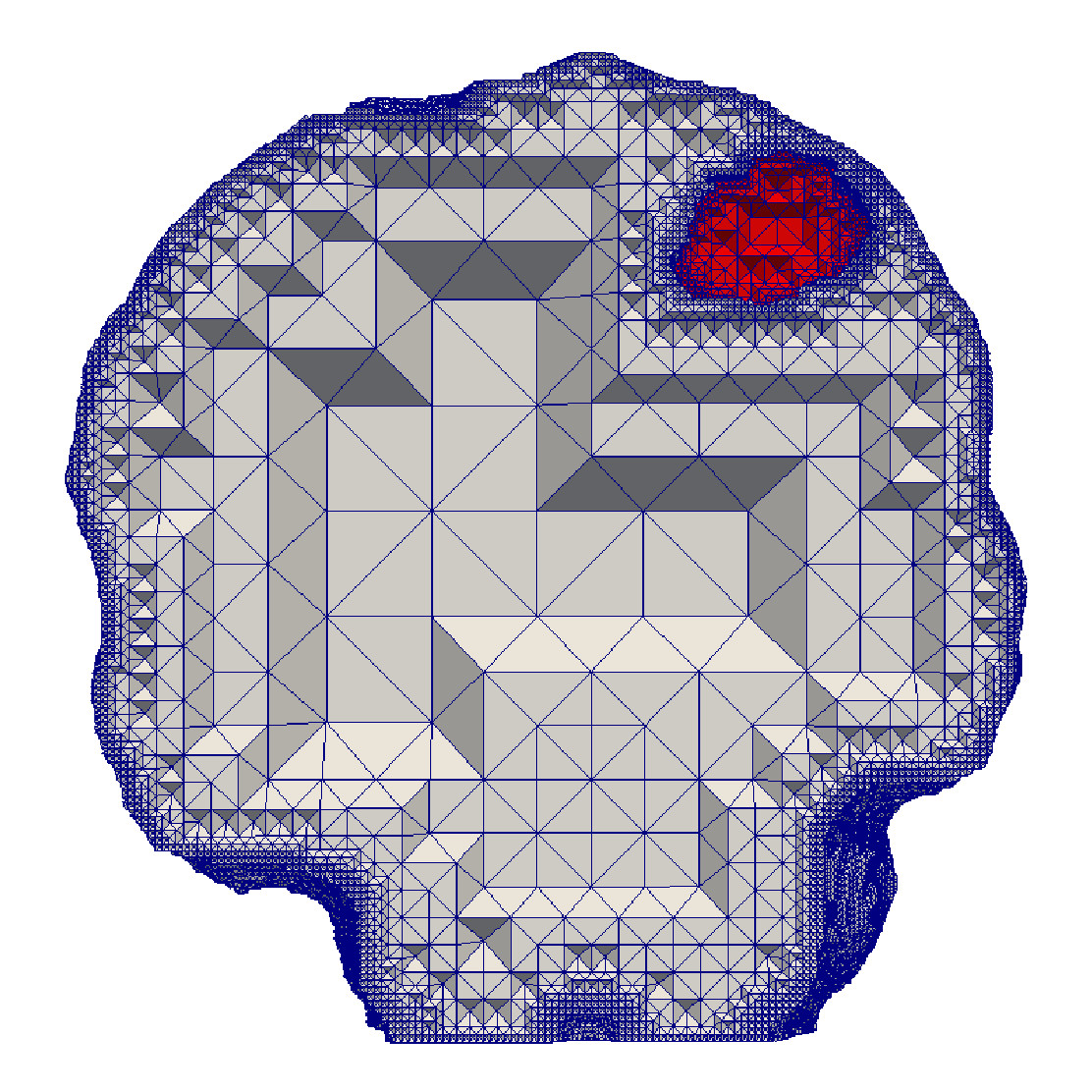}}
\subfloat{
\label{brainTumorLD}
\includegraphics[width=0.185\textwidth]{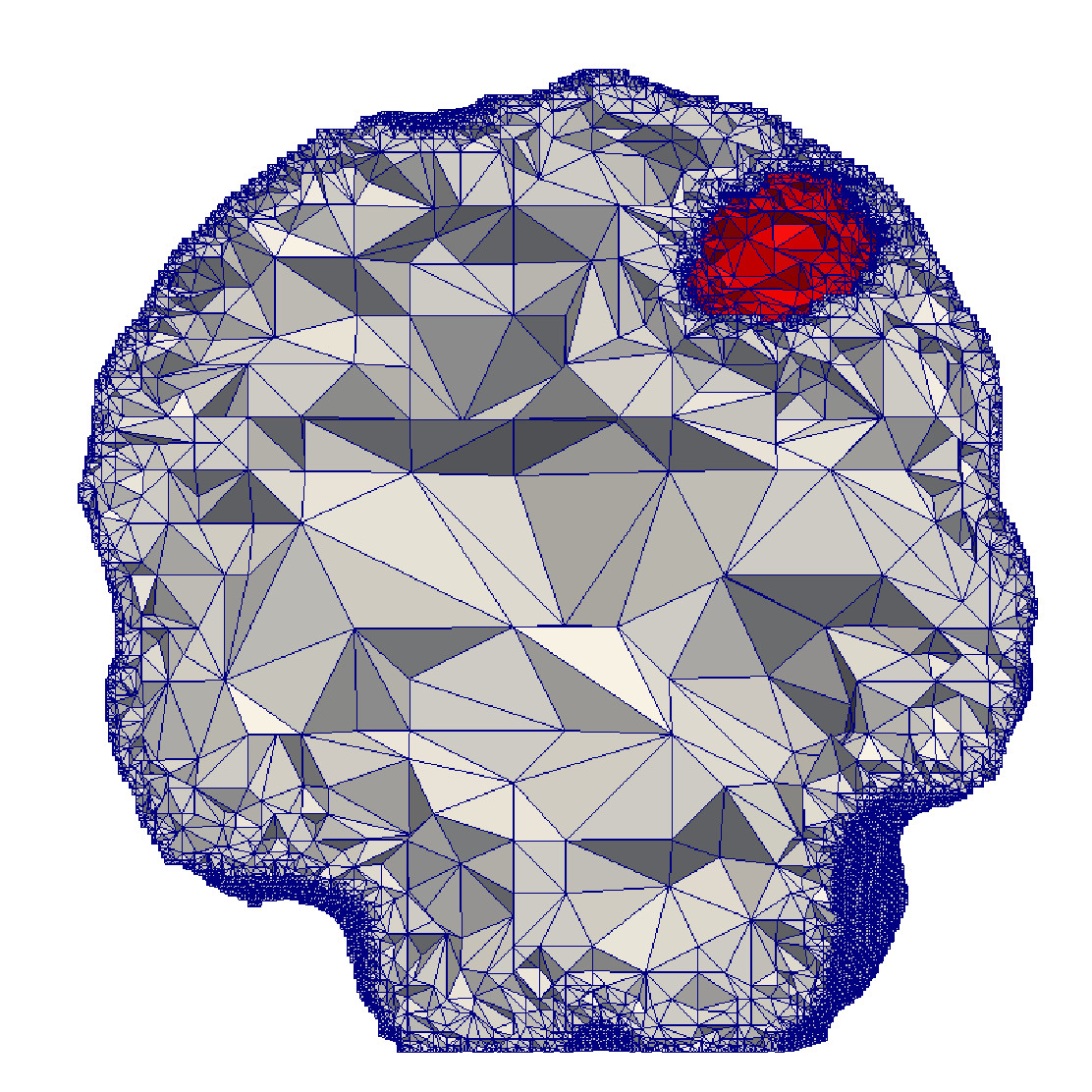}}
\subfloat{
\label{brainTumorPODM}
\includegraphics[width=0.185\textwidth]{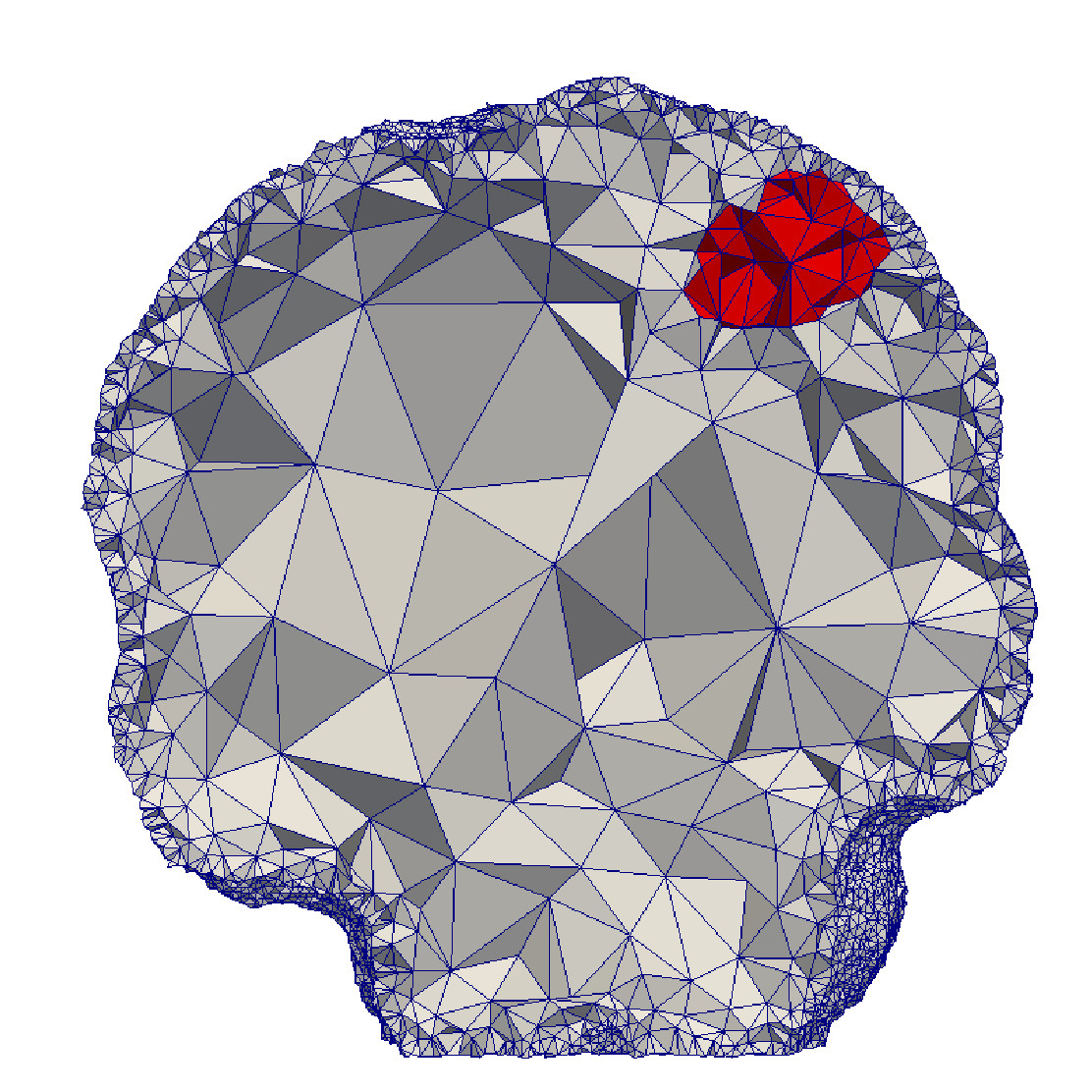}}\

\subfloat[CBC3D]{
\label{brainAVMCBC3D}
\includegraphics[width=0.185\textwidth]{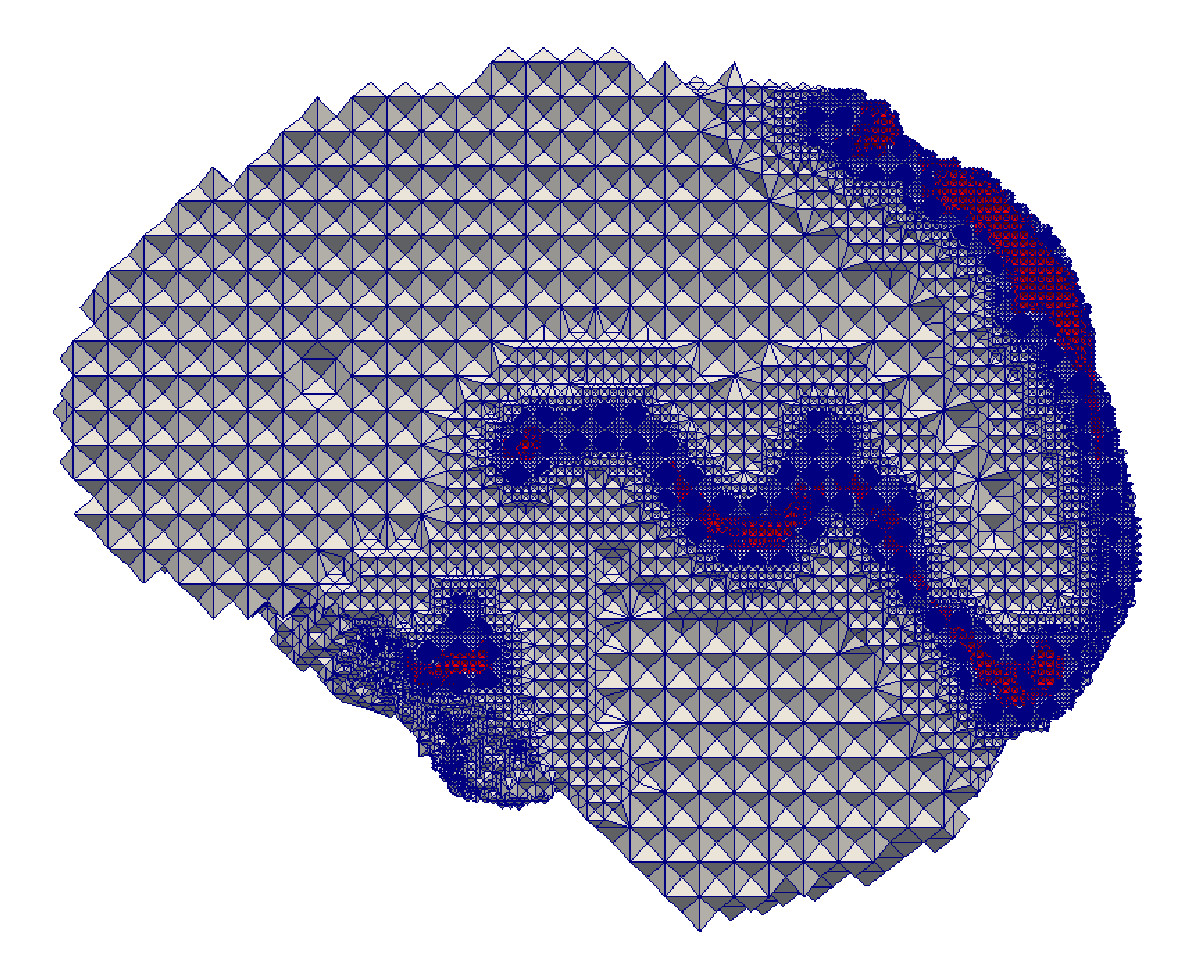}}
\subfloat[CGAL]{
\label{brainAVMCGAL}
\includegraphics[width=0.185\textwidth]{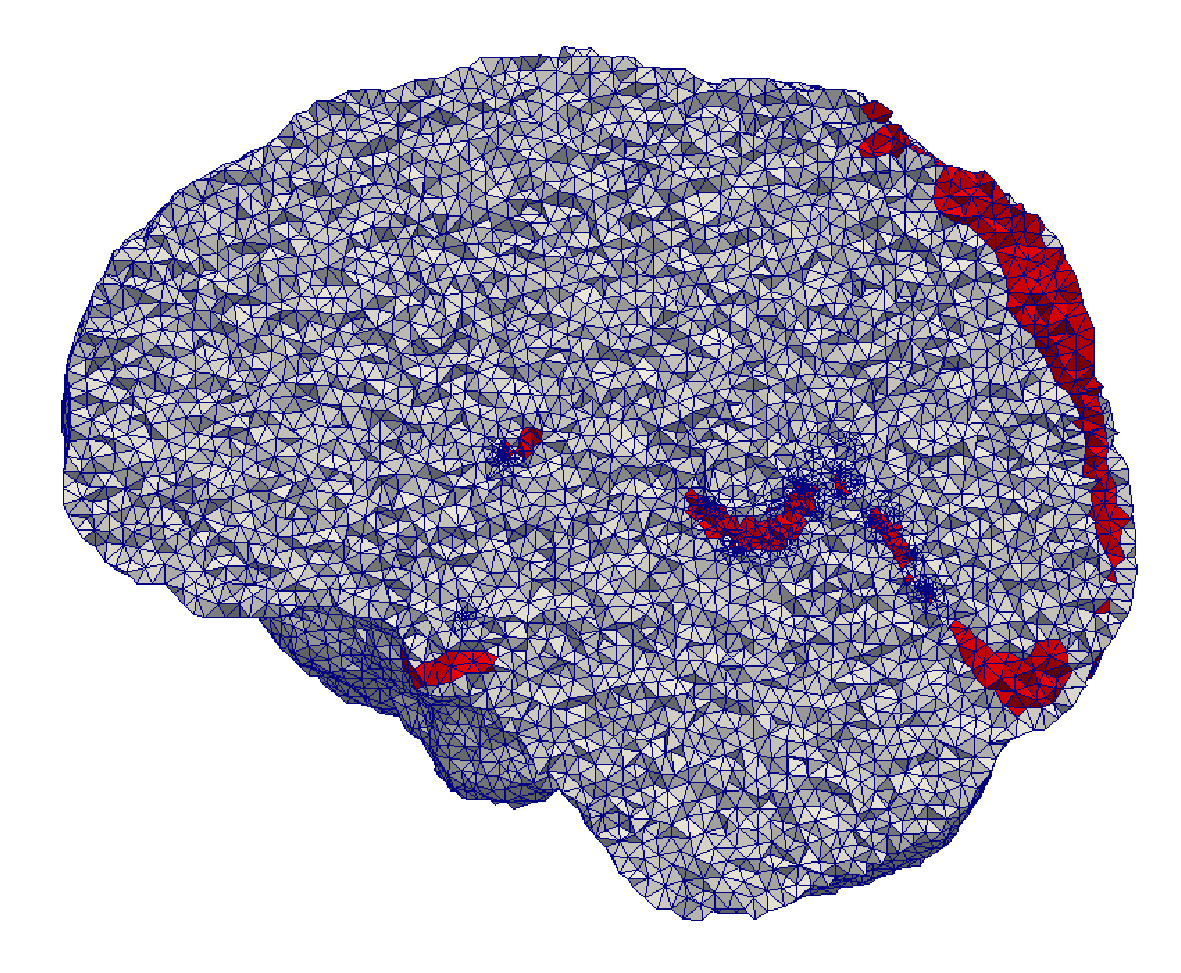}}
\subfloat[CLEAVER]{
\label{brainAVMCLEAVER}
\includegraphics[width=0.185\textwidth]{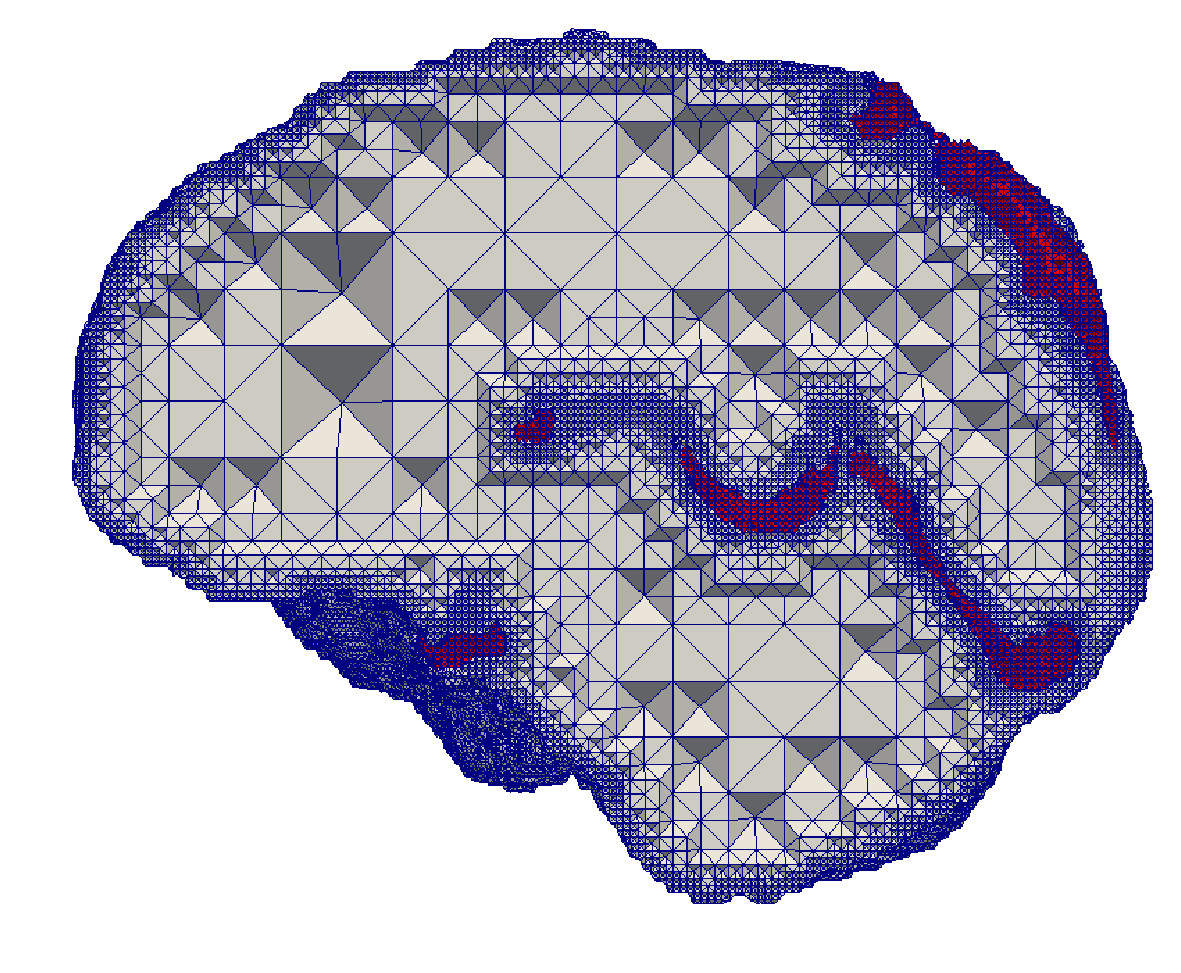}}
\subfloat[LD]{
\label{brainAVMLD}
\includegraphics[width=0.185\textwidth]{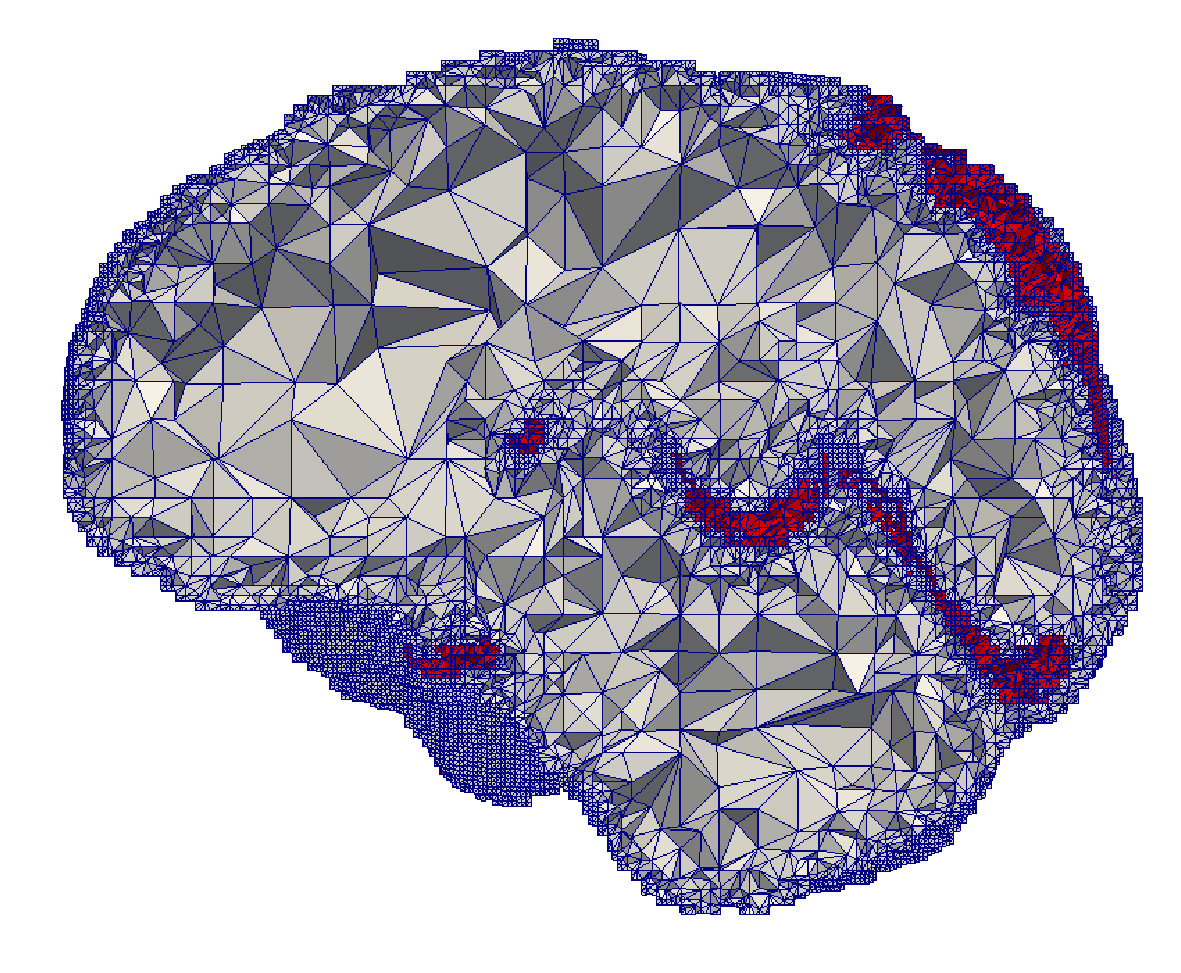}}
\subfloat[PODM]{
\label{brainAVMPODM}
\includegraphics[width=0.185\textwidth]{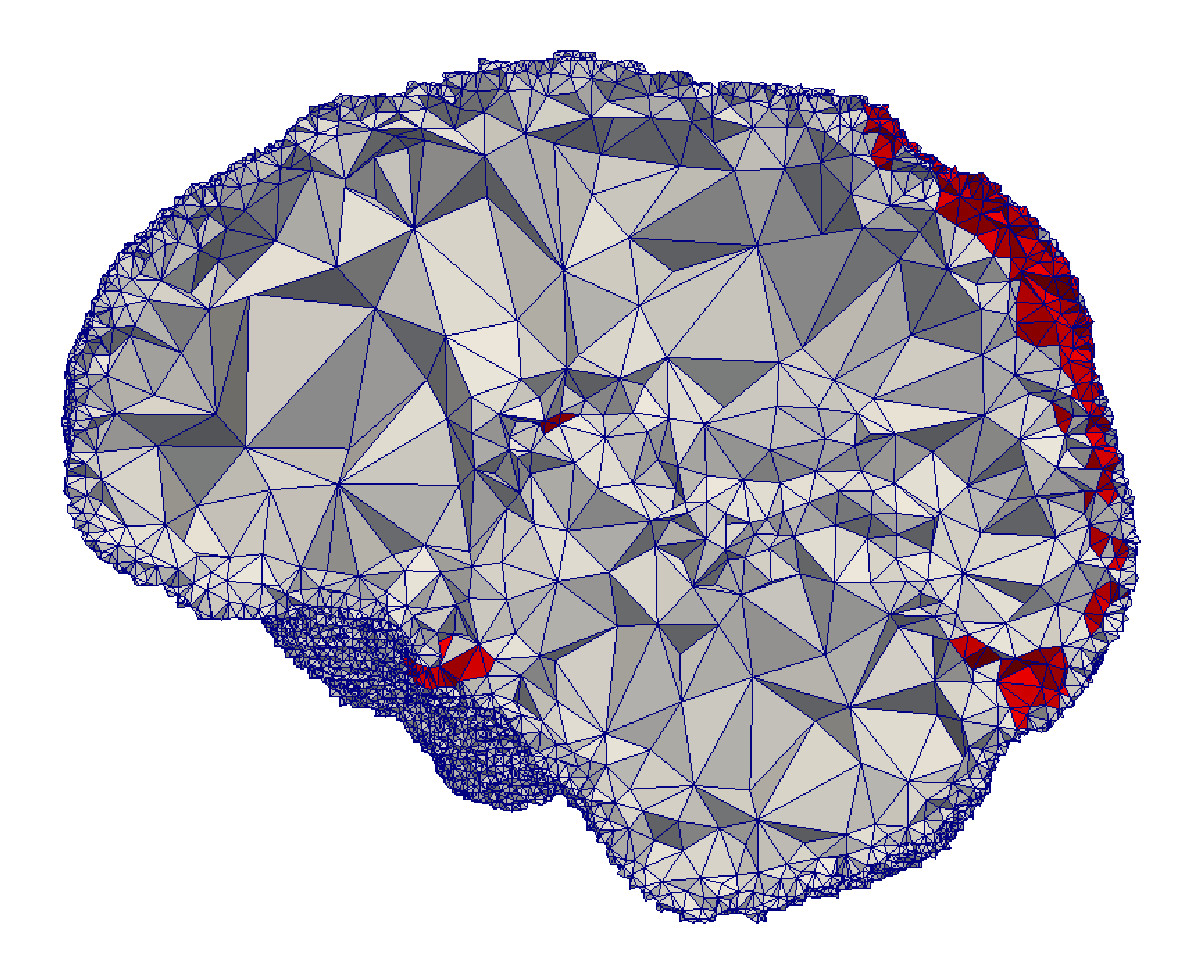}}
\caption{Cuts of the generated tetrahedral meshes. The top, middle, and bottom row correspond to cases 1, 2, and 3, respectively. 
Each column depicts meshes that are generated with a single method. Identical cut section planes are used for all the meshes in a single case.
The growth from small to large elements varies among the methods. The quality of these meshes is 
evaluated using a min/max dihedral angle metric and an element angle distribution in 5-deg increments.}
\label{cutSections}
\end{figure}

\begin{figure}[htb]
\centering	
\subfloat[CBC3D]{
\label{cutSectionsLVISOne}
\includegraphics [width=0.48\textwidth]{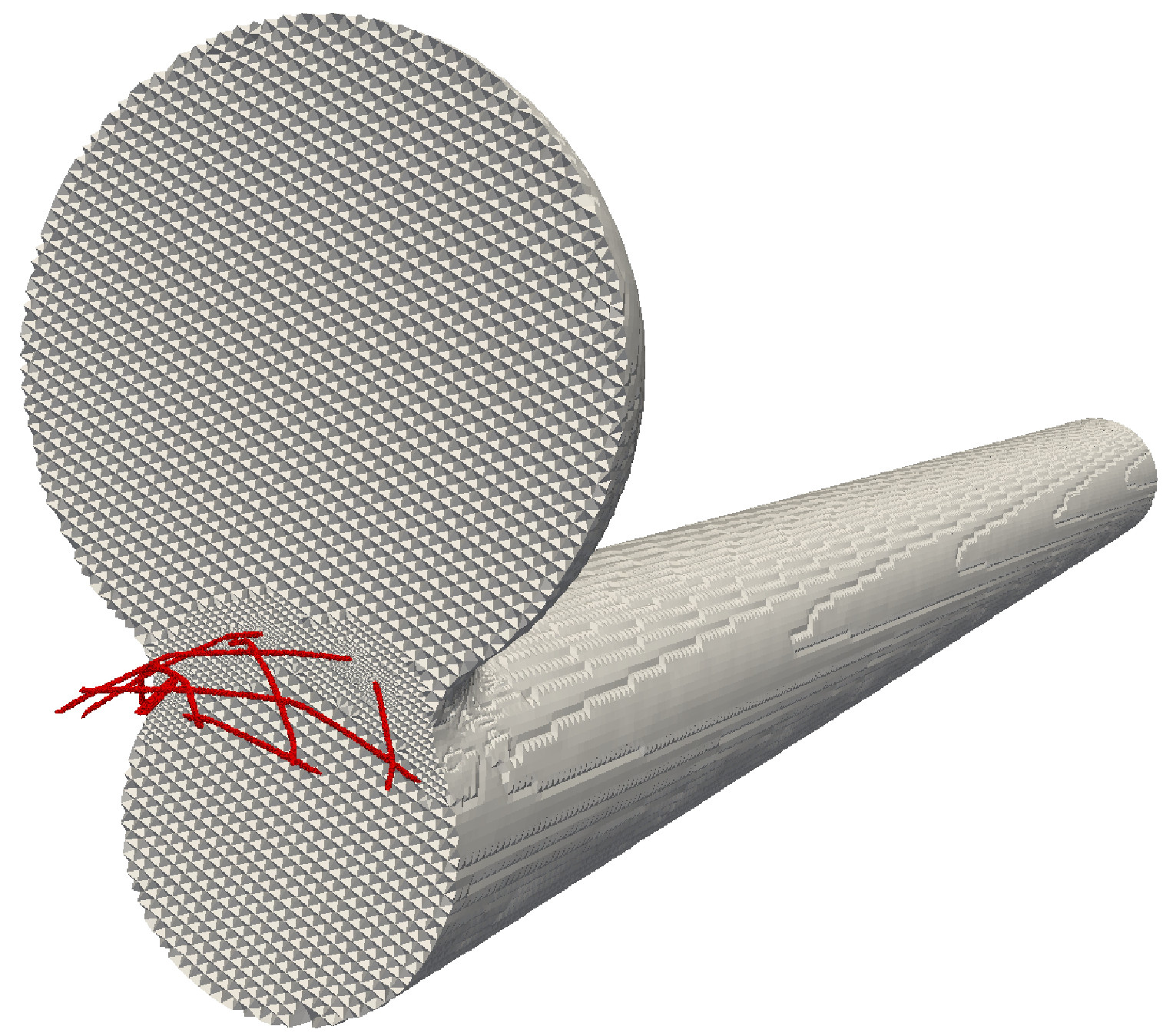}}
\subfloat[PODM]{
\label{cutSectionsLVISTwo}
\includegraphics [width=0.48\textwidth]{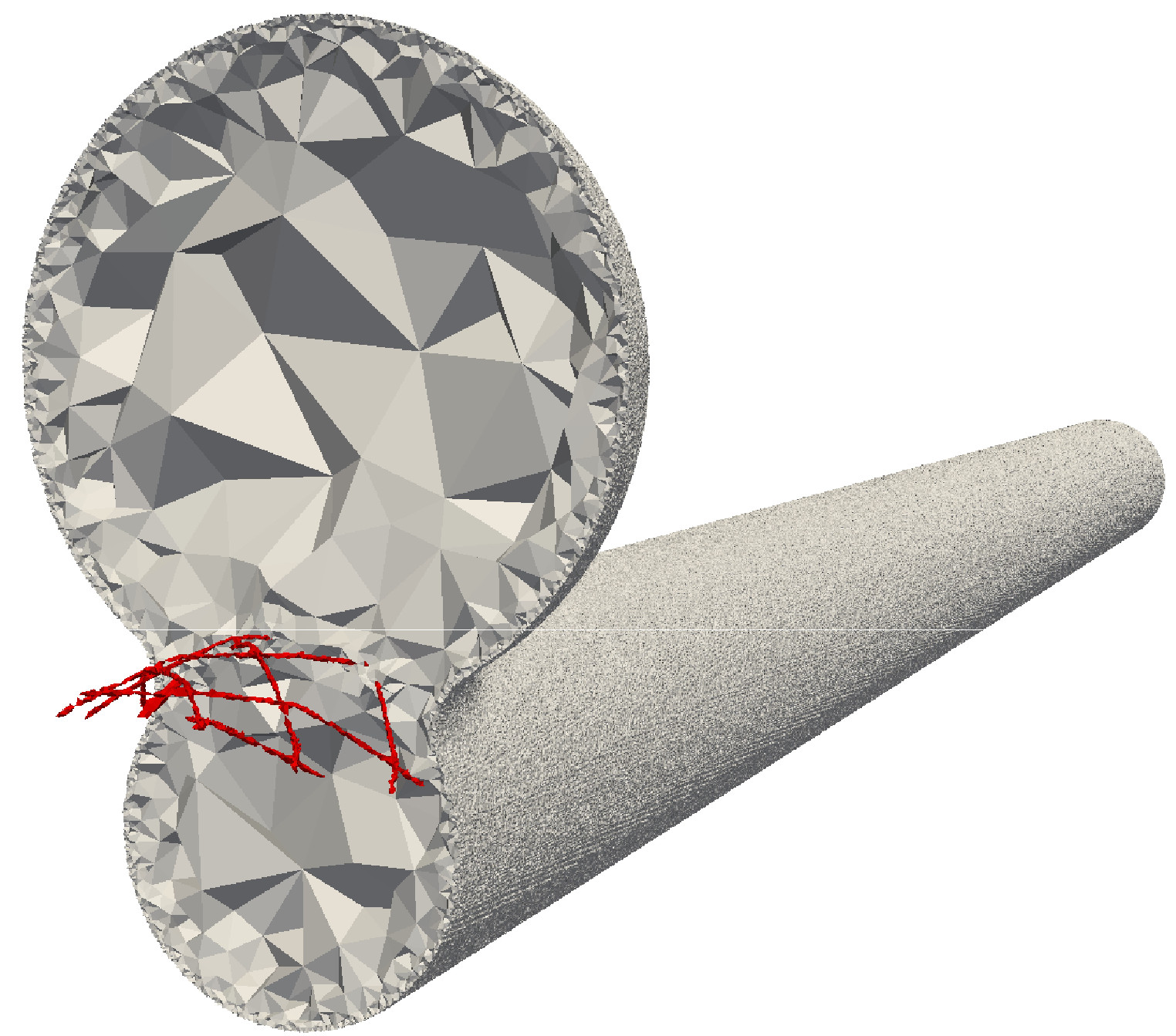}}
\caption{Cuts of the generated tetrahedral meshes of the Lumen-LVIS Stent.}
\label{cutSectionsLVIS}
\end{figure}

Figures \ref{cutSections} and \ref{cutSectionsLVIS} depict cuts of the generated meshes.
CLEAVER exhibits the most rapid element gradation among all the methods.
CGAL, CLEAVER, and LD fail to generate a mesh for the Lumen-LVIS stent case (Table \ref{imagePropertiesCBC}).
CGAL's image reader cannot read the input image. CLEAVER terminates the program with a 
message: "Cleaver Tet Mesher terminated with an unknown exception".
LD throws an instance of std::bad\_alloc because it exceeds the physical memory of the system (757 GB).
For comparison purposes, CBC3D and PODM allocate 354 GB and 141 GB, respectively.

\begin{figure}[htb]
\centering	
\subfloat[CBC3D]{
\label{cutSectionsStentsOne}
\includegraphics [width=0.48\textwidth]{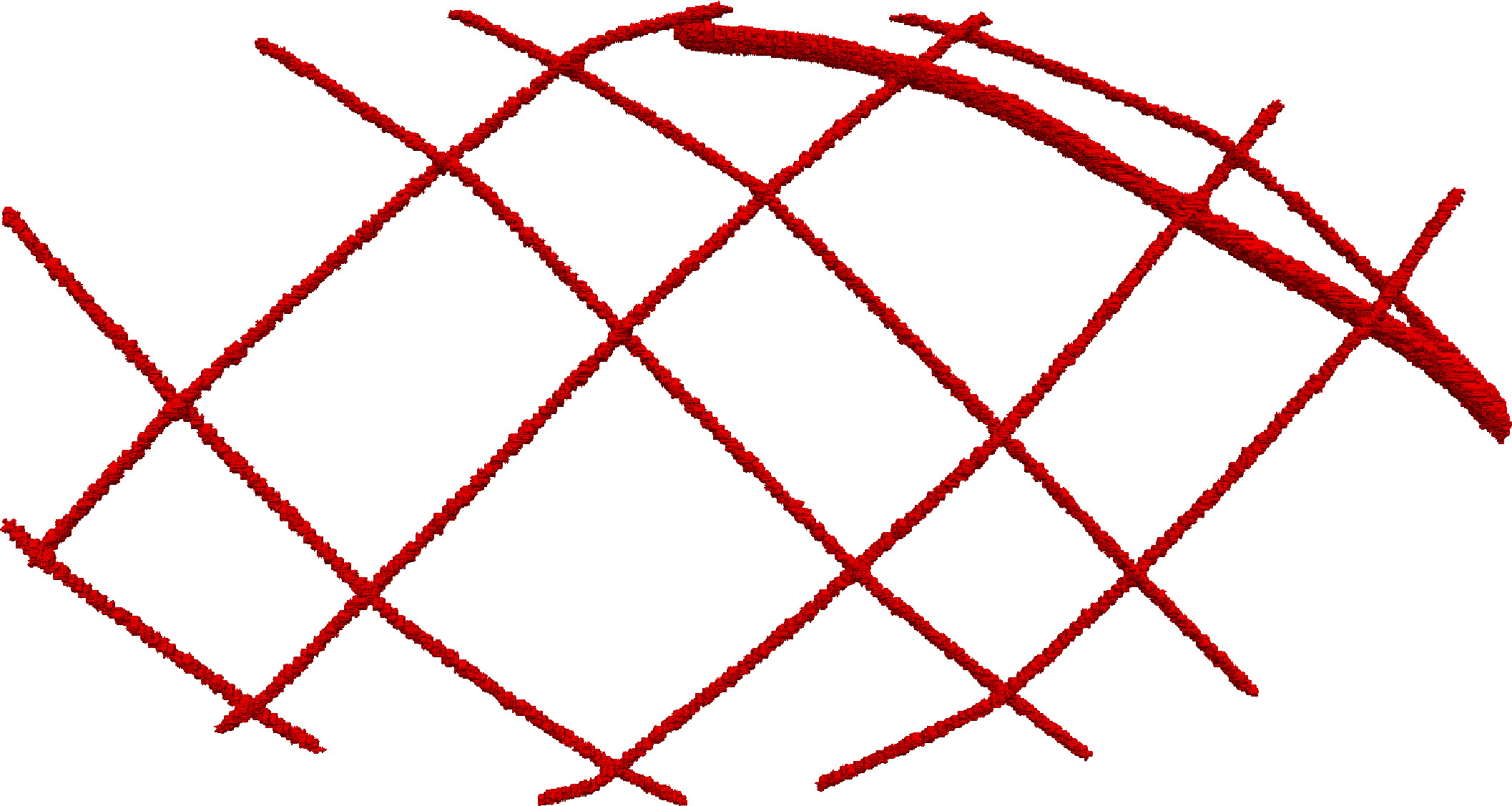}}
\subfloat[PODM]{
\label{cutSectionsStentsTwo}
\includegraphics [width=0.48\textwidth]{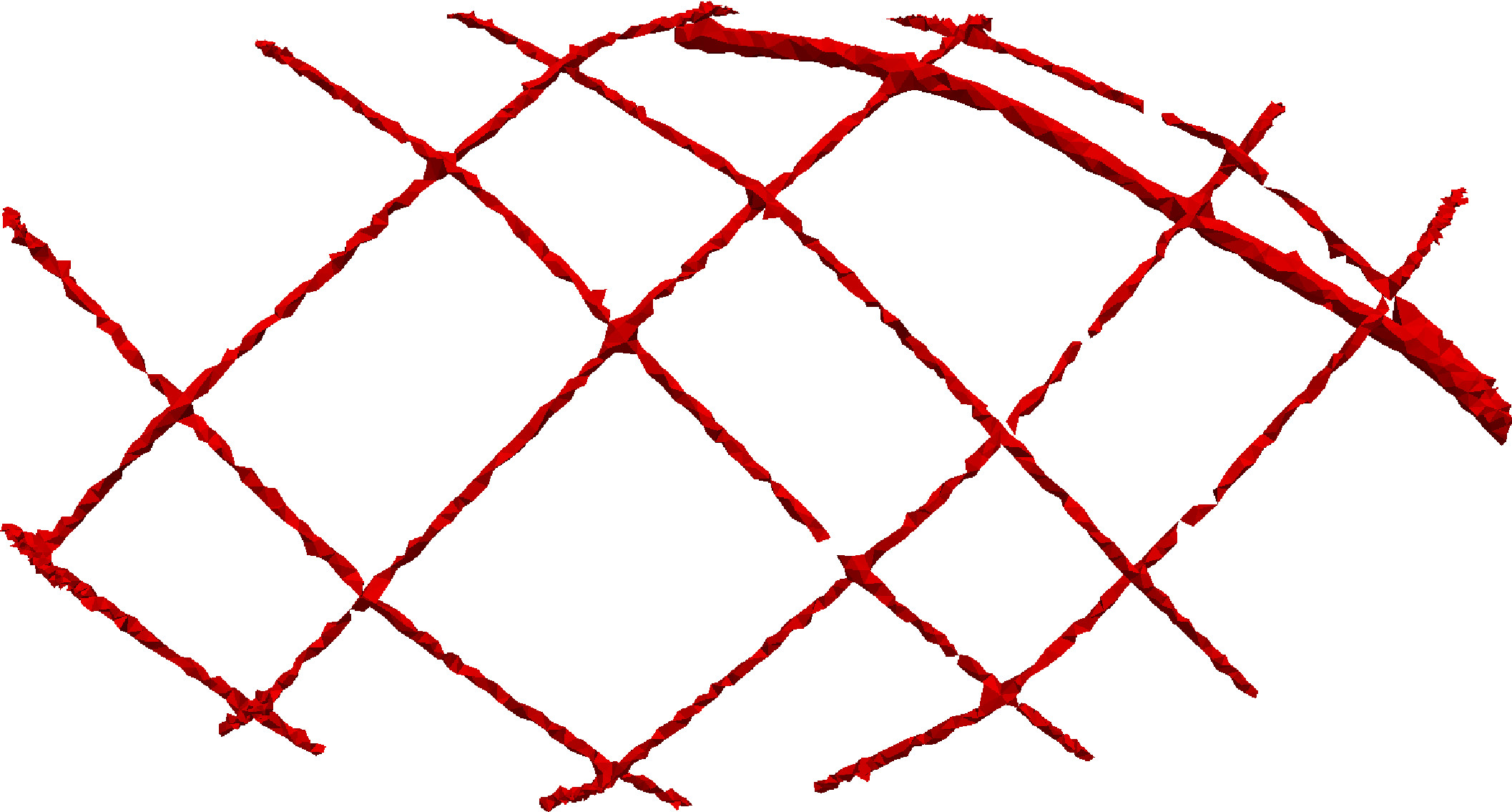}}
\caption{Extracted mesh of the LVIS stent.}
\label{cutSectionsStents}
\end{figure}

Tables \ref{quantitativeResultsCBCtets} - \ref{quantitativeResultsCBCangles} report quantitative results on element count and element quality.
The lattice-based methods (CBC3D, CLEAVER, and LD) result in larger meshes compared to the Delaunay-based methods (CGAL and PODM) mainly 
because the templates impose stricter rules for element subdivisions. 
Nevertheless, the largest mesh in this study is generated by PODM (15.85 M) because a small element size ($\delta=0.02$)
is specified to resolve the LVIS stent. Figure \ref{cutSectionsStents} compares the meshes of the LVIS stent.
PODM does not adequately resolve the features of the stent, hence the mesh appears to be disconnected (Figure \ref{cutSectionsStentsTwo}).

CLEAVER and LD generate meshes with high-quality elements, although no experimental results are available for case 4 (Tables \ref{quantitativeResultsCBCtets} - \ref{quantitativeResultsCBCangles}).
LD imposes a quality bound in the generated mesh (i.e., minimum dihedral angle of $15^\circ$) by adjusting the element count appropriately. 
CBC3D produces meshes of well-shaped elements. Deformation does not significantly affect case 3, hence the angle extrema are similar to those in the adaptive lattice.
The Delaunay-based methods generate meshes of reasonably well-shaped elements.

\begin{table}[htb]
\caption{Evaluation results on element count.}
\centering
\begin{tabular}{c c c c c c}
\hline
\multirow{2}{*}{Case} & \multicolumn{5}{c}{\#Tetrahedra} \\
 & CBC3D & CGAL & CLEAVER & LD & PODM \\\hline
$1$& $272$K& $300$K & $4.33$M & $776$K& $370$K \\ 
$2$& $578$K& $517$K & $3.20$M & $2.54$M& $244$K \\ 
$3$ & $5.05$M& $1.68$M& $3.59$M & $1.48$M & $1.03$M\\
$4$ & $12.98$M & - & -& - & $15.85$M \\\hline
\end{tabular}
\label{quantitativeResultsCBCtets}
\end{table}

\begin{table}[htb]
\caption{Evaluation results on element quality. The minimum and maximum dihedral angles are reported in degrees $\in (0^\circ,180^\circ)$.
The larger the minimum angle and the smaller the maximum angle, the higher the quality.}
\centering
\begin{tabular}{c c c c c c}
\hline
\multirow{2}{*}{Case} & \multicolumn{5}{c}{Dihedral angle (min, max)} \\
 & CBC3D & CGAL & CLEAVER & LD & PODM \\\hline
$1$ & $(5.07^\circ, 171.73^\circ)$& $(12.04^\circ, 162.23^\circ)$& $(25.06^\circ, 126.94^\circ)$ & $(15.00^\circ, 171.05^\circ)$ & $(4.54^\circ, 170.13^\circ)$ \\ 
$2$ & $(8.89^\circ, 166.67^\circ)$& $(12.00^\circ, 162.73^\circ)$& $(11.34^\circ, 153.15^\circ)$ & $(15.00^\circ, 171.86^\circ)$ & $(4.90^\circ, 170.26^\circ)$ \\ 
$3$ & $(29.94^\circ, 116.67^\circ)$& $(1.41^\circ, 176.76^\circ)$& $(5.81^\circ, 157.66^\circ)$& $(15.00^\circ, 170.55^\circ)$& $(4.40^\circ, 170.24^\circ)$\\
$4$ & $(4.95^\circ, 173.10^\circ)$ & - & - & - & $(2.23^\circ, 176.44^\circ)$ \\\hline
\end{tabular}
\label{quantitativeResultsCBCangles}
\end{table}

\begin{figure}[htbp]
\centering	
\includegraphics [width=0.85\textwidth]{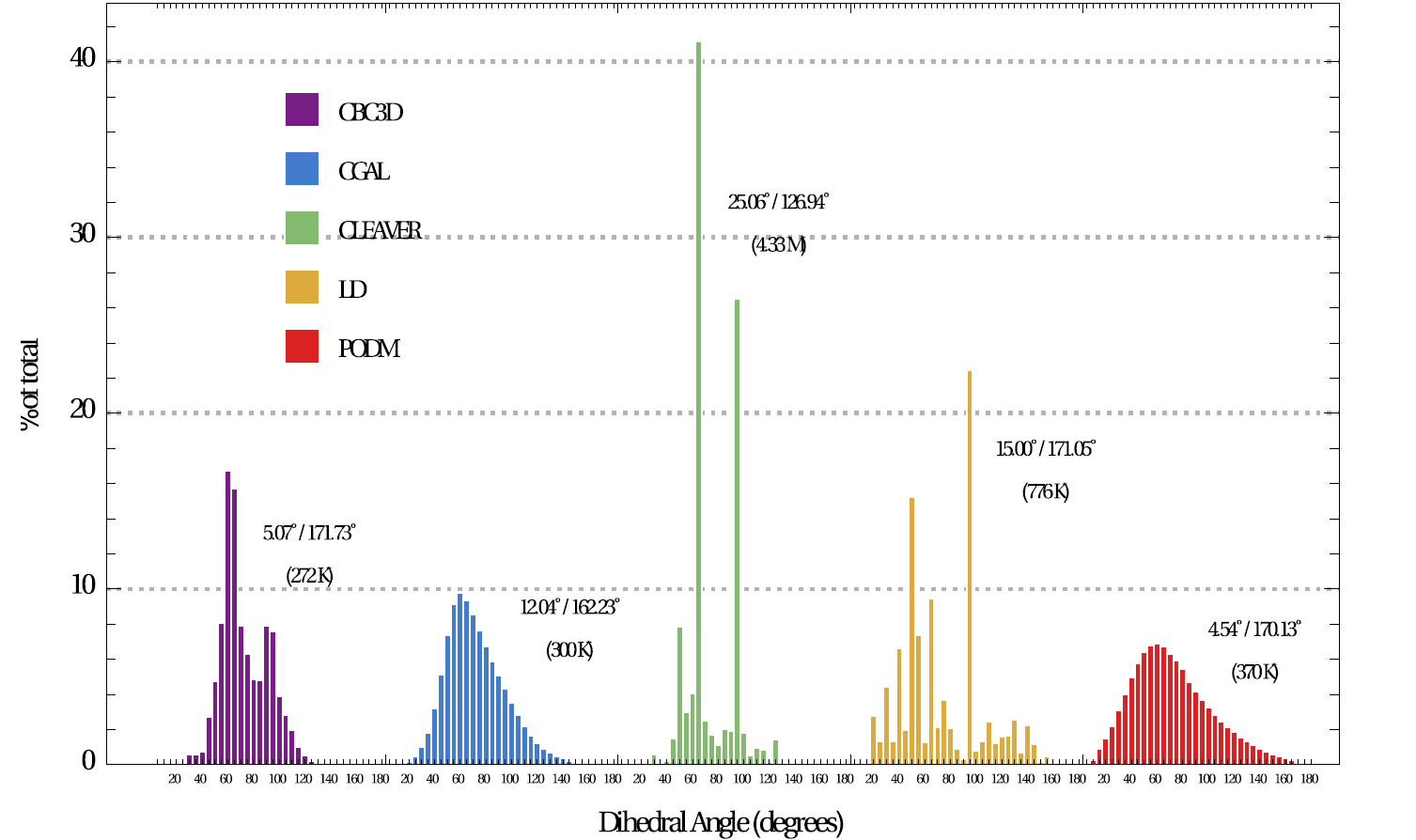}
\caption{Element angle distribution (in 5-deg increments) of Cavernous Aneurysm meshes. The min/max dihedral angles and the element count are reported for each method.}
\label{histogramAneurysm}
\end{figure}

\begin{figure}[htbp]
\centering	
\includegraphics [width=0.85\textwidth]{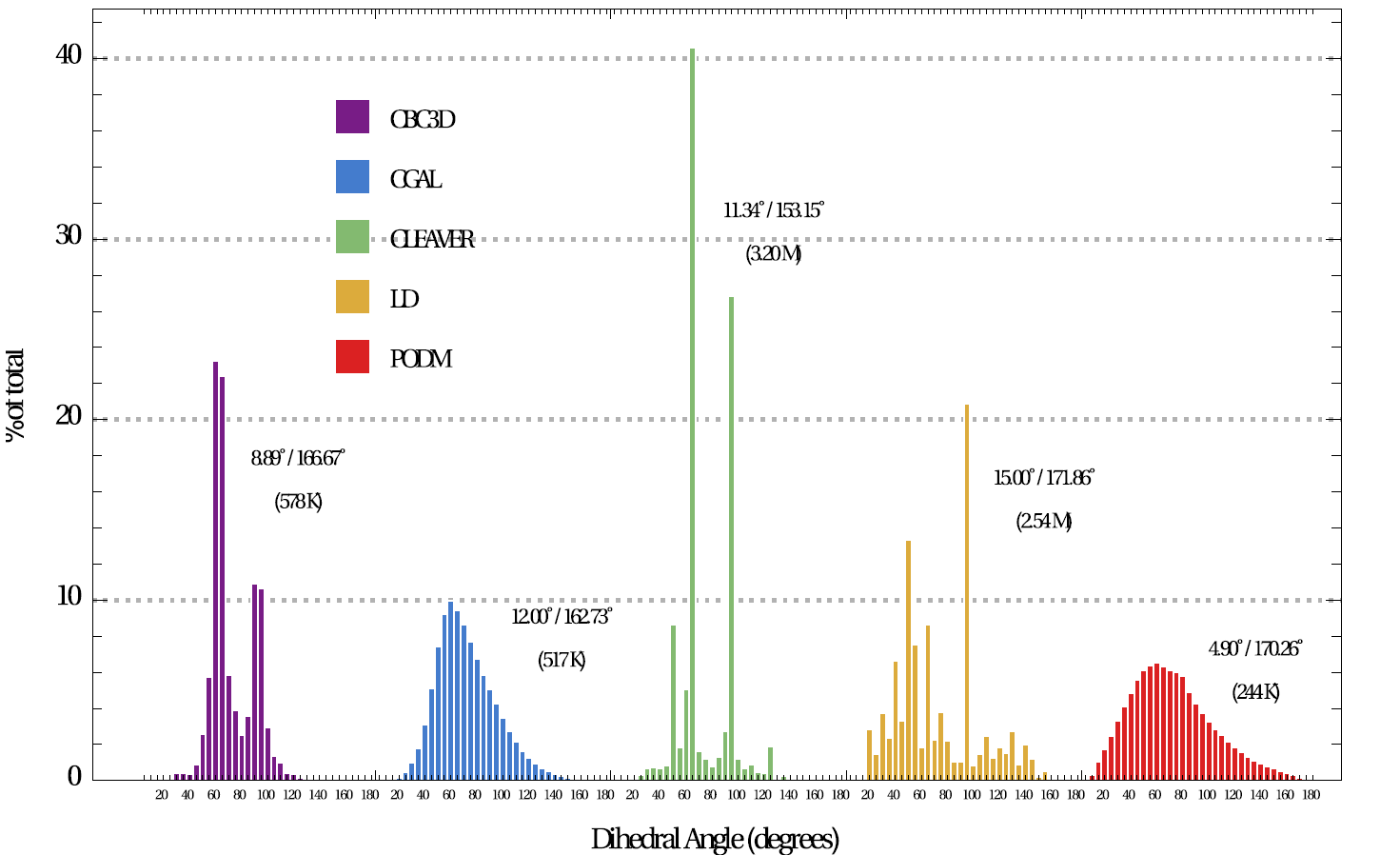}
\caption{Element angle distribution (in 5-deg increments) of Brain-Tumor meshes. The min/max dihedral angles and the element count are reported for each method.}
\label{histogramBrainTumor}
\end{figure}

\begin{figure}[htbp]
\centering	
\includegraphics [width=0.85\textwidth]{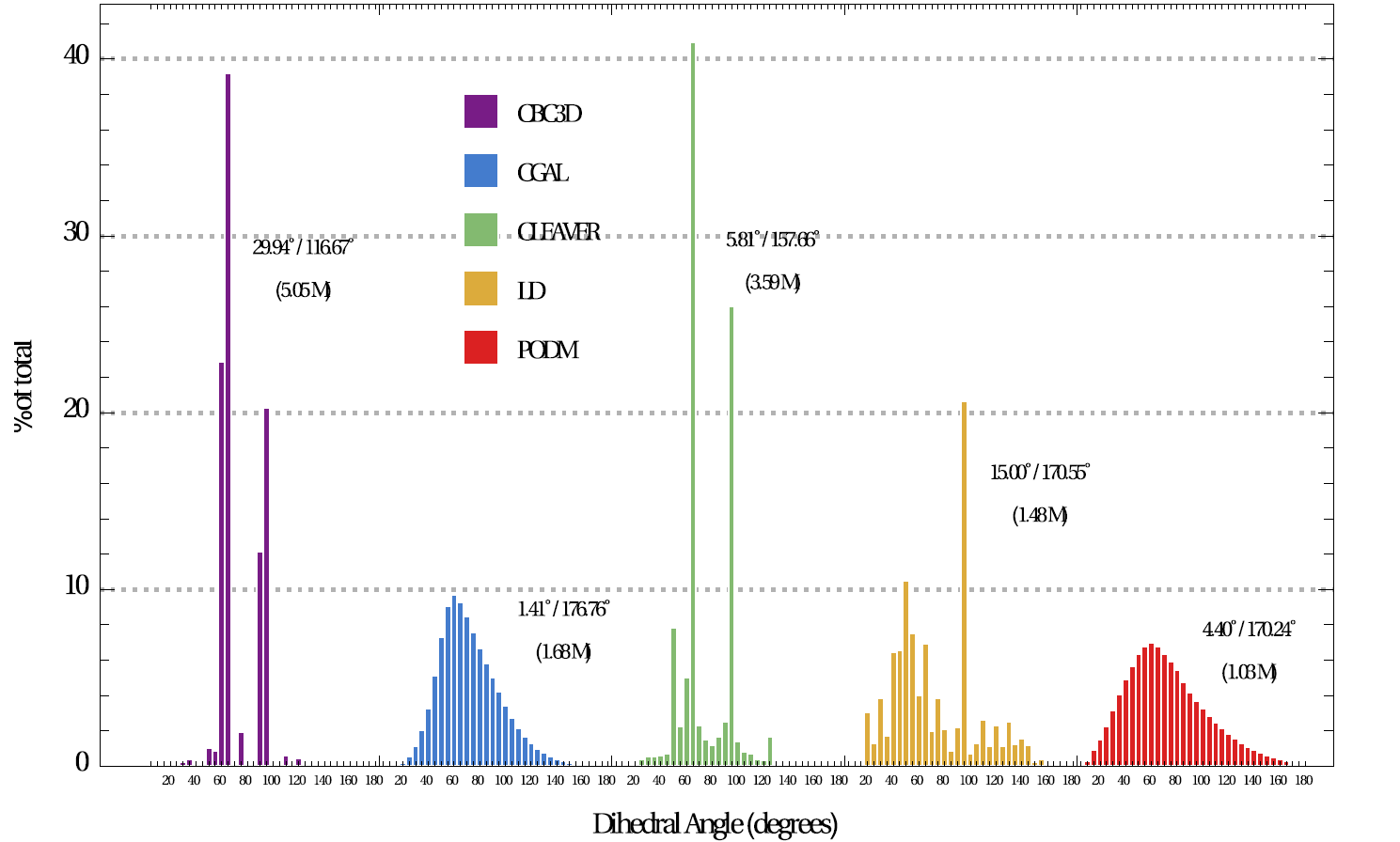}
\caption{Element angle distribution (in 5-deg increments) of Brain-AVM meshes. The min/max dihedral angles and the element count are reported for each method.}
\label{histogramBrainAVM}
\end{figure}

\begin{figure}[htbp]
\centering	
\includegraphics [width=0.85\textwidth]{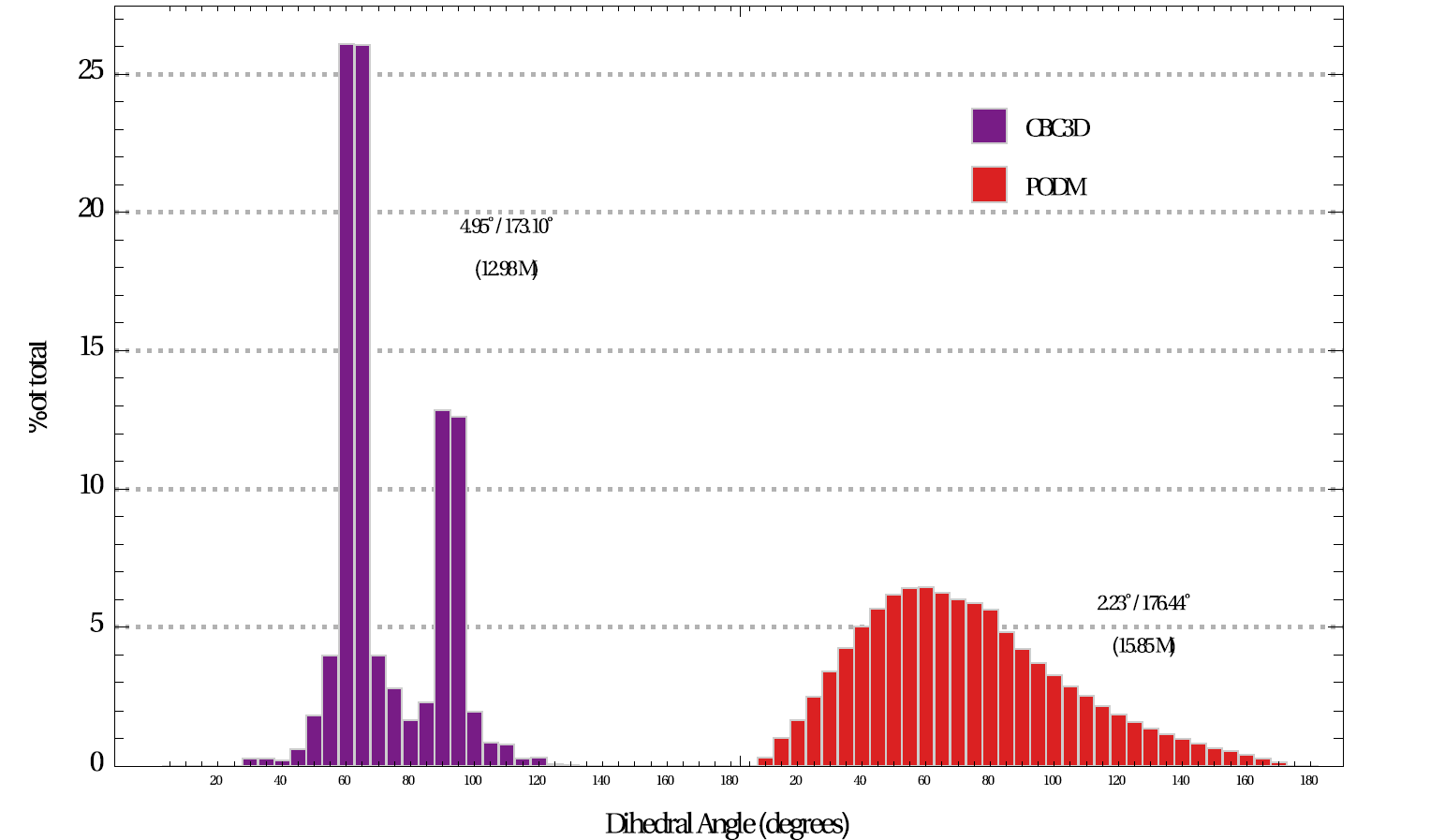}
\caption{Element angle distribution (in 5-deg increments) of Lumen-LVIS stent meshes. The min/max dihedral angles and the element count are reported for each method.}
\label{histogramLVIS}
\end{figure}

Figures \ref{histogramAneurysm}-\ref{histogramLVIS} illustrate the element angle distribution of the meshes.
The Delaunay-based methods exhibit a Gaussian distribution of dihedral angles.
The lattice-based methods exhibit a different distribution due to a more structured connectivity.
For example, in CLEAVER, about $41\%$ of the dihedral angles are between $60^\circ-65^\circ$, and about $26\%$ are $90^\circ-95^\circ$.
In CBC3D, about $50\%$ of the dihedral angles are between $55^\circ-65^\circ$, and about $20\%$ are between $85^\circ-95^\circ$.
In LD, about $15\%$ of the angles are between $45^\circ-50^\circ$, and about $20\%$ are between $90^\circ-95^\circ$.

\begin{figure}[htb]
\centering	
\subfloat[AVM image]{
\label{AVMFidelityZero}
\includegraphics[width=0.32\textwidth]{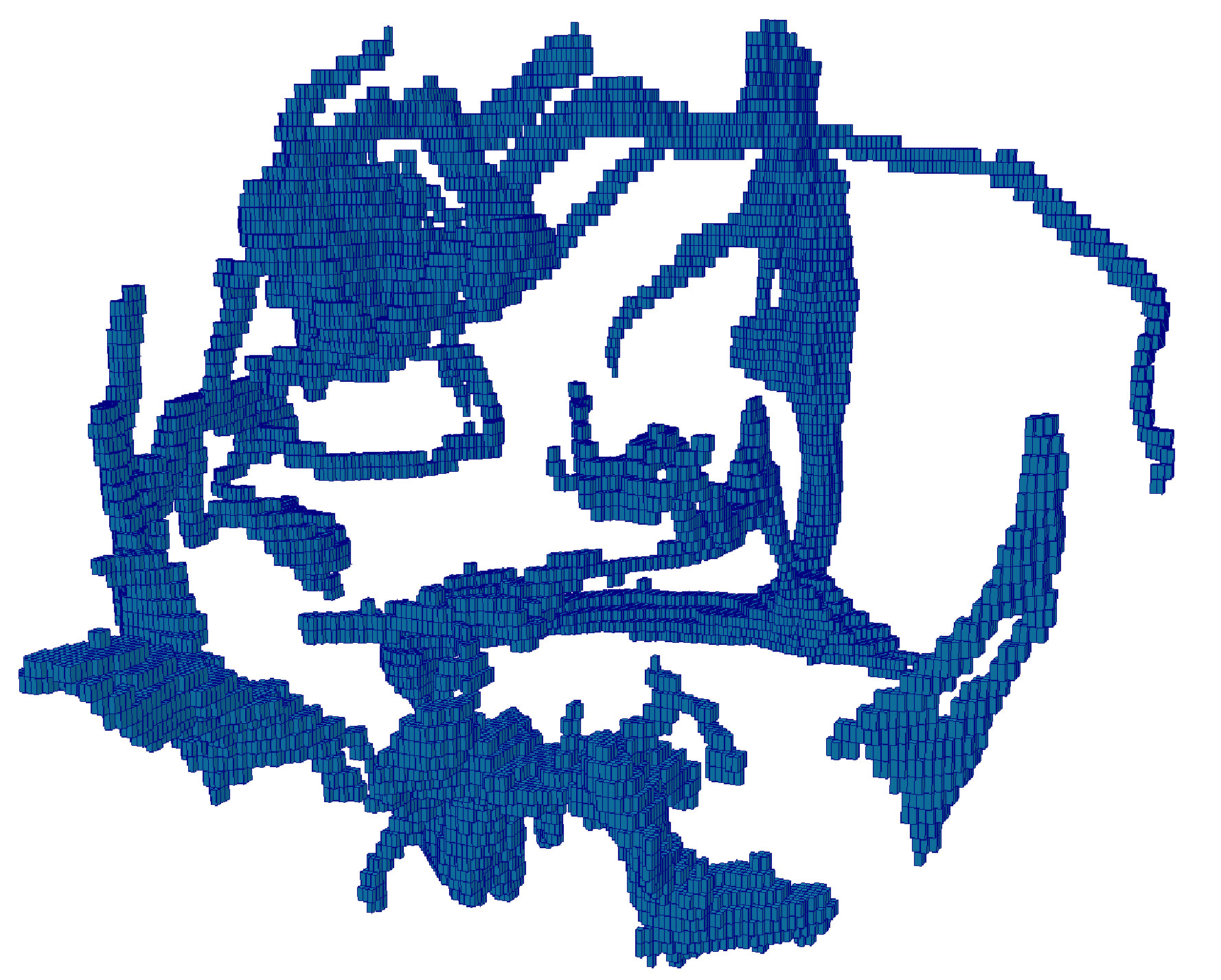}}
\subfloat[CBC3D]{
\label{AVMFidelityOne}
\includegraphics[width=0.32\textwidth]{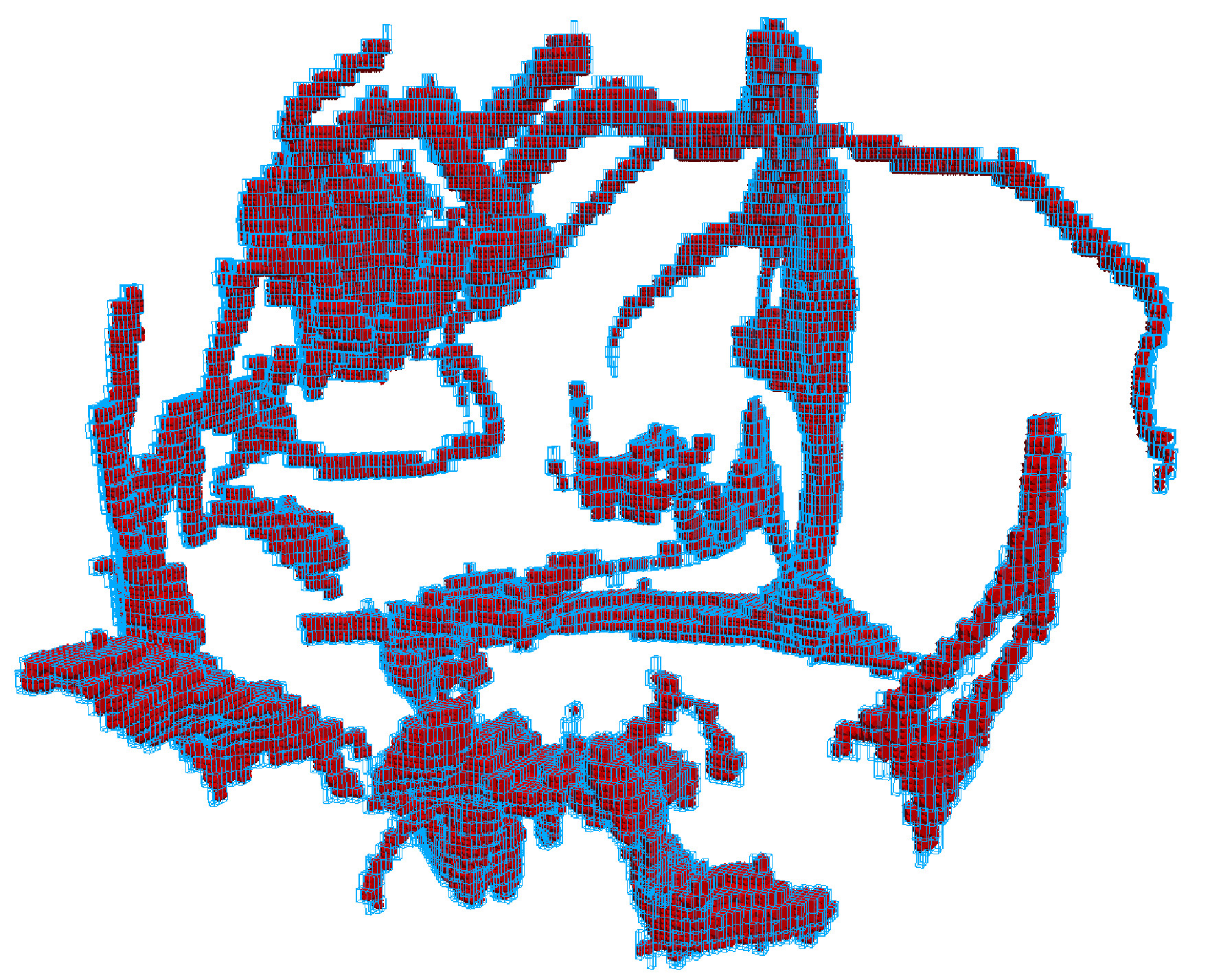}}
\subfloat[CGAL]{
\label{AVMFidelityTwo}
\includegraphics[width=0.32\textwidth]{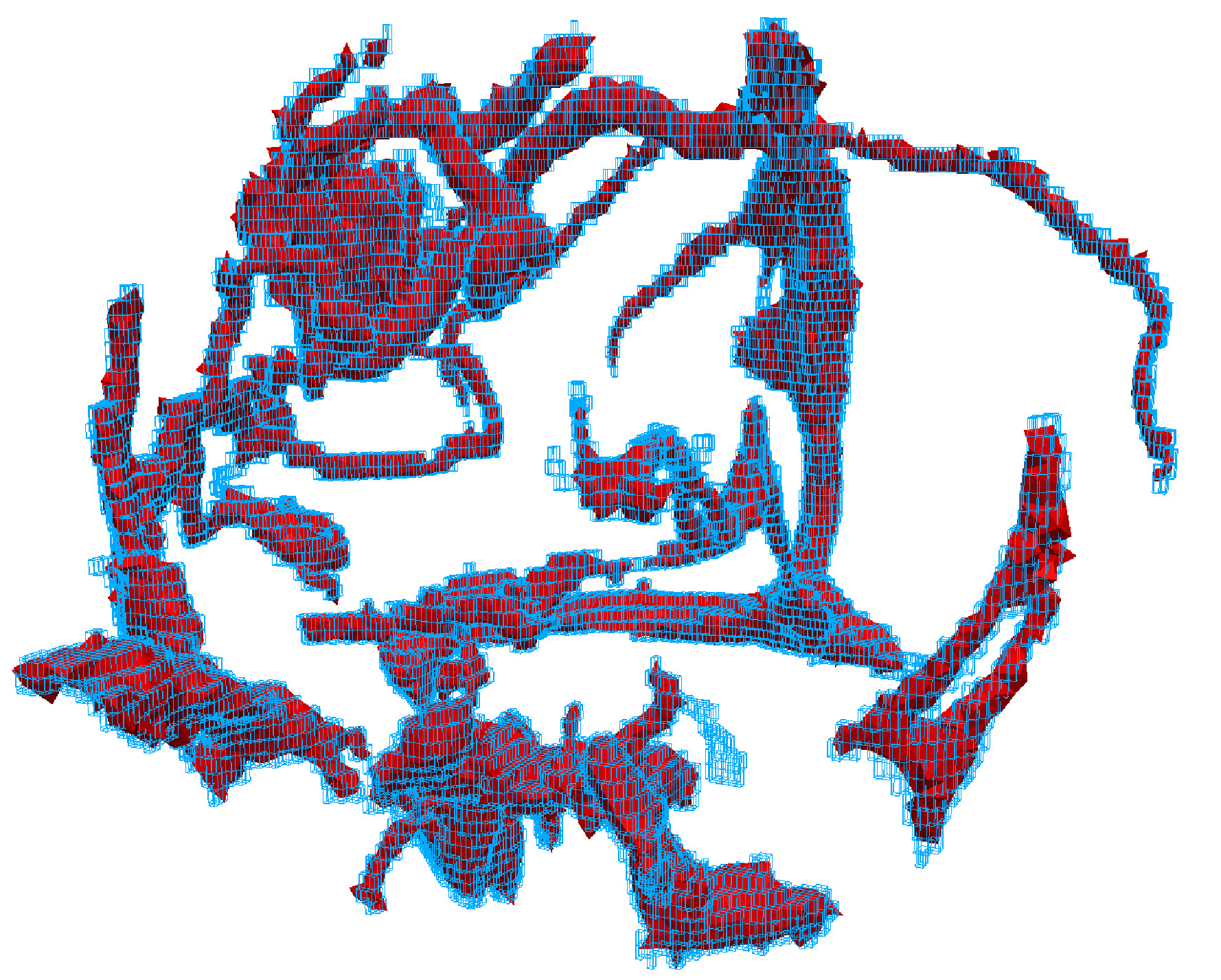}}

\subfloat[CLEAVER]{
\label{AVMFidelityThree}
\includegraphics[width=0.32\textwidth]{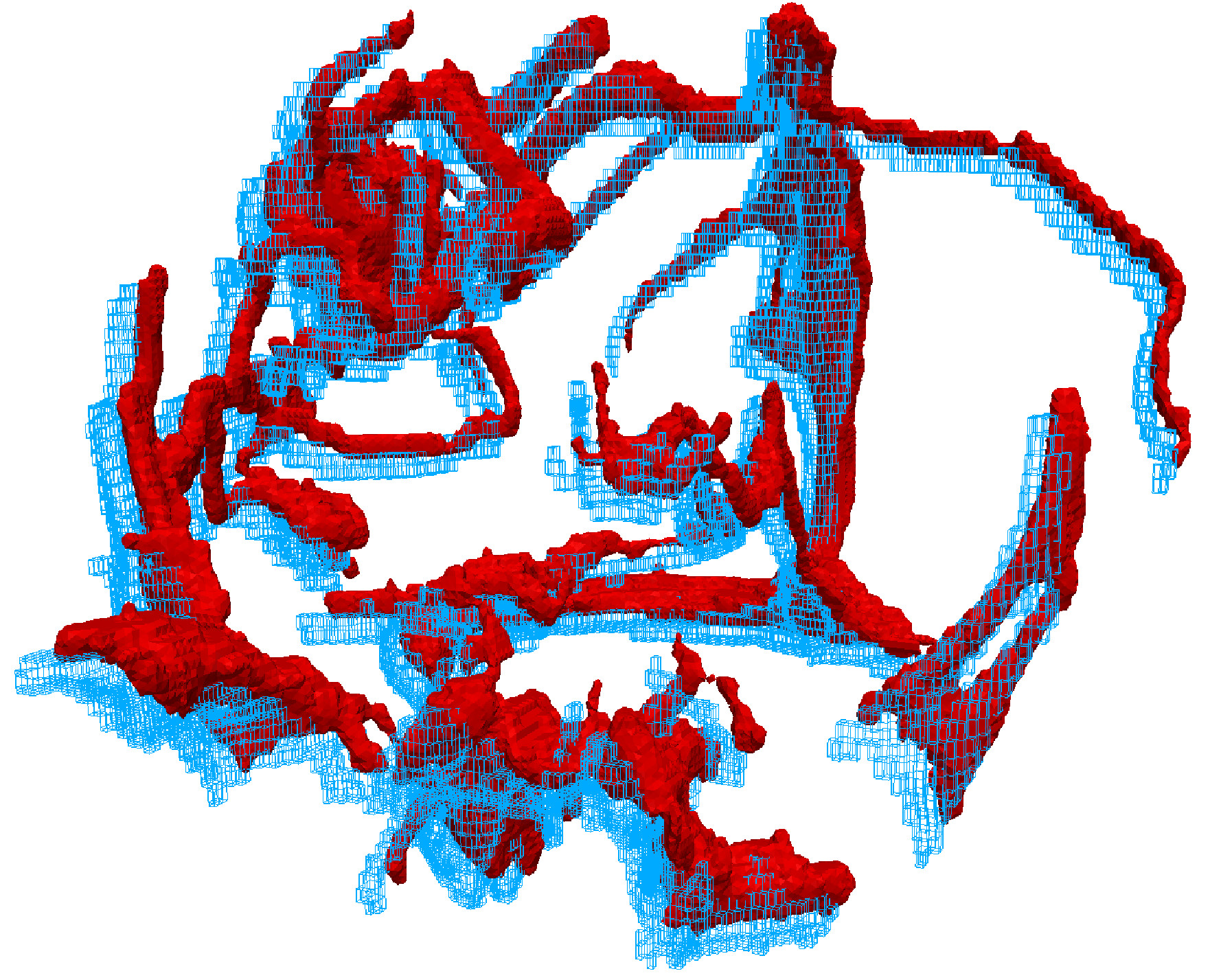}}
\subfloat[LD]{
\label{AVMFidelityFour}
\includegraphics[width=0.32\textwidth]{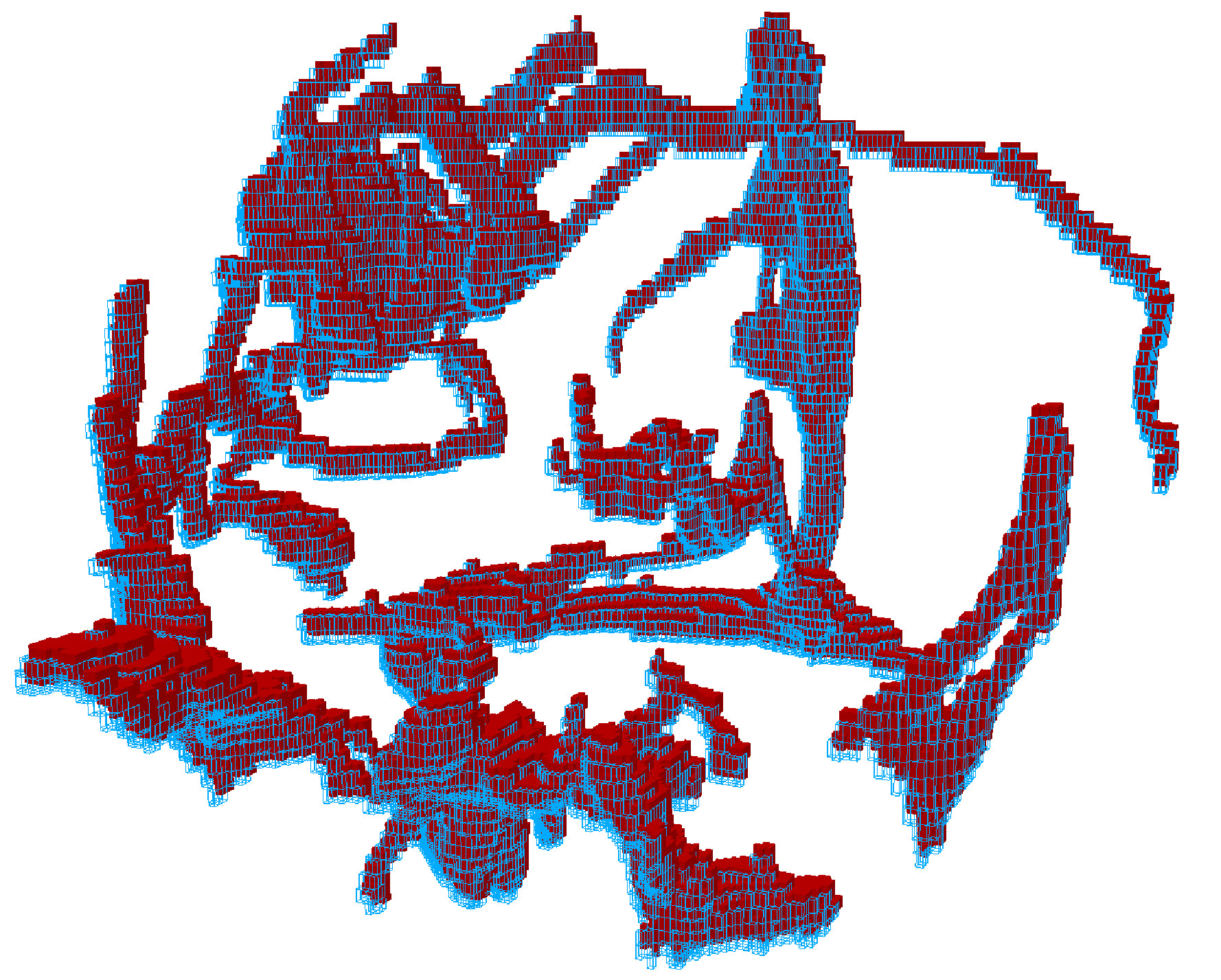}}
\subfloat[PODM]{
\label{AVMFidelityFive}
\includegraphics[width=0.32\textwidth]{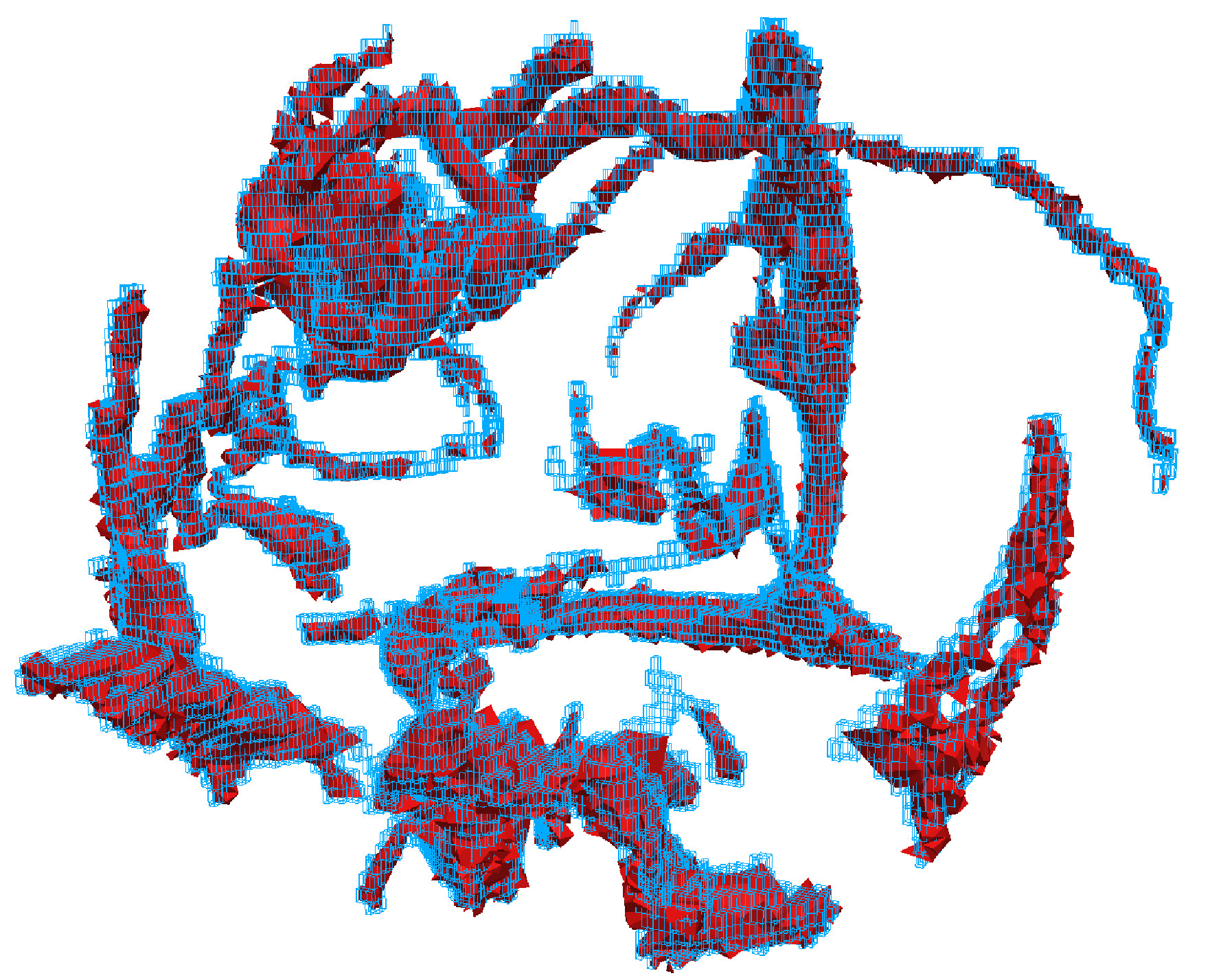}}
\caption{Qualitative evaluation on the fidelity of AVM mesh. Figures \protect\subref{AVMFidelityOne}- \protect\subref{AVMFidelityFive} depict the AVM mesh (red) superimposed on the AVM segmentation (blue).
The closer the mesh surface is to the boundary of the segmented material, the higher the fidelity.}
\label{AVMFidelity}
\end{figure}

The mesh fidelity is qualitatively evaluated on AVM data (case 3). For this purpose, the AVM mesh is first extracted from the 
multi-material mesh and then superimposed on the AVM segmentation (Figure \ref{AVMFidelity}). 
The closer the mesh surface is to the boundary of the segmented AVM image, the higher the fidelity.
The LD method achieves high fidelity because it completely resolves the vessels (it creates a voxelized mesh surface). 
Nevertheless, Figure \ref{AVMFidelityFour} indicates a small shift in the output mesh in relation to the input 
image, most likely due to image resampling. CLEAVER resolves most of the vessel structures, but the generated mesh is noticeably shifted (Figure \ref{AVMFidelityThree}).
CBC3D achieves satisfactory fidelity. CBC3D's mesh topological checks are turned off to avoid further red-green subdivisions, keeping the element count low.

\begin{figure}[htb]
\centering	
\subfloat[HD metric]{
\label{plotsHDTIMEOne}
\includegraphics [width=0.48\textwidth]{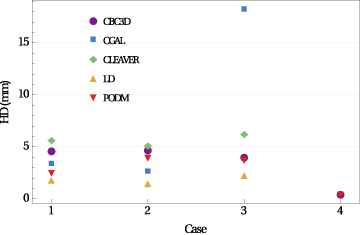}}
\subfloat[Mesh generation time]{
\label{plotsHDTIMETwo}
\includegraphics [width=0.48\textwidth]{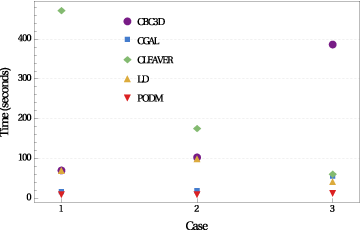}}
\caption{Plots of the results in Tables \ref{resultsHD} - \ref{resultsTime}. \protect\subref{plotsHDTIMETwo} does not include case 4 ($3932.98$, and $750.02$ seconds for CBC3D and PODM, respectively).}
\label{plotsHDTIME}
\end{figure}

Quantitative evaluation on mesh fidelity is performed using a two-sided Hausdorff Distance 
metric: $\text{HD} = \max\{ \text{HD}_{\text{I}\rightarrow \text{M}}, \text{HD}_{\text{M}\rightarrow \text{I}}\}$, where $\text{HD}_{\text{I}\rightarrow \text{M}}$ 
is the value of the metric from the image to the mesh,
and $\text{HD}_{\text{M}\rightarrow \text{I}}$ is the value of the metric from the mesh to the image.
The lower the HD error, the higher the fidelity. HD is computed between two point sets. The first point set contains the coordinates of the vertices located on the mesh surface.
The second point set contains the coordinates of the voxels located on the boundaries of the segmented material (i.e., where the EDT value is zero).
HD is computed for each material and the maximum value is reported. 
An open-source implementation is employed to calculate the HD metric \cite{Commandeur}.
Tables \ref{resultsHD} - \ref{resultsTime} report the HD values and the end-to-end running time for mesh generation.
PODM is a parallel multi-threaded code. CBC3D is sequential but computes the EDTs, source, and target points in parallel.
The remaining codes are sequential. Figure \ref{plotsHDTIME} depicts plots of the results. 

\begin{table}[htb]
\caption{Results on mesh fidelity calculated using a two-sided Hausdorff 
Distance metric. The lower the HD value, the higher the fidelity.}
\centering
\begin{tabular}{c c c c c c}
\hline
\multirow{2}{*}{Case} & \multicolumn{5}{c}{HD (mm)}\\
 & CBC3D & CGAL & CLEAVER & LD & PODM \\\hline
$1$ & $4.51$ & $3.26$ & $5.56$ & $1.73$  & $2.42$ \\
$2$ & $4.59$ & $2.51$ & $5.03$& $1.41$ &  $3.87$\\
$3$ & $3.91$ & $18.09$ & $6.14$ & $2.19$ & $3.65$\\
$4$ & $0.34$ & - & - &  - & $0.33$\\
\hline
\end{tabular}
\label{resultsHD}
\end{table}

\begin{table}[htb]
\caption{Results on end-to-end running time. The time for writing the mesh is not included. T represents the number of threads utilized.}
\centering
\begin{tabular}{c c c c c c}
\hline
\multirow{2}{*}{Case} & \multicolumn{5}{c}{Mesh Generation Time (seconds)}\\
 & CBC3D & CGAL & CLEAVER & LD & PODM \\\hline
$1$ & $68.24$ $(8\text{T})$ & $12.33$ & $470.22$ & $68.44$  & $7.86$ $(8\text{T})$ \\
$2$ & $101.03$ $(8\text{T})$ & $14.92$ & $173.85$& $97.76$ &  $7.93$ $(8\text{T})$\\
$3$ & $384.86$ $(24\text{T})$ & $51.59$ & $59.13$ & $40.19$ & $10.98$ $(24\text{T})$\\
$4$ & $3932.98$ $(24\text{T})$ & - & - & - & $750.02$ $(24\text{T})$\\
\hline
\end{tabular}
\label{resultsTime}
\end{table}

It should be noted that the error in Table \ref{resultsHD} is relative to the image spacing. 
An error approximately 3-4 times the spacing may be satisfactory, depending on the application.
LD achieves the highest fidelity among all the methods, but no experimental results are available for case 4. 
PODM and CBC3D exhibit a reasonably good fidelity.

A fine mesh that accurately conforms to a segmented image typically has a voxelized appearance.
However, the voxelized segmentation is itself inaccurate. The true surface is smooth and a binary segmentation 
does not capture this fact. On the other hand, a smooth surface mesh is easier to obtain with a lower fidelity.
Therefore, fidelity must be compromised in favor of a more realistic appearance.
Applications such as interactive surgical simulations require volume meshes with smooth surfaces. 
Additionally, a bumpy surface can deteriorate the accuracy of a CFD solution. 

\begin{figure}[htb]
\centering	
\subfloat{
\label{smoothingEvaluationOne}
\includegraphics[angle=-90,width=0.185\textwidth]{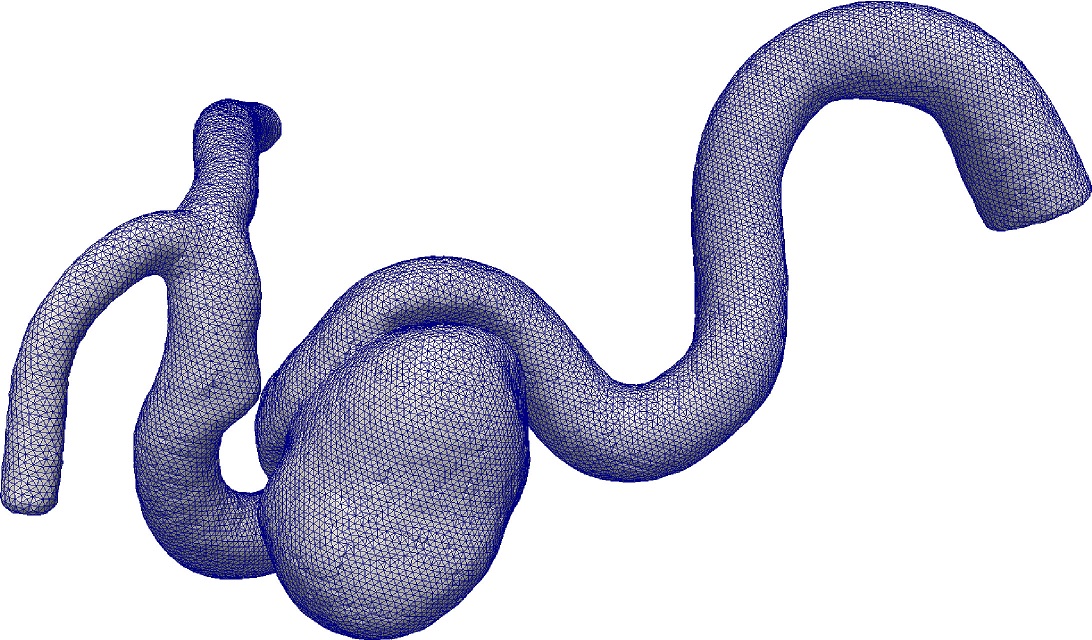}}
\subfloat{
\label{smoothingEvaluationTwo}
\includegraphics[angle=-90,width=0.185\textwidth]{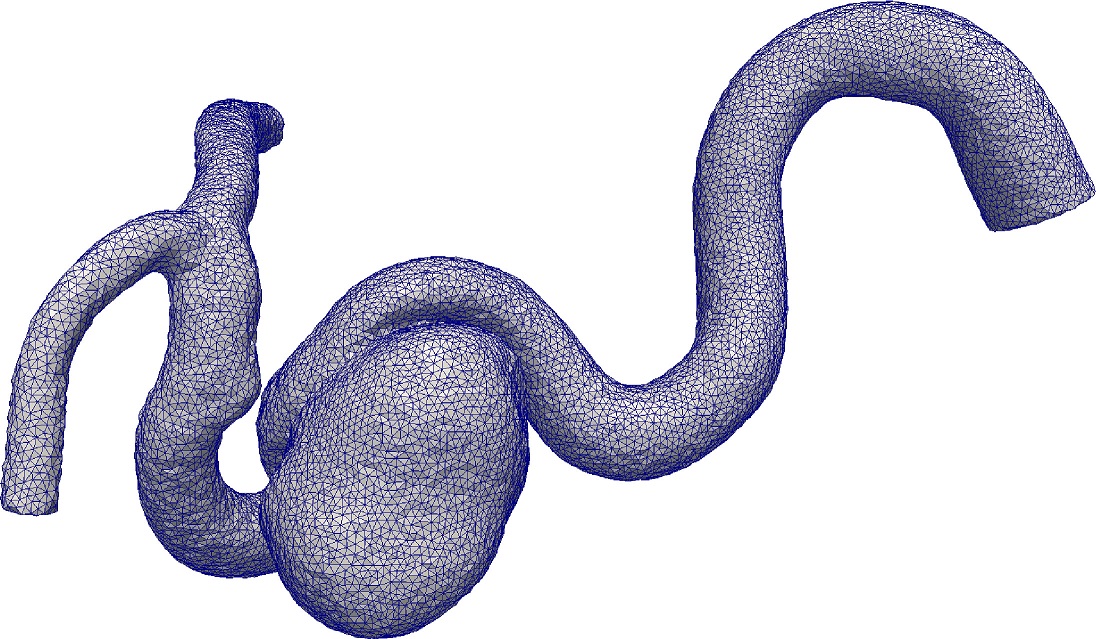}}
\subfloat{
\label{smoothingEvaluationThree}
\includegraphics[angle=-90,width=0.185\textwidth]{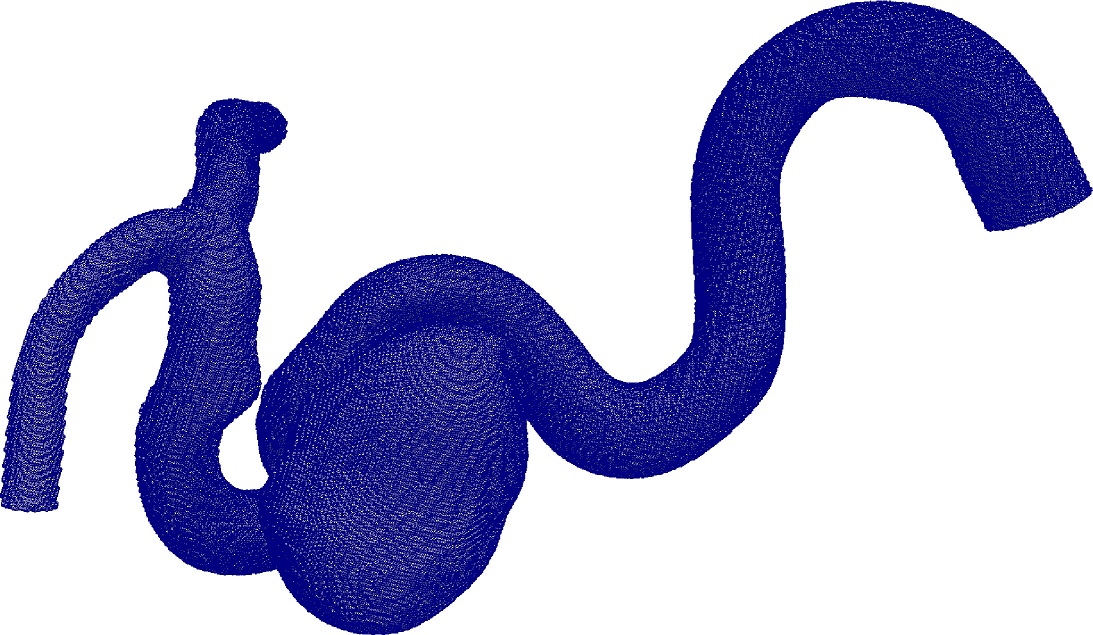}}
\subfloat{
\label{smoothingEvaluationFour}
\includegraphics[angle=-90,width=0.185\textwidth]{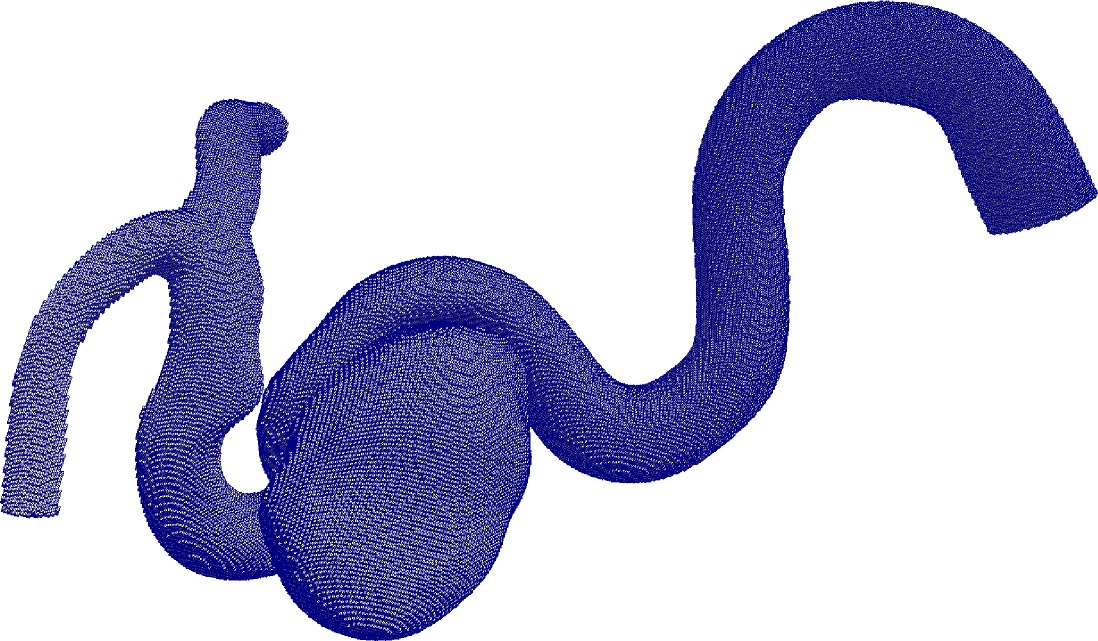}}
\subfloat{
\label{smoothingEvaluationFive}
\includegraphics[angle=-90,width=0.185\textwidth]{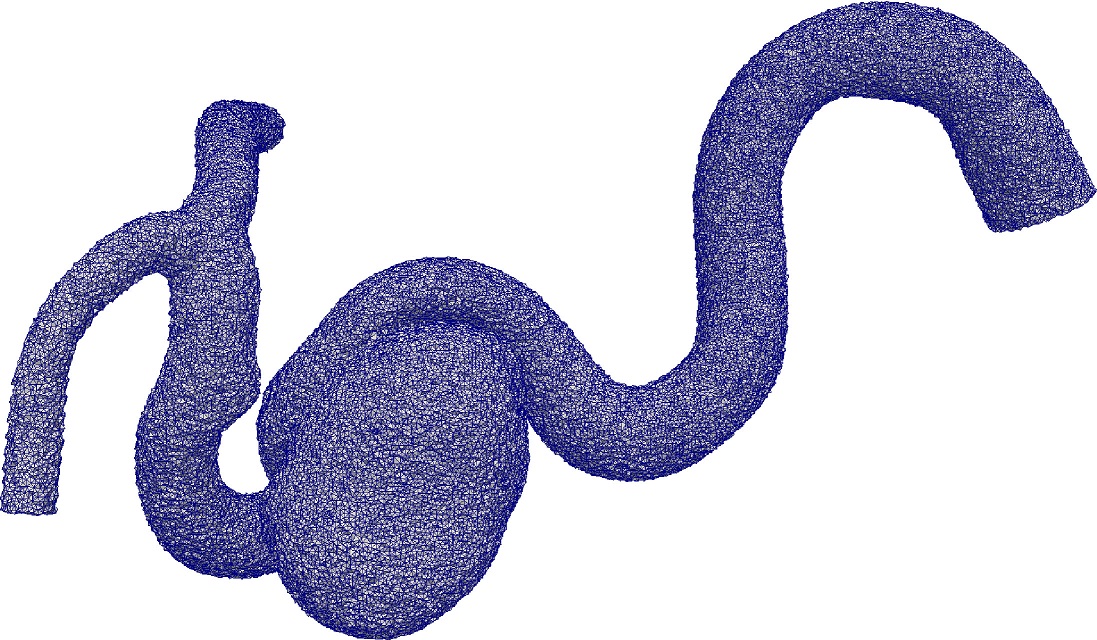}}
\setcounter{subfigure}{0}
\subfloat[CBC3D]{
\label{smoothingEvaluationSix}
\includegraphics[angle=90,width=0.185\textwidth]{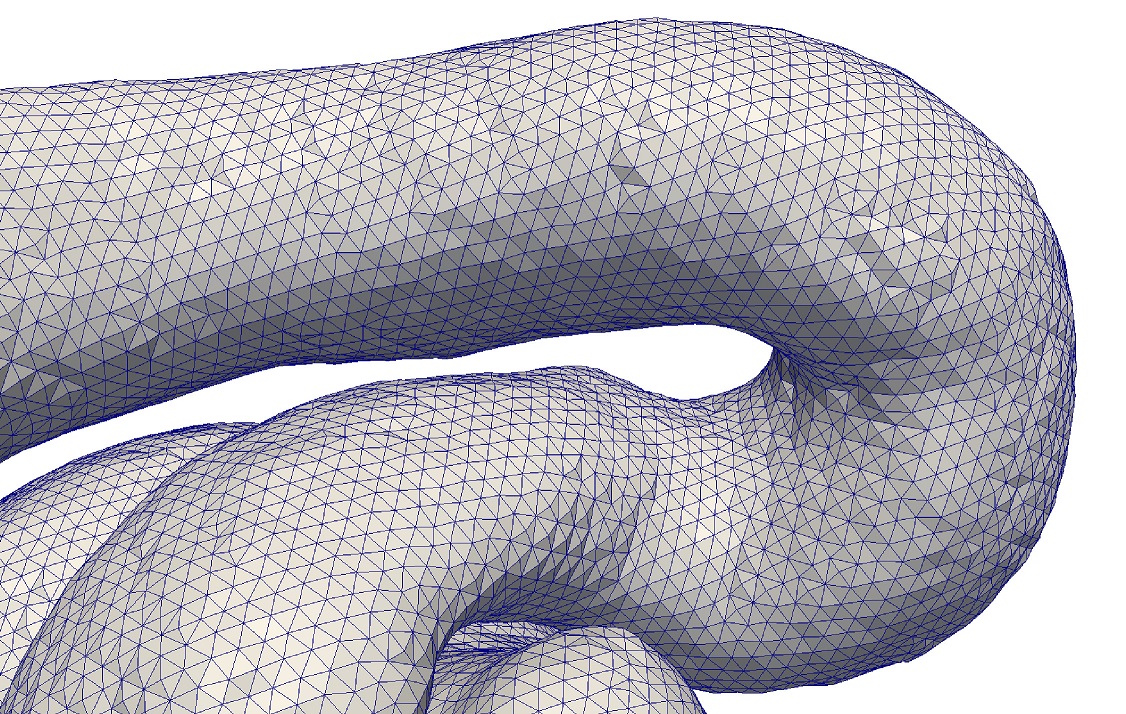}}
\subfloat[CGAL]{
\label{smoothingEvaluationSeven}
\includegraphics[angle=90,width=0.185\textwidth]{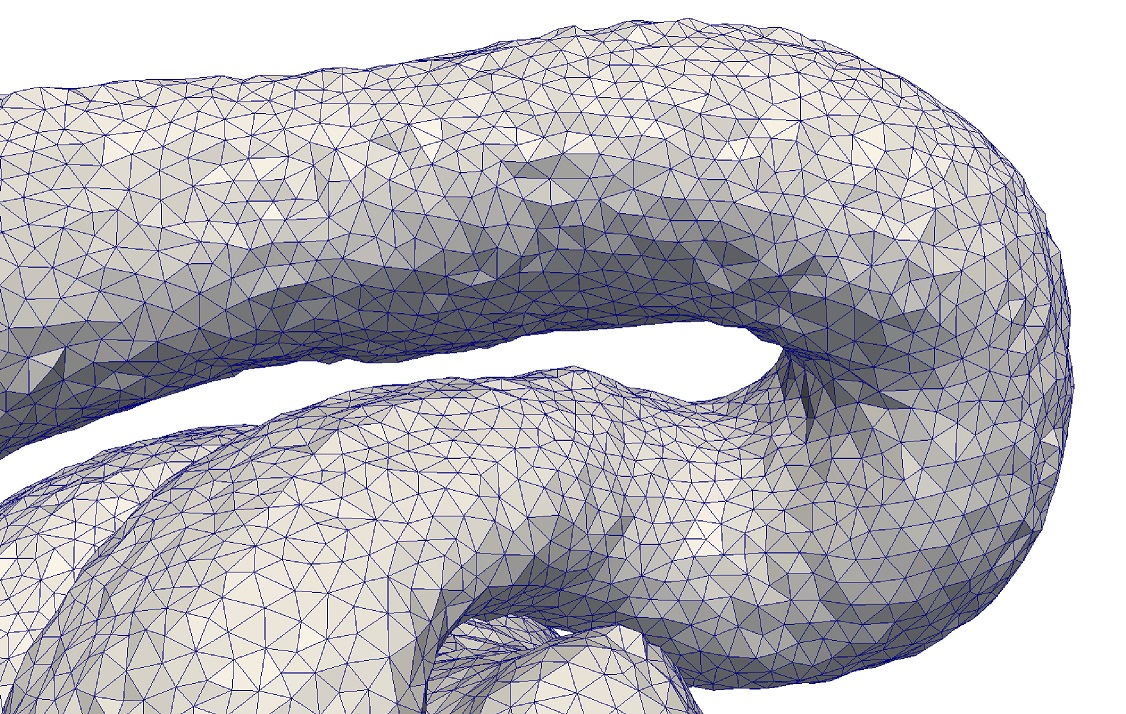}}
\subfloat[CLEAVER]{
\label{smoothingEvaluationEight}
\includegraphics[angle=90,width=0.185\textwidth]{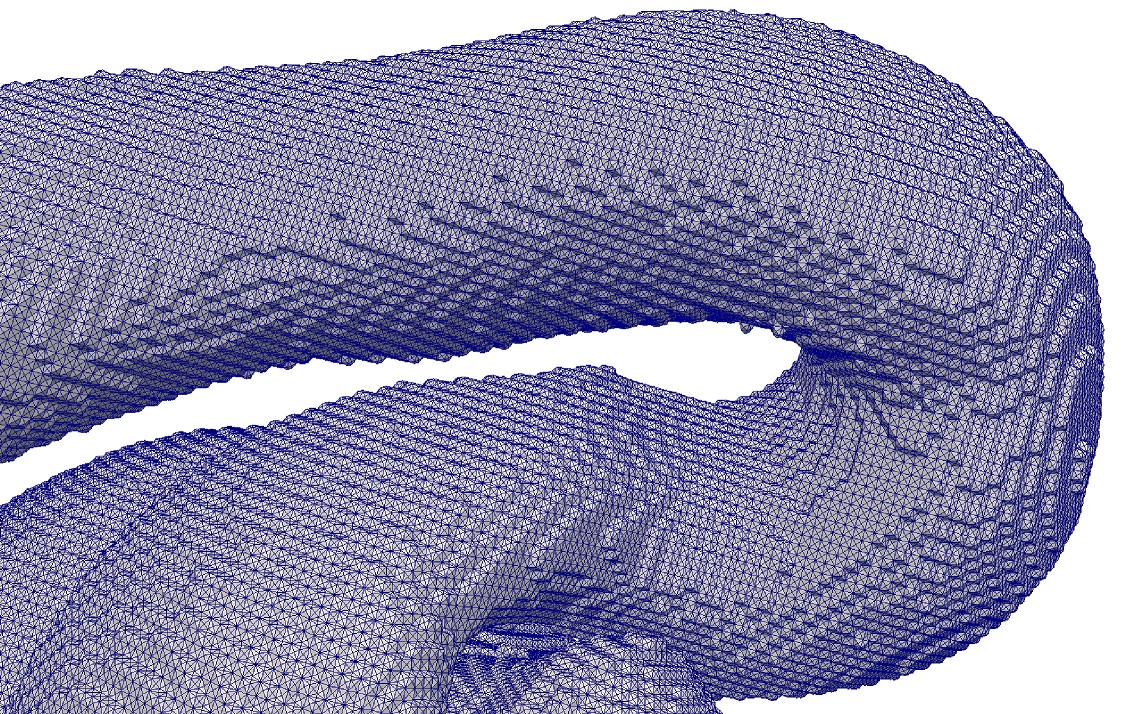}}
\subfloat[LD]{
\label{smoothingEvaluationNine}
\includegraphics[angle=90,width=0.185\textwidth]{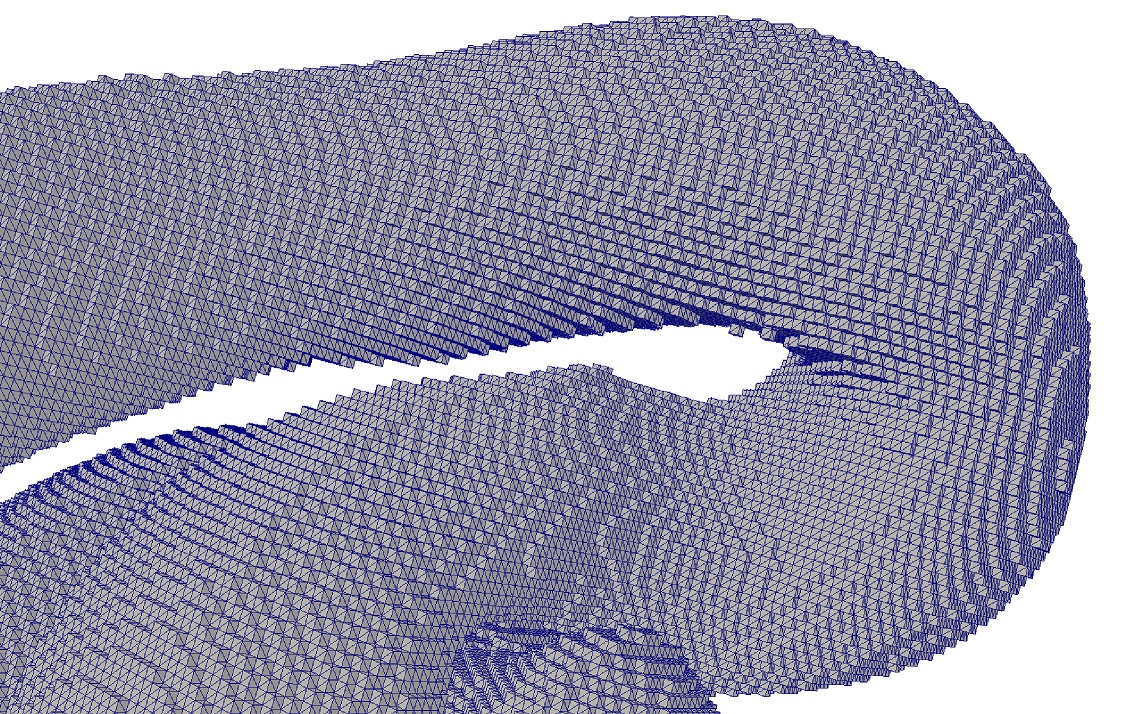}}
\subfloat[PODM]{
\label{smoothingEvaluationTen}
\includegraphics[angle=90,width=0.185\textwidth]{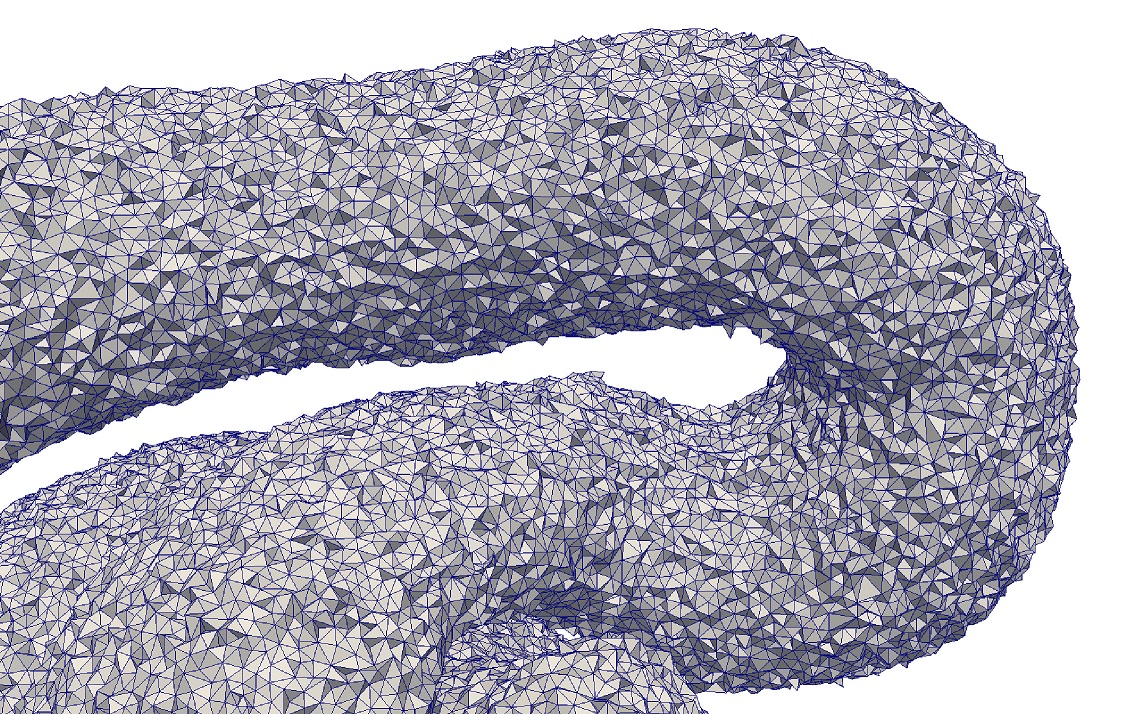}}
\caption{Comparison between the surface meshes of case 1 (Cavernous Aneurysm). Each column corresponds to a single method. 
The bottom row depicts a closer view of the surface. Among the methods, only CBC3D approximates the voxelized segmentation with a smooth surface that reflects a certain degree of visual reality.}
\label{smoothingEvaluation}
\end{figure}

Figure \ref{smoothingEvaluation} compares the surface meshes of a Cavernous Aneurysm.
CBC3D exhibits the smoothest surface among all the methods. 
CGAL creates a relatively smooth surface as it follows a two-step approach; it first reconstructs the surface 
and then generates a volume mesh given the surface as an input. 
CLEAVER and LD generate voxelized surfaces. PODM generates a bumpy surface (Figure \ref{smoothingEvaluation}).

\begin{figure}[htbp]
\centering	
\subfloat[\#(pts, tets): (63.6, 272.2)\;K; $\alpha_{min}$: $\text{5.07}^\circ$]{
\label{comparisonTetMixedOne}
\includegraphics [width=0.48\textwidth]{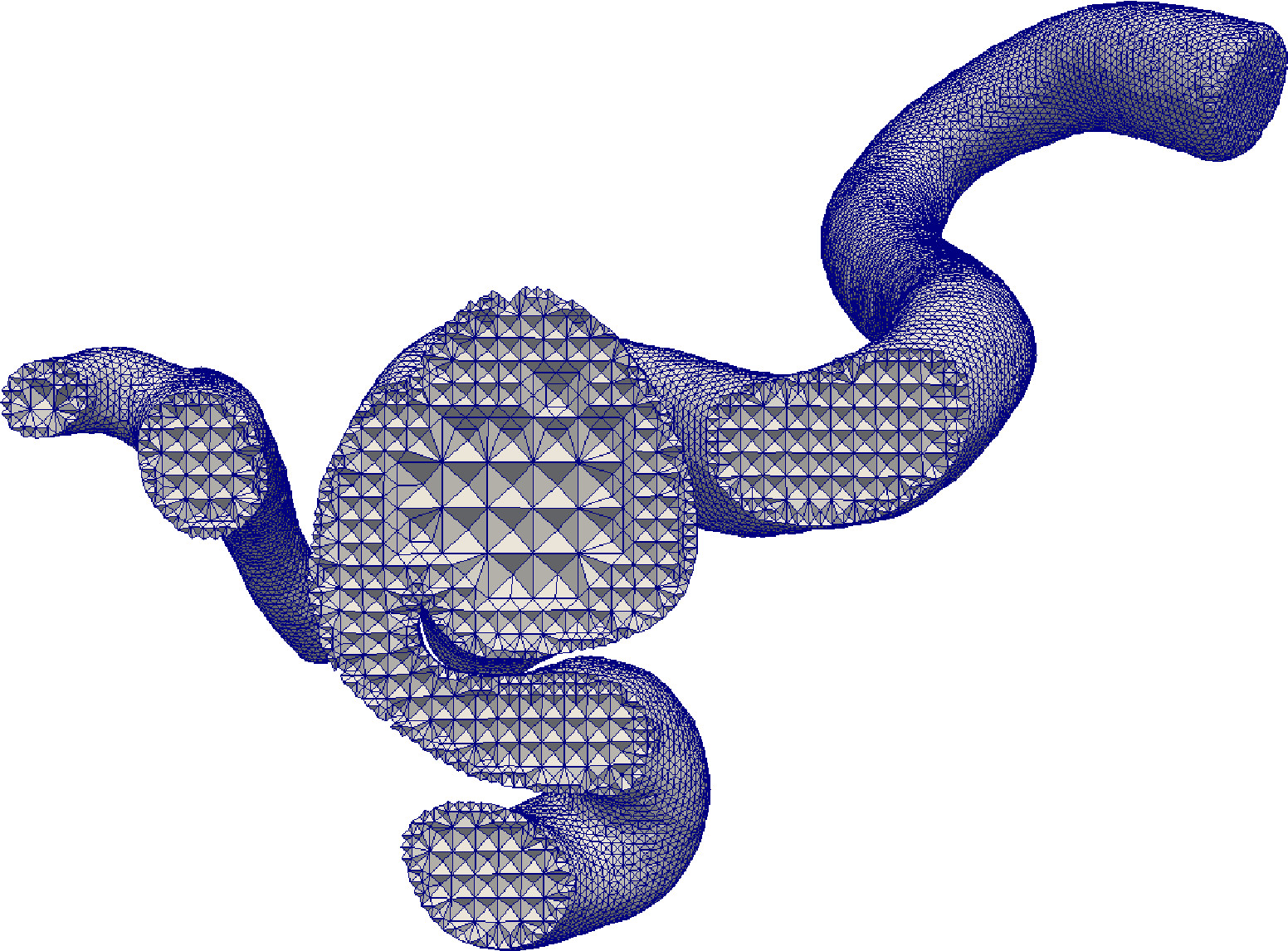}}
\subfloat[\#(pts, tets, pyrs, hexs): (62.1, 246.7, 4.0, 1.4)\;K; $\alpha_{min}$: $\text{5.31}^\circ$; $J_{min}$: 0.47]{
\label{comparisonTetMixedTwo}
\includegraphics [width=0.48\textwidth]{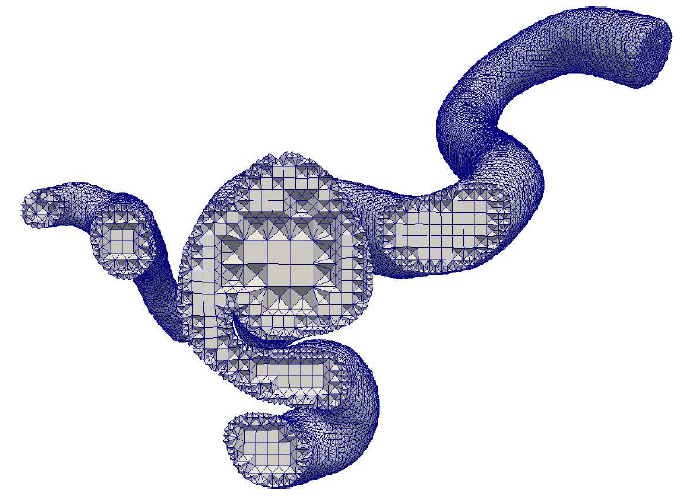}}

\subfloat[\#(pts, tets): (101.4, 578.3)\;K; $\alpha_{min}$: $\text{8.89}^\circ$]{
\label{comparisonTetMixedThree}
\includegraphics [width=0.4\textwidth]{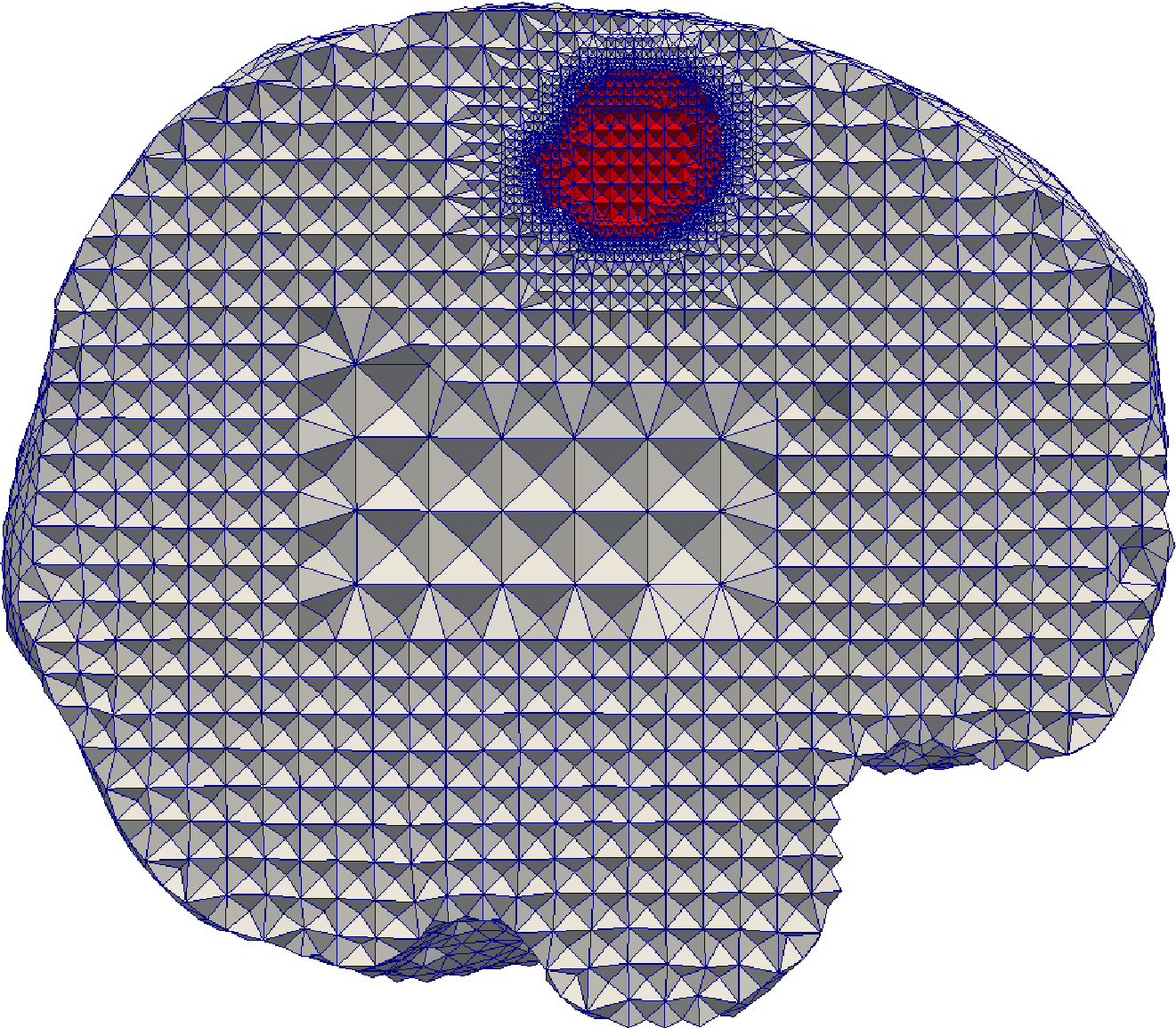}}\hspace{0.28cm}
\subfloat[\#(pts, tets, pyrs, hexs): (78.8, 226.8, 40.2, 22.5)\;K; $\alpha_{min}$: $\text{5.19}^\circ$; $J_{min}$: 0.55]{
\label{comparisonTetMixedFour}
\includegraphics [width=0.4\textwidth]{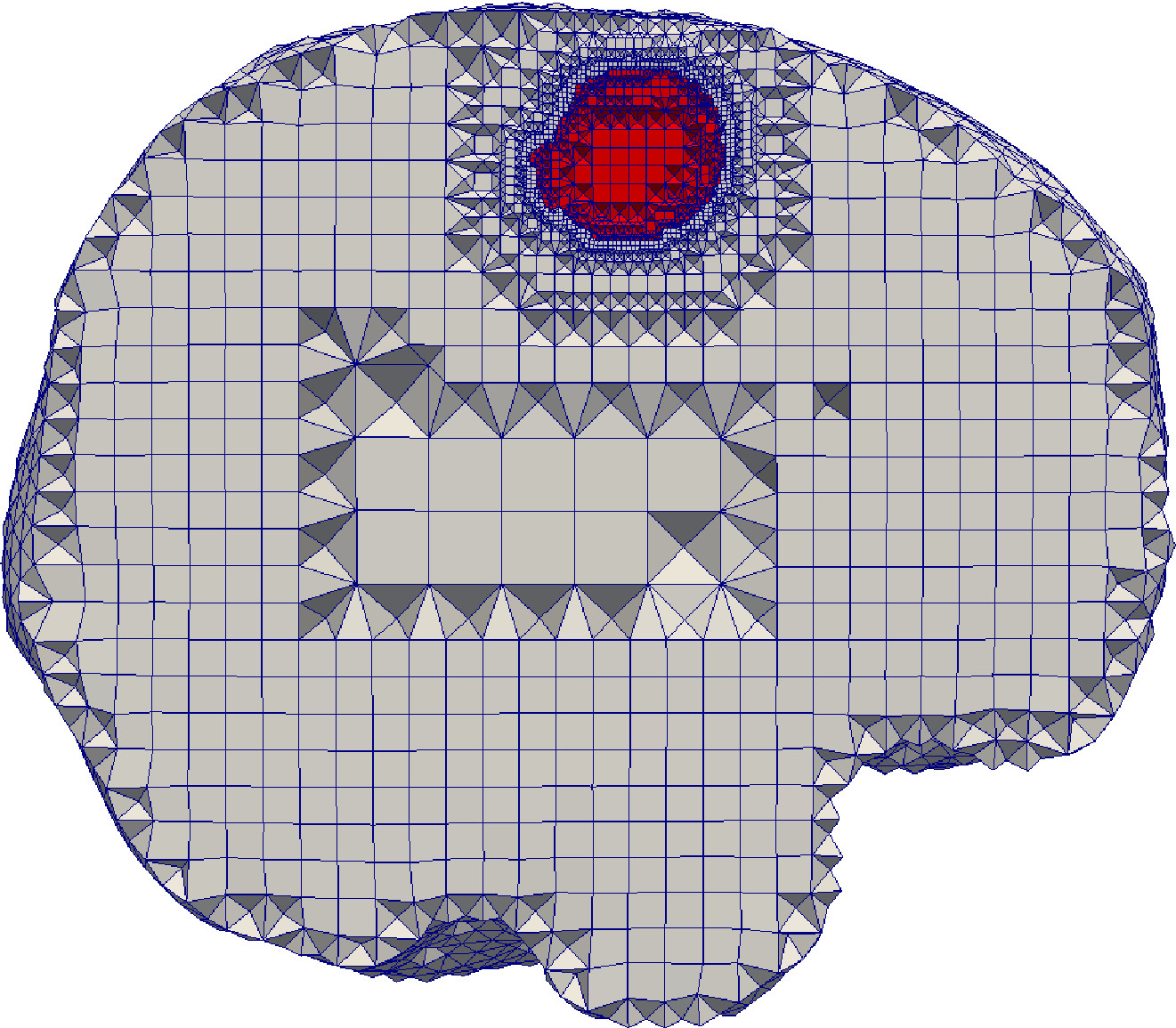}}

\subfloat[\#(pts, tets): (887.0, 5059.9)\;K; $\alpha_{min}$: $\text{29.94}^\circ$]{
\label{comparisonTetMixedFive}
\includegraphics [width=0.4\textwidth]{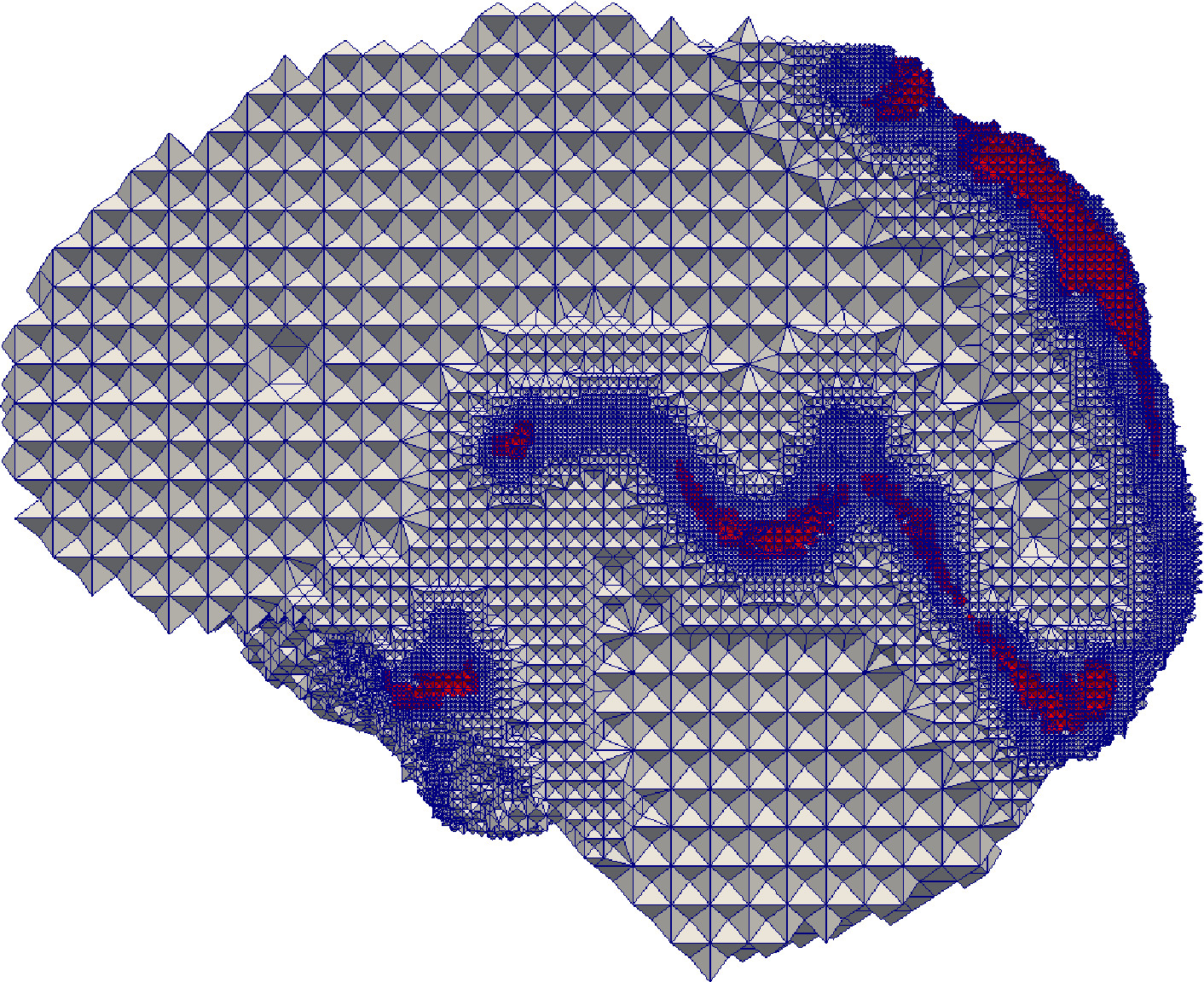}}\hspace{0.28cm}
\subfloat[\#(pts, tets, pyrs, hexs): (679.4, 1839.1, 364.9, 207.5)\;K; $\alpha_{min}$: $\text{29.18}^\circ$; $J_{min}$: 0.98]{
\label{comparisonTetMixedSix}
\includegraphics [width=0.4\textwidth]{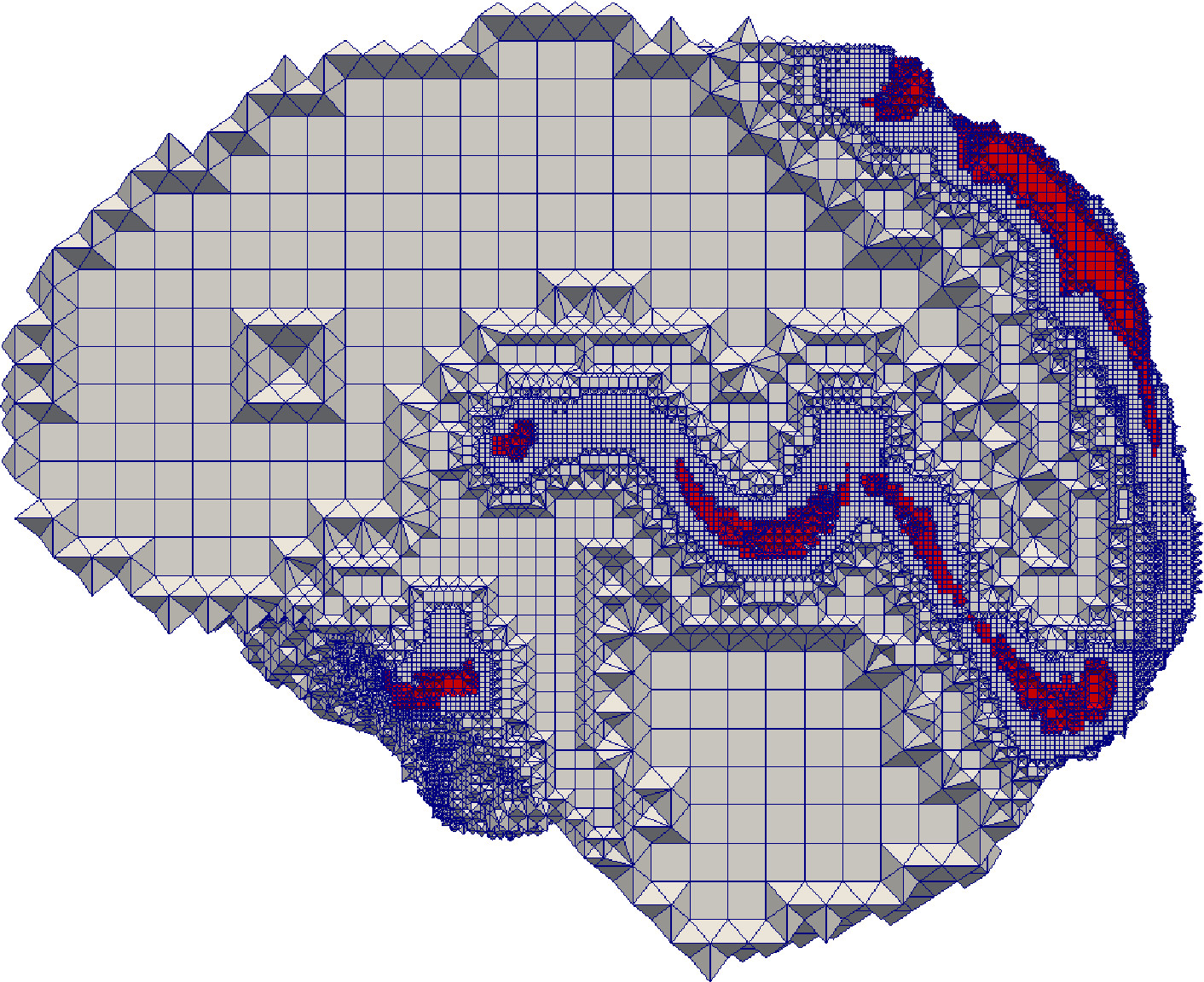}}
\caption{Comparison between adaptive terahedral meshes and their corresponding mixed meshes all generated with CBC3D. The top, middle, and bottom row 
correspond to cases 1, 2, and 3, respectively. $\alpha_{min} \in [0, 180]$: minimum dihedral angle (for tetrahedra); $J_{min} \in [0, 1]$: minimum scaled Jacobian (for pyramids and hexahedra).}
\label{comparisonTetMixed}
\end{figure}

\begin{figure}[htbp]
\centering	
\subfloat[\#(pts, tets): (2.25, 12.98)\;M; $\alpha_{min}$: $\text{4.95}^\circ$]{
\label{comparisonTetMixedLVISOne}
\includegraphics [width=0.49\textwidth]{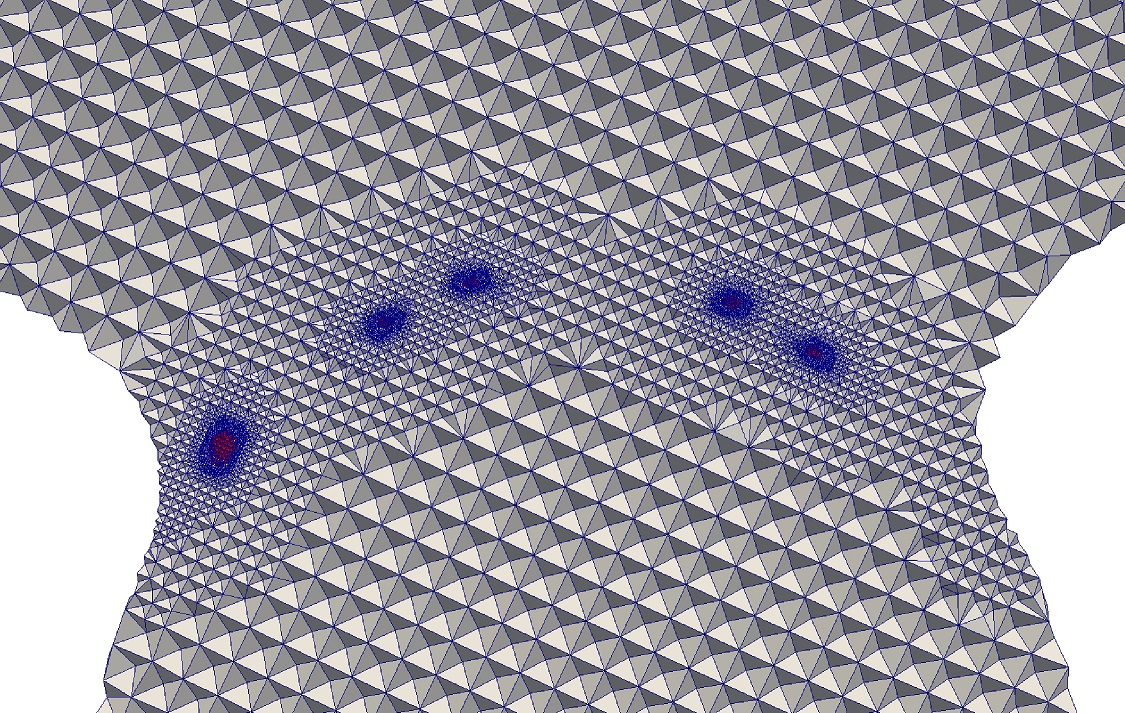}}
\subfloat[\#(pts, tets, pyrs, hexs): (1.57, 3.54, 0.65, 0.67)\;M; $\alpha_{min}$: $\text{5.04}^\circ$; $J_{min}$: 0.40]{
\label{comparisonTetMixedLVISTwo}
\includegraphics [width=0.49\textwidth]{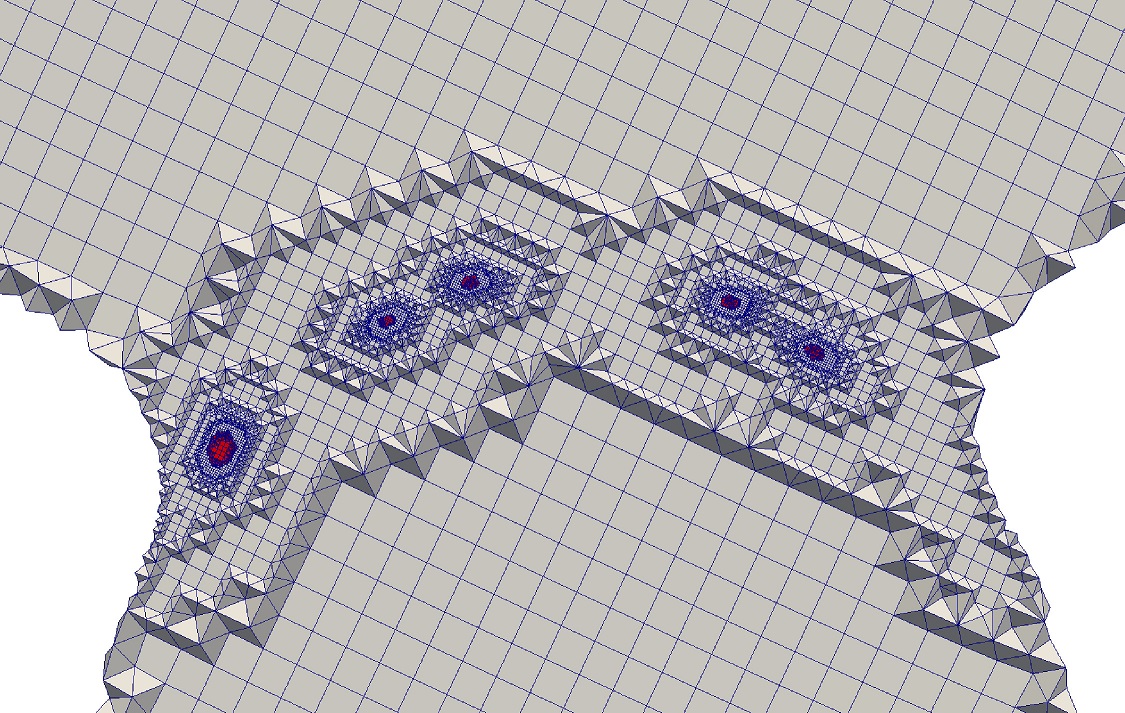}}
\caption{Comparison between adaptive terahedral mesh and its corresponding mixed mesh for case 4 (Lumen-LVIS Stent). The meshes were generated with CBC3D.
$\alpha_{min} \in [0, 180]$: minimum dihedral angle (for tetrahedra); $J_{min} \in [0, 1]$: minimum scaled Jacobian (for pyramids and hexahedra).}
\label{comparisonTetMixedLVIS}
\end{figure}

CBC3D's tetrahedral meshes are converted to mixed meshes.
Figures \ref{comparisonTetMixed} and \ref{comparisonTetMixedLVIS} depict the results. 
A mixed mesh reduces the number of mesh vertices, as well as the subsequent memory and CPU requirements for the solver without any loss of accuracy.
The number of vertices is reduced by $2.35\%$, $22.28\%$, $23.40\%$, and $30.22\%$ for cases 1, 2, 3, and 4, respectively. 
The mixed meshes exhibit qualities and smooth surfaces similar to the adaptive tetrahedral meshes.

\

\section{Discussion and Future Work} \label{discussion}
CBC3D noticeably does not exhibit a good element size gradation, such as that generated by Cleaver in Figure \ref{cutSections} or PODM in Figure \ref{cutSectionsLVIS}. This is due to the specified lattice spacing parameter. We specify these values so that CBC3D can generate elements of good quality, and then maintain that quality while the mesh is deformed to improve fidelity. As mentioned previously in section \ref{quality_control}, the 3D region in which the displacement vector from equation \eqref{deformationEq} is calculated during deformation is limited by the local element size. Additionally, we quantitatively evaluate the similarity between a mesh that corresponds to a material and the image region of this material (as specified in section \ref{adaptive_refinement}) based on metrics also utilized in \cite{CBC3DYixun}, which were shown to satisfy the requirements of non-rigid registration solvers for brain images \cite{meshEvaluationNRR}. We will, however, utilize different evaluation criteria in the future and expand CBC3D's capability to provide accurate geometric and topologic representation for a more robust range of test cases.

In its pre-processing stage, CBC3D implements a relabeling algorithm that processes disconnected regions in the image as different materials (section \ref{relabeling_disconnected_regions}), leading to a more localized refinement per material. A more efficient approach to handling non-manifold voxel connectivity would be to solve the topological issues in the meshing stage (i.e., ensuring that the generated mesh does not contain any non-manifold edges or vertices). Since the red-green templates utilized by CBC3D's adaptive refinement do not guarantee a manifold mesh, this problem requires further consideration and will be addressed in the future. It should also be noted that CBC3D's output is non-deterministic due to: (i) the randomness introduced in the relabeling operations used for eliminating non-manifold voxel connectivity (section \ref{EliminatingNonManifoldImageTopology}) and (ii) the parallelized computations of EDTs, source points, and target points. While this may be an issue for certain applications, the targeted biomedical simulations do not require deterministic output from the mesher. Instead, they focus on the aforementioned requirements - fidelity, quality, real-time performance, and smoothness.

Currently, the smoothness of the surface meshes generated from data with small features (e.g., Micro-CT imaging of stents) is not satisfactory (Figure \ref{cutSectionsStentsOne}).
The smoothness can be improved by incorporating a semi-analytic technique and by using a priori knowledge of the radii of the stent wires.
Also, the mechanical properties (i.e., Young's modulus, Poisson ratio) of the materials adjacent to the mesh interfaces can be appropriately adjusted 
to improve smoothness. The accuracy of CFD simulations used to study localized alterations in hemodynamics due to 
different stent devices may be improved further with boundary layer meshes \cite{dyedov2009variational} and anisotropic meshes \cite{AnisotropicCerebralAneurysms2013}. 

In table \ref{resultsTime}, PODM exhibits better performance than CBC3D (leading to its integration within a non-rigid registration method for image-guided neurosurgery \cite{FotisAPBNRRDeepLearning2021}); however, CBC3D is proven to generate meshes of better quality than PODM \cite{meshEvaluationNRR} (CBC3D also exhibits a runtime of approximately 5-15 minutes when meshing smaller geometries \cite{CBC3DYixun}). While CBC3D parallelizes the calculation of EDTs during adaptive refinement and computes the source and target points in parallel during deformation (achieving better runtime performance compared to its earlier versions \cite{CBC3DFedorov, CBC3DYixun}), more work must be done to parallelize the method fully. A Structured Adaptive Mesh Refinement (SAMR) scheme \cite{SAMR} can improve the performance 
and the gradation of the sequential CBC3D method. 
SAMR initially discretizes the image domain with 
a uniform mesh, and then generates finer sub-meshes (components) 
near the areas where the physics are ``changing.'' SAMR uses an independent Partial Differential Equation 
(PDE) solver in each mesh component and assesses the validity of the numerical results. 
If the numerical solution satisfies the analysis requirements, then the refinement 
stops; otherwise, SAMR discretizes the mesh with a finer resolution until 
it obtains an acceptable solution. Each mesh component has its own solution vector, which is 
computed independently from the solution of the other mesh components.
This is important for the parallelization of the SAMR scheme, where each mesh component 
is associated with a single core/node, and the under-utilized cores/nodes (e.g., those that compute the 
PDE solution in a smaller number of cells or mesh points) can automatically request additional work 
from the remaining cores/nodes.

A previous multi-tissue version of CBC3D \cite{CBC3D2015} is available as a stand-alone ITK library and is integrated \cite{Drakopoulos15T} within an interactive simulator for neurosurgical procedures involving brain Arteriovenous Malformations (AVM) developed in SOFA, a framework for real-time medical simulations \cite{faure2012sofa}. A single-material version is available within the 3D Slicer package for visualization and image analysis. As mentioned previously, the current version of the code utilizes shared-memory processing cores during the smoothing component of its algorithm -- this is outside the scope of this paper.  For more information on integration and parallelization, see \cite{FotisDissertation}.

\section{Conclusion} \label{conclusion}
An adaptive, image-to-mesh conversion method called CBC3D is presented. This method builds upon previous work \cite{CBC3DFedorov, CBC3DYixun, CBC3D2015}, maintaining the ability to generate meshes of good quality that were shown to enhance non-rigid registration solver performance and reduce error when compared to other image-to-mesh conversion methods \cite{meshEvaluationNRR}.
CBC3D initially discretizes a segmented, labeled image with a uniform BCC lattice of high-quality tetrahedra. It then employs red-green templates to subdivide the lattice near the segmented boundaries while ensuring mesh conformity (i.e., manifold connectivity between mesh regions representing different tissues). Finally, the generated surfaces are deformed to their corresponding tissue boundaries to improve fidelity while maintaining quality. 

CBC3D offers several attributes 
: (i) it provides accurate geometric and topologic representation for the cases presented in our evaluation,
(ii) it provides material-dependent mesh resolution to reduce element count,
(iii) it maintains good element quality during mesh deformation,
(iv) it reduces subsequent memory and CPU requirements for the solver by introducing mixed elements and further reduces the size of the mesh,
and (v) it improves the overall reliability and portability of the code as it builds 
upon the ITK open-source, cross-platform system. The CBC3D software is available as (i) a stand-alone ITK library (previous multi-tissue version based on \cite{CBC3D2015}), (ii) a 3D Slicer\footnote{https://www.slicer.org} 
extension (single-tissue version), and (iii) a SOFA plugin within an interactive simulator (multi-tissue version with limited features) for neurosurgical procedures involving 
brain Arteriovenous Malformations (AVM).

The present method is compared with four image-to-mesh conversion codes commonly used in industry and academia: 
CGAL's 3D Mesh Engine (v4.5.2), CLEAVER \cite{LatticeCleaving} (v1.5.4),
Lattice-Derefinement (LD) \cite{LD_Chernikov}, and PODM \cite{foteinos2014high}.
Previous work on lattice-based meshing \cite{LD_Chernikov} show that quality, fidelity, and size criteria conflict with one another, as it is challenging to balance all three satisfactorily. Our evaluation results indicate that the CBC3D meshes (i) exhibit high fidelity, 
(ii) keep the element count reasonably low, and (iii) exhibit good element quality.

\section{Acknowledgments}
This work is in part funded by NSF grant no. CCF-1439079, the NASA Transformational Tools and Technologies Project (NNX15AU39A) of the Transformative Aeronautics Concepts Program under the Aeronautics Research Mission
Directorate, Richard T. Cheng Endowment, the Modeling and Simulation fellowship of Old Dominion University, and the National Institute of General Medical Sciences of the National Institutes of Health under Award Number 1T32GM140911-03. This paper describes objective technical results and analysis. Any subjective views or opinions expressed in the paper do not necessarily represent the views of the National Institutes of Health or the United States Government. 

\bibliography{references}

\end{document}